%% file: topology_revised.tex
\documentclass{aastex63}

\usepackage{amsmath}

\submitjournal{ApJ}

\shorttitle{3D Cloud Structure}
\shortauthors{Zucker et al.}

\begin{document}

\title{On the Three-Dimensional Structure of Local Molecular Clouds}

\correspondingauthor{Catherine Zucker}
\email{catherine.zucker@cfa.harvard.edu}

\author[0000-0002-2250-730X]{Catherine Zucker}
\affiliation{Center for Astrophysics $\mid$ Harvard \& Smithsonian, 
60 Garden St., Cambridge, MA, USA 02138}

\author[0000-0003-1312-0477]{Alyssa Goodman}
\affiliation{Center for Astrophysics $\mid$ Harvard \& Smithsonian, 
60 Garden St., Cambridge, MA, USA 02138}

\author[0000-0002-4355-0921]{Jo\~{a}o Alves}
\affiliation{University of Vienna, Department of Astrophysics, T{\"u}rkenschanzstra{\ss}e 17, 1180 Vienna, Austria}

\author[0000-0002-0404-003X]{Shmuel Bialy}
\affiliation{Center for Astrophysics $\mid$ Harvard \& Smithsonian, 
60 Garden St., Cambridge, MA, USA 02138}

\author[0000-0001-9605-780X]{Eric W. Koch}
\affiliation{University of Alberta, Department of Physics, 4-183 CCIS, Edmonton AB T6G 2E1, Canada}
\affiliation{Center for Astrophysics $\mid$ Harvard \& Smithsonian, 
60 Garden St., Cambridge, MA, USA 02138}

\author[0000-0003-2573-9832]{Joshua S. Speagle}
\affiliation{Department of Statistical Sciences, University of Toronto, Toronto, ON M5S 3G3, Canada}
\affiliation{David A. Dunlap Department of Astronomy \& Astrophysics, University of Toronto, Toronto, ON M5S 3H4, Canada}
\affiliation{Dunlap Institute for Astronomy \& Astrophysics, University of Toronto, Toronto, ON M5S 3H4, Canada}
\affiliation{Center for Astrophysics $\mid$ Harvard \& Smithsonian, 
60 Garden St., Cambridge, MA, USA 02138}

\author[0000-0002-6747-2745]{Michael M. Foley}
\affiliation{Center for Astrophysics $\mid$ Harvard \& Smithsonian, 
60 Garden St., Cambridge, MA, USA 02138}

\author[0000-0003-2808-275X]{Douglas Finkbeiner}
\affiliation{Center for Astrophysics $\mid$ Harvard \& Smithsonian, 
60 Garden St., Cambridge, MA, USA 02138}

\author[0000-0002-1640-6772]{Reimar Leike}
\affiliation{Max Planck Institute for Astrophysics, Karl-Schwarzschildstra\ss e 1, 85748 Garching, Germany}
\affiliation{Ludwig-Maximilians-Universit\"at, Geschwister-Scholl Platz 1, 80539 Munich, Germany}

\author[0000-0001-5246-1624]{Torsten En${\ss}$lin}
\affiliation{Max Planck Institute for Astrophysics, Karl-Schwarzschildstra\ss e 1, 85748 Garching, Germany}
\affiliation{Ludwig-Maximilians-Universit\"at, Geschwister-Scholl Platz 1, 80539 Munich, Germany}

\author[0000-0003-4797-7030]{Joshua E.G. Peek}
\affiliation{Space Telescope Science Institute, 3700 San Martin Dr, Baltimore, MD 21218, USA}
\affiliation{Department of Physics \& Astronomy, Johns Hopkins University, Baltimore, MD 21218, USA}

\author[0000-0003-3122-4894]{Gordian Edenhofer}
\affiliation{Max Planck Institute for Astrophysics, Karl-Schwarzschildstra\ss e 1, 85748 Garching, Germany}
\affiliation{Ludwig-Maximilians-Universit\"at, Geschwister-Scholl Platz 1, 80539 Munich, Germany}



\begin{abstract}
We leverage the 1 pc spatial resolution of the \citet{Leike_2020} 3D dust map to characterize the three-dimensional structure of nearby molecular clouds ($d \lesssim 400$ pc). We start by ``skeletonizing" the clouds in 3D volume density space to determine their ``spines," which we project on the sky to constrain cloud distances with $\approx 1\%$ uncertainty. For each cloud, we determine an average radial volume density profile around its 3D spine and fit the profiles using Gaussian and Plummer functions. The radial volume density profiles are well-described by a two-component Gaussian function, consistent with clouds having broad, lower-density outer envelopes and narrow, higher-density inner layers. The ratio of the outer to inner envelope widths is $\approx 3:1$. We hypothesize that these two components may be tracing a transition between atomic and diffuse molecular gas or between the unstable and cold neutral medium.  Plummer-like models can also provide a good fit, with molecular clouds exhibiting shallow power-law wings with density, $n$, falling off like $n^{-2}$ at large radii. Using Bayesian model selection, we find that parameterizing the clouds' profiles using a single Gaussian is disfavored. We compare our results with 2D dust extinction maps, finding that the 3D dust recovers the total cloud mass from integrated approaches with fidelity, deviating only at higher levels of extinction ($A_V \gtrsim 2 - 3$ mag). The 3D cloud structure described here will enable comparisons with synthetic clouds generated in simulations, offering unprecedented insight into the origins and fates of molecular clouds in the interstellar medium. 
\end{abstract}


\section{Introduction}
Much of our understanding of the structure and properties of molecular clouds is obtained via either two-dimensional (2D) observations of dust in emission or extinction \citep[see][]{Andre_2010, Lombardi_2009, Planck_2011} or three-dimensional (3D) spectral-line observations of gas \citep[see][]{Dame_2001}, where the third dimension is not distance, but radial velocity. Until very recently, any insights into the true 3D spatial structure of molecular clouds were available only in numerical simulations, and were limited by the extent to which these simulations could approximate the observed properties of real clouds. Even then, detailed comparisons between observations and simulations could, again, only be made either in 2D projected space, or via spectral line maps, where the correspondence of intensity features in \textit{position-position-velocity (p-p-v)} space to ``real" density features in \textit{position-position-position (p-p-p)} space has been shown to be fraught with complications \citep{Beaumont_2013}. 

 Interstellar dust reddens starlight, so by \textit{simultaneously} modeling the distance and extinction to large numbers of stars, it is possible to infer a self-consistent 3D distribution of stars and dust that produces the observed distribution of stellar colors.  Prior to the advent of \textit{Gaia} \citep{Gaia_2018} in 2018, accurate (e.g. astrometric) measurements of stars' distances independent from measures of their colors were scarce, so  transforming the 2D sky into a 3D physical picture of the distribution of stars and  clouds that form them was extraordinarily challenging  statistically. Several path-breaking 3D dust maps were created prior to \textit{Gaia} \citep[][]{Marshall_2006, Sale_2014, Green_2015}, but  \textit{Gaia}'s astrometric constraints on stellar distances have allowed for significant gains in the spatial resolution of 3D dust maps.   And the fidelity of 3D maps has improved even more as a variety of photometric catalogs, which offer billions of stellar color measurements needed to constrain extinction in fits,  have been crossmatched with \textit{Gaia}, \citep[see e.g.][]{Green_2019, Lallement_2019, Rezaei_2018, Chen_2019, Hottier_2020, Leike_2019}. Customization of the 3D dust mapping framework, where distance resolution can be enhanced at the expense of spatial resolution, has yielded highly accurate distance measurements for nearby molecular clouds, with typical distance uncertainties of $\approx 5\%$ \citep{Zucker_2019, Zucker_2020, Chen_2020, Yan_2019}, a factor of five improvement over typical distance uncertainties in the pre-\textit{Gaia} era. 

While these new catalogs of molecular cloud distances have shown good agreement with maser distances \citep{Zucker_2020}, revealed new Galactic-scale features \citep[e.g.\ the \textit{Radcliffe Wave;}][]{Alves_2020}, and even mapped distance gradients \textit{within} an individual molecular cloud \citep{Zucker_2019, Zucker_2018a}, up until now, catalogs have not had the resolution to resolve the detailed 3D structure (e.g. thickness, orientation) of clouds on scales comparable to numerical simulations. Most existing approaches group batches of stars into pixels \textit{on the sky}, and then fit stellar distance and extinction measurements in each pixel to obtain the distribution of dust along the line of sight. A 3D map can then be created by combining 1D line-of-sight information for all 2D sky pixels. However, the line-of-sight dust distributions suffer from a distortion effect, where the angular resolution on the sky is always superior to the distance resolution along the line of sight, resulting in dust clouds appearing to  extend radially outward from the Sun in an unphysical manner. 

Recently \citet{Leike_2020} presented a new 3D dust map of the solar neighborhood out to a distance of $\approx 400$ pc with a distance resolution of 1 pc. \citet{Leike_2020} overcome the ``fingers of god" effect by directly modeling the dust density in 3D \textit{x-y-z} cartesian space. Their method combines metric Gaussian variational inference \citep{Knollmuller_2019} with Gaussian Processes \citep[e.g.][]{Rasmussen_2006} in the context of information field theory \citep[see][]{Ensslin_2009, Ensslin_2019, Arras_2019, Ensslin_2011}. The integrated extinction for a star known to lie in a given 3D voxel with some uncertainty provides information on the integrated dust density along that given line of sight, acting as an ``anchor" point that can constrain the smooth 3D distribution of dust. By inferring the spatial correlation of the dust on different scales, one can connect these anchor points, and infer the dust distribution at arbitrary points in space, including along lines of sight where no stars are observed. Specifically, \citet{Leike_2020} model the dust as a log-normal Gaussian process, while \textit{simultaneously} inferring the correlation kernel of the process, corresponding to the physical spatial correlation power spectrum of the dust. In this way, \citet{Leike_2020} use information on the distance and extinction to 5 million stars in the vicinity of the Sun \citep[obtained from the Starhorse algorithm; see][]{Anders_2019} to reconstruct the 3D spatial distribution of dust on a fixed cartesian grid with a spatial resolution of 1 pc. The  \citet{Leike_2020} results offer the first clear 3D spatial view of the local interstellar medium.\footnote{The methodology is similar to that summarized in \citet{Rezaei_2017, Rezaei_2018} and implemented in \citet{Rezaei_2020} to obtain a 3D cartesian view of the Orion A molecular cloud at a spatial resolution of approximately 10 pc. We will compare the \citet{Leike_2020} results to the \citet{Rezaei_2020} results in Orion A in \S \ref{exposition}.}

In this work, we utilize the high spatial resolution of the \citet{Leike_2020} 3D dust map to analyze the 3D spatial structure and thicknesses of famous nearby star forming regions for the first time. Specifically, we extend two existing algorithms geared towards 2D data --- \texttt{FilFinder} \citep{Koch_2015} and \texttt{RadFil} \citep{Zucker_2018b} --- into 3D. We apply the \texttt{FilFinder3D} algorithm to determine the topological skeletons of nearby clouds. We then use \texttt{RadFil3D} to determine their average radial volume density profiles, and characterize the shapes of these profiles via both Gaussian and Plummer fitting. We then validate our results by creating synthetic 2D dust extinction maps from the 3D density distributions, and comparing these maps to independent results \citep[derived from the NICEST algorithm;][]{Lombardi_2009, Lombardi_2011} often used in star formation studies of nearby molecular clouds. 

In \S \ref{data}, we present the datasets used in this work, including the \citet{Leike_2020} dust map and the NICEST 2D extinction maps. In \S \ref{methods}, we detail the methodology we use to derive the topological spines of the clouds, and to construct radial volume density profiles for each region. We also discuss the derivation of our 2D extinction maps. In \S \ref{results}, we present our results in the context of a custom 3D visualization environment, including the \textit{x-y-z-l-b-d} values (the Heliocentric Galactic Cartesian coordinates, the Galactic coordinates, and the distance) for each point in the spine of every cloud, and their thicknesses derived from the radial profile fitting. We also compare the 2D dust extinction maps derived from the projected \citet{Leike_2020} 3D dust distribution with NICEST 2D extinction maps, compute cloud masses from each method, and quantify the dynamic range in density that the \citet{Leike_2020} dust map is able to probe.  In \S \ref{discussion}, we discuss the implications of our results, including potential evidence of a phase transition in our radial density profiles. Finally, we conclude in \S \ref{conclusion}. 

\section{Data} \label{data}

\subsection{\citet{Leike_2020} 3D Dust Map} \label{density}
\citet{Leike_2020} compute the 3D dust distribution on a cartesian grid centered on the Sun, extending $x$ = [-370 pc, 370 pc], $y$ = [-370 pc, 370 pc], and $z$ = [-270 pc, 270 pc], where $x$, $y$, and $z$ are the Heliocentric Galactic Cartesian coordinates. As such, the distance range of the map varies from $d=370$ pc in the midplane, up to $d\approx$ 590 pc at the corners of the grid. This distance range encompasses about a dozen well-studied nearby star-forming regions, and in this work we target the Chamaeleon, Ophiuchus, Orion B, Orion A, $\lambda$ Orionis, Lupus, Taurus, Perseus, Musca, Pipe, and Cepheus molecular clouds for our radial profile analyses. We also provide the topological skeleton for one additional cloud --- Corona Australis --- but do not fit for its width, since it does not meet the density threshold required for inclusion in our study (see \S \ref{filfinder}). We emphasize that three clouds -- Orion A, Orion B, and $\lambda$ Orionis -- lie at the very edge of the \citet{Leike_2020} grid, and the results for these clouds are subject to additional scrutiny, as discussed further in \S \ref{results}.

\subsubsection{Converting to Gas Volume Density}
The \citet{Leike_2020} 3D dust map is provided in terms of the differential extinction, specifically the optical depth in the \textit{Gaia} G-band,  $\tau_G$,  per one parsec. \citet{Leike_2020} generate twelves samples of their 3D dust map, and we adopt the mean of the samples in our main analysis. To ensure that uncertainties in the underlying dust reconstruction are not affecting our results, we  will additionally repeat our radial profile analysis using all twelve realizations of the dust map, as discussed further in the Appendix.\footnote{For more details, see the public release of the \citet{Leike_2020} map available on Zenodo \citep{leike_zenodo}. The mean dust map is available in the \texttt{mean\_std.h5} file, while the twelve samples are available in \texttt{samples.h5}.}

There are several steps needed to convert the data values of the mean map --- referred to as the extinction density $s_{x}$ in \citet{Leike_2020} --- to gas volume density. First, we use the relationship between $\frac{A_\lambda}{N_{\rm H}}$ (the ratio of the extinction $A$ at some wavelength $\lambda$ over the hydrogen column density $N_{\rm H}$) and inverse wavelength from \citet[][see Figure 1]{Draine_2009}. Assuming a mean wavelength for the \textit{Gaia} G-band of 673 nm \citep[$\lambda^{-1} = 1.49 \; \micron^{-1}$;][]{Jordi_2010}, we obtain $\frac{A_G}{N_{\rm H}} =  4\times 10^{-22} \rm \; cm^{2} \; mag$ from \citet{Draine_2009}.\footnote{\citet{Draine_2009} is assuming an $R_V$ = 3.1. Thus, variation in $R_V$ could induce changes in our adopted conversion to gas volume density.} Given the known relationship between extinction $A_{\rm G}$ and optical depth ($A_G = 1.086 \times \tau_G$) we obtain $N_{\rm H} = 2.71 \times 10^{21} \; \tau_G \; \rm cm^{-2}$. Finally,  given that the \citet{Leike_2020} map is provided in the form of $\frac{\tau_G}{1 \; \rm{pc}}$ (rather than the usual $\tau_G$), we obtain $n_{\rm H} = 880 \; {\rm cm^{-3}} \times s_x$. Thus, to convert the \citet{Leike_2020} 3D dust map to volume density of hydrogen, we simply multiply the entire volume by a factor of 880. We emphasize that this volume density is the total volume density of hydrogen nuclei, independent of phase ($n = n_{\rm HI} + 2n_{\rm H_2}$) and is used to derive all results reported in \S \ref{results}. Volume density cubes for each cloud, with WCS information in a cartesian reference frame, are available on the Harvard Dataverse (see \url{https://doi.org/10.7910/DVN/IADP7W}).

\subsubsection{Systematic Effects} \label{systematics}

While \citet{Leike_2020} provide their results on a 1 pc grid, they argue that the functional resolution of the map is 2 pc, and only features which manifest at spatial scales greater than 2 pc should be considered reliable. Accordingly, for the main analysis, we  exclude the inner 2 pc of each dust cloud's radial volume density profile when inferring their structure, while reporting results including the inner 2 pc in the Appendix. 

By performing tests on synthetic mock data, \citet{Leike_2019, Leike_2020} also find that there are a small number of outliers in their 3D dust distribution, in which the differential dust extinction values in certain voxels is significantly larger than predicted, to a degree that cannot be captured by the reported uncertainties. While \citet{Leike_2019} and \citet{Leike_2020} use different underlying stellar data for their statistical reconstruction, the methodology is similar, and we find the same high-differential-extinction outliers in the newer \citet{Leike_2020} map. In this analysis, we will be averaging over many voxels and will find that our results are insensitive to these outliers. 

We emphasize that the \textit{3D volume densities reported here are likely lower limits}. We observe inhomogeneities in column density maps of nearby molecular clouds at spatial scales smaller than 2 pc, the effective spatial resolution of the \citet{Leike_2020} 3D dust map \citep[see e.g.][]{Zari_2016}. Thus, much of the dense gas in these clouds is unresolved, not just due to the map's spatial resolution, but due to the inability of the 3D dust mapping technique to probe high density gas, stemming from the lack of stellar data constraining the 3D dust reconstruction in this regime. 

Finally, one additional consideration is the effect that YSOs have on the 3D dust reconstruction in \citet{Leike_2020} and whether this would bias any 3D volume density estimates. The stellar catalog underpinning \citet{Leike_2020} is based on the StarHorse algorithm, presented in \citet{Anders_2019}. \citet{Anders_2019} does not employ any pre-main sequence models when determining stellar properties. Thus, as argued in \citet{Anders_2019}, any young stars detected with \textit{Gaia} and included in the StarHorse sample will have poor fits, triggering a reliability flag. Since \citet{Leike_2020} only use high-quality stellar data with no questionable reliability flags set, there should be very little contamination from YSOs in the 3D dust map.

\subsection{2D NICEST Extinction Maps} \label{nicest}
To validate our 3D cloud structure results (see \S \ref{methods}) and quantify any ``missing mass" in the 3D maps, we compute 2D maps of the $K$ band extinction, $A_K$, using the NICEST algorithm \citep[][A Near-Infrared Color Excess Method Tailored for Small-Scale Structures]{Lombardi_2009} for each cloud. \citet{Goodman_2009} showed that dust extinction estimates based on the near-infrared color excess are relatively free from bias and provide more robust estimates of the column density, at higher dynamic range, than gas-based approaches. NICEST, and its predecessor NICER \citep{Lombardi_2001}, rely on the fact that the difference between the observed and intrinsic colors of a star (background to a cloud) provide information on the cloud's extinction along the line of sight to that star. By measuring the unreddened colors of stars in a nearby, low extinction control field, it is possible to model the color excess on a star-by-star basis. Because there is a small dispersion in the intrinsic near-infrared colors of stars (and photometric errors), the technique requires averaging the extinction estimates for many individual stars in batches across the cloud, and pixel-by-pixel extinction estimates are computed as spatial averages via the adoption of a Gaussian smoothing kernel. Compared to NICER, NICEST removes contamination by foreground stars, which by their nature do not probe the cloud's extinction, and better characterizes small-scale inhomogeneities (i.e. filaments) present in the dust. 

Like \citet{Lombardi_2011}, to compute the extinction, we leverage the near-infrared $J$, $H$, and $K$ band colors of millions of stars detected in 2MASS \citep{Skrutskie_2006} for each field. Using the 2MASS photometry, NICEST creates two sets of stellar samples: main sequence stars and YSOs, defined using their respective \textit{K}-band magnitude distributions. In practice, given a star, the algorithm computes the intrinsic \textit{K}-band magnitude (i.e. removes the extinction) and compares this with the \textit{K}-band number counts of field stars (typically an exponential distribution) and YSOs (typically a Gaussian-like curve). This comparison is used to assign to each star a probability of being a YSO. The functionality to produce custom cloud-by-cloud NICEST extinction maps is available online \citep{Lombardi_2001, Lombardi_2009}\footnote{See \href{http://interstellarclouds.fisica.unimi.it/html/index.html}{http://interstellarclouds.fisica.unimi.it/html/index.html}} and we adopt the default map making and cluster parameters. The default option is to include all stars in the extinction calculation, including YSOs. To check that YSOs are not affecting our mass calculations, we compute two masses for Orion A both with and without including YSOs in the sample. We find that including the YSOs in the extinction calculation only increases the mass by 2\%. Since Orion A contains the largest number of YSOs in the solar neighborhood, our decision to compute the extinction maps using all stars should have no effect on our results. The configuration files used to produce the NICEST maps are available online at the Harvard Dataverse; the NICEST maps for all clouds targeted in this study are likewise available online (see \url{https://doi.org/10.7910/DVN/WONVUH} for both). 

\section{Methods} \label{methods}

\subsection{FilFinder 3D} \label{filfinder}
To determine the topological skeletons of nearby molecular clouds, we employ a modified 3D version of the FilFinder package from \citet{Koch_2015}. \texttt{FilFinder} was originally developed in 2D to constrain the filamentary structure in $\rm H_2$ column density maps obtained as part of the Herschel Gould Belt Survey \citep{Andre_2010}. \texttt{FilFinder3D} uses similar morphological operations to the existing 2D version, but now employs the \texttt{skan} package for efficient operations on large morphological skeletons \citep{NunezIglesias_2018_skan}. A full description of the 3D algorithm will be presented in Koch et al. (2021, in preparation).

To compute the skeletons, we start by defining the same density threshold for all clouds, equivalent to $n=35\; \rm cm^{-3}$, based on the volume density distribution computed in \S \ref{density}.\footnote{Corona Australis has no significant extended density features above a level of $n = 35 \; \rm cm^{-3}$, so we skeletonize it at a level of $n = 5\; \rm cm^{-3}$ to provide distance information. It is included in Table \ref{tab:skeletonization} but not in Tables \ref{tab:radial_profiles_excl2} and \ref{tab:mass}.} Hierarchical structure finding algorithms for feature identification do exist for 3D volumes, the most common being dendrograms \citep{Rosolowsky_2008}, which are frequently applied to spectral-line cubes of CO emission in \textit{position-position-velocity} space \citep{Rice_2016}. However, it is well known that dendrograms are not able to capture filamentary structure, with rarely any structures exceeding an aspect ratio of $2:1$ \citep[see][]{Rice_2016}. Since we find that nearby star-forming regions are indeed filamentary, even at lower densities, we forgo dendrograms for this analysis and defer to future work. Before settling on a fixed threshold of $n = \rm 35 \; cm\; ^{-3}$, we test a variety of thresholds, ranging from \rm $20 - 50 \; \rm cm^{-3}$. Thresholds significantly higher than $n = \rm 35 \; cm\; ^{-3}$ were too patchy to skeletonize, while thresholds significantly lower than $n = \rm 35 \; cm\; ^{-3}$ meant that the peak radial densities were often offset from the main spine.\footnote{In previous analyses based on 2D dust emission maps \citep{Zucker_2018c}, we implemented a ``shift" option in the \texttt{RadFil} package to account for dense clumps offset from the main spine, since \texttt{FilFinder} does not take into the density range \textit{inside} each mask when defining the spine. The ``shift" option corrects for this by shifting each cut so that its peak density is centered at a radial distance of 0 pc. However, we forgo shifting in this work due to the small handful of noisy outliers at very high differential extinction, which would bias the density results at small radii.} Our choice of density threshold is validated in \S \ref{results}, since the spines defined in 3D tend to align very well with the spines of filaments seen in 2D dust emission from Planck \citep{Planck_2011}, and the radial density profiles peak near a radial distance of zero. Overall we find that all clouds are well-defined at $n = \rm 35 \; cm\; ^{-3} $ and we can produce the most unbiased widths possible by not tweaking feature identification on a cloud-by-cloud basis. 

After defining this initial $n = \rm 35 \; cm\; ^{-3} $ threshold, we morphologically close the 3D mask by performing a dilation, followed by an erosion. This procedure removes small voids in the mask, rendering it suitable for skeletonization.\footnote{More details on mathematical morphology, including the erosion, dilation, and closing of features in 2D or 3D space, can be found in the \href{https://scikit-image.org/}{scikit-image} documentation, which we use to perform these operations \citep{scikit_image}.} We then skeletonize the 3D mask using the Medial Axis Transform, convert the skeleton to a graph,  and run a longest path algorithm on the resulting skeleton to obtain the main trunk of each density feature. While we only use the spine computed from the longest path algorithm to derive the widths, we also preserve and report the full skeleton without pruning applied, in order to retain distance information on all branches. In this scheme, nearby molecular clouds can either be defined by a single, or multiple, trunks, depending on whether they are connected in density space above a threshold of $n=35\; \rm cm^{-3}$. Thus, this skeletonization process is performed between one and four times, depending on how many isolated 3D features there are after performing the initial thresholding. For clouds like Perseus, which manifest as a single feature in density space, we identify only a single spine. However, clouds like Taurus are actually composed of several distinct features in density space. The skeletonization and longest path algorithms are individually run on each isolated mask, to produce a set of trunks defining the cloud.\footnote{The two exceptions are the Musca filament and the Pipe nebula. We find that Musca is connected to the Chamaeleon cloud in 3D density space at a level of $n=35\; \rm cm^{-3}$. The same is true for the Pipe nebula, which is connected to Ophiuchus. Because these clouds have always been treated as individual entities in the literature, and obtaining widths for the Pipe nebula and Musca (as opposed to the Pipe-Ophiuchus, and Musca-Chamaeleon complexes) is beneficial for comparison with prior work, we mask out the Pipe nebula when obtaining widths for Ophiuchus, and the Musca cloud when obtaining widths for Chamaeleon (and vice versa). In reality, the complexes are connected by only very thin tendrils, so this procedure does not affect the morphology of the derived masks, or the resulting widths.}

Finally, we emphasize again that the \texttt{FilFinder} algorithm is designed to work on elongated, \textit{filamentary} structures, with aspect ratios of at least $3:1$. We find that the vast majority of our masks tend to be elongated with aspect ratios of at least $3:1$, but there are a few sub-features in clouds which are better described as ellipsoids than filaments, some of which may be unresolved in the \citet{Leike_2020} map. Ellipsoids do not dominate the set of density features, either across the sample, or within an individual cloud, so this has a negligible effect on the results. The volume density distributions (\url{https://doi.org/10.7910/DVN/IADP7W}), 3D masks (\url{https://doi.org/10.7910/DVN/NMRLP1}), and 3D spines (\url{https://doi.org/10.7910/DVN/YQYBRD}) are available online with full WCS information at the Harvard Dataverse. We refer interested readers there to explore our masking and skeletonization results on a cloud-by-cloud basis in more detail.

\subsection{RadFil 3D}

\subsubsection{Building Profiles} \label{profile_build}
To characterize the thicknesses of local clouds, we employ the radial profile builder \texttt{RadFil} \citep{Zucker_2018b}. \texttt{RadFil} was designed in 2D to build radial column density profiles using filament spines determined by the \texttt{FilFinder} package \citep{Zucker_2018b, Koch_2015, Zucker_2018c}. \texttt{RadFil} builds profiles by taking radial cuts across the spine. The same logic for constructing a radial profile in 2D can easily be extended into 3D. We take the 3D skeleton returned by \texttt{FilFinder3D}, and fit a B-spline to obtain a smoothed continuous version of the spine and the first derivative along its length. Using functionality within the Python package \texttt{PyVista} \citep{Sullivan_2019}, we generate a slice through the volume density cube at each point along the spine given the normal vector obtained from the first derivative. Then, for each point in the slice, we determine its radial distance from the spine point and its volume density via linear interpolation. On a slice by slice basis, we compute the median density in forty radial distance bins, spaced half a parsec apart, out to 20 pc. Given the full set of profiles, an average profile is obtained by computing the median volume density in each radial distance bin. Before taking the average, we remove the small minority of profiles in which the peak radial density is less than the density threshold used to define the mask ($n=35\; \rm cm^{-3}$).  For clouds composed of multiple isolated density features, the average is taken using slices across all density features, to compute a representative width for each cloud.

This averaged radial profile is used to perform the Gaussian and Plummer fitting described in \S \ref{fitting}. Because of the 1 pc voxel size of the 3D dust map, and the fact that \citet{Leike_2020} recommend that scales less than 2 pc are suspect, we remove the inner 2 pc of each averaged profile when fitting the profiles. However, in the Appendix, we report results for the full profile fits (including the inner 2 pc) and find that our choice to remove radial distances between 0 and 2 pc does not significantly affect our results.

\subsubsection{Fitting Profiles} \label{fitting}
We fit three functions to the averaged radial volume density profile computed for each cloud in \S \ref{profile_build} --- a single-component Gaussian, a two-component Gaussian, and a Plummer function. One important consideration is the modeling of the scatter in the averaged profile. In our case, the scatter in the volume density within a bin at fixed radial distance is not a true measurement error, but rather encompasses real physical variation in the radial density profile taken at different slices along the spine of a cloud. For this reason, we choose to model the scatter on the density as part of our fitting procedure, introducing it as an additional free parameter in all three of our models. We perform our inference in a Bayesian framework, adopting a simple Gaussian likelihood \citep[see e.g.][]{Hogg_2010} and flat priors on all our parameters. None of our parameters are prior-dominated, so this choice of flat prior has no effect on our results. For computational expediency, we minimize the negative log-likelihood, in lieu of maximizing the likelihood. Our Gaussian log-likelihood is of the form: 

\begin{equation}
\ln\, p(n \big| \theta, \sigma_n^2) = - \frac{1}{2}   \sum_i \left[ \frac{[n_i - n_{\theta,i}]^{2}}{\sigma_n^2} + \ln(2\pi \sigma_n^{2})  \right]
\end{equation}

\noindent where $n_i$ is the measured density in the $i$th radial distance bin, $n_{\theta,i}$ is the modeled density in the $i$th radial distance bin (defined by the free parameters $\theta$), and $\sigma_n^2$ is the square of the scatter in the density across all bins that we also infer as part of our modeling.

For the single-component Gaussian fit, our model for the volume density of hydrogen nuclei $n$ as a function of radial distance $r$ is of the form:

\begin{equation}
n(r)=a \,\exp{\left(\frac{{-r}^2}{2\sigma^2}\right)}
\end{equation}

\noindent Our free parameters $\theta$ are the amplitude ($a$) and standard deviation ($\sigma$). The mean of the Gaussian is fixed to zero. We adopt a flat prior between $0-80$ $\rm cm^{-3}$ for $a$, between $0-15$ pc for $\sigma$ and between $0-100$ $\rm cm^{-3}$ for $\sigma_n^2$.

To better model the significant tails of the distribution, we additionally fit a two-component Gaussian, again with the mean of each component fixed to zero: 

\begin{equation}
n(r)=a_1 \,\exp{\left(\frac{{-r}^2}{2\sigma_1^2}\right)} + a_2 \,\exp{\left(\frac{{-r}^2}{2\sigma_2^2}\right)}
\end{equation}

\noindent where the amplitudes ($a_1, a_2$) and their standard deviations ($\sigma_1, \sigma_2$) are the free parameters in the modeling. We adopt a flat prior between $20-80$ $\rm cm^{-3}$ for $a_1$, between $0 - 6$ pc for $\sigma_1$, between $2-20$ $\rm cm^{-3}$ for $a_2$, between $6 - 25$ pc for $\sigma_2$, and between $0-100$ $\rm cm^{-3}$ for $\sigma_n^2$. The offset in the prior range between $a_1$ and $a_2$ is adopted to avoid degeneracies in the modeling.  

Finally, we fit a Plummer function, which we parameterize as:

\begin{equation} \label{eq:plummer}
n(r) = \frac{n_0}{[{1+({\frac{r}{R_{\mathrm{flat}}})^2}]}^{\; \frac{p}{2}}}
\end{equation}
\noindent where our free parameters are $n_0$ (peak profile height), $R_\mathrm{flat}$ (the flattening radius), and $p$ (index of the density profile). Equation \ref{eq:plummer} is the same ubiquitous Plummer function applied to column density maps of smaller-scale filaments \citep[widths a few tenths of a parsec, lengths a few parsecs;][]{Arzoumanian_2019,Panopoulou_2017} from the Herschel Gould Belt Survey \citep{Andre_2010}, except we have transformed the model from column density space back to volume density space \citep[see Equation 1 of][]{Arzoumanian_2011}. We adopt a flat prior between $20-170$ $\rm cm^{-3}$ for $n_0$, between $0 - 15$ pc for $R_\mathrm{flat}$, between $0-6$ for $p$, and between $0-100$ $\rm cm^{-3}$ for $\sigma_n^2$.

We sample for all model parameters using the nested sampling code \texttt{dynesty} \citep{Speagle_2020}. We adopt the default parameters of the nested sampler, with a stopping criterion of \texttt{dlogz = 0.1}. We utilize the logarithm of the evidence $\ln(Z)$ extracted from the \texttt{dynesty} chains to compute a Bayes Factor for the purpose of model comparison, as discussed further in \S \ref{results}. 

\subsection{Creating 2D Dust Extinction Maps from 3D Dust Data} \label{leikenicest}
To compare the \citet{Leike_2020} 3D dust map with traditional 2D approaches (see \S \ref{nicest}), we create projected 2D dust extinction maps from the 3D volume density distributions used in the construction of the radial profiles. To do so, we utilize the \texttt{FitsOffAxisProjection} functionality from the Python package \texttt{yt} \citep{Turk_2011}. The \texttt{FITSOffAxisProjection} function integrates the volume density $n$ along a line of sight $\hat{l}$, where $\hat{l}$ is determined by the vector between the Sun and the median cartesian coordinates of the points in a cloud's skeleton, $(x_{\rm cen},y_{\rm cen}, z_{\rm cen})$. The resulting projection corresponds to the total hydrogen column density distribution $N$ = $N_{\rm HI} + 2N_{\rm H_2}$ (in $\rm{cm^{-2}}$) as seen from the Sun. In order to compare with the NICEST maps of the $K$ band extinction $A_{\rm K}$ (from \S \ref{nicest}), we convert from total hydrogen column density to $A_K$ by assuming a constant conversion factor from \citet{Lada_2009} of $\frac{N({\rm HI}) + 2N({\rm H_{\rm 2}})}{A_K} = 1.67 \times 10^{22}  \rm  \; cm^{-2} \; mag^{-1}$.

To assign sky coordinates to each projection, we convert from the skeleton's central coordinates $(x_{\rm cen},y_{\rm cen},z_{\rm cen})$ to sky coordinates $(l_{\rm cen},b_{\rm cen},d_{\rm cen})$, where $l_{\rm cen}$ and $b_{\rm cen}$ become the central Galactic Longitude and Latitude of the projected image on the plane of the sky, and $d_{\rm cen}$ is the distance between the Sun and the central point of the cloud's skeleton. The pixel scale of the image is computed by determining the angular size spanned by a 1 pc voxel (the voxel size of the original 3D dust grid) at the central distance of the cloud's skeleton $d_{\rm cen}$. These projected $A_K$ maps are available for download at the Harvard Dataverse (see \url{https://doi.org/10.7910/DVN/GXXKHD}).

\subsection{Mass Calculation} \label{mass}
To quantify the fraction of cloud mass being recovered with 3D dust mapping, we compute cloud masses using both traditional NICEST 2D dust extinction maps (from \S \ref{nicest}) and the projected $A_K$ maps we derive from the 3D dust in \S \ref{leikenicest}. In both cases, we adopt a consistent threshold of $A_K$=0.1 mag. This threshold is the same used in previous literature to calculate cloud masses from the NICEST 2D dust extinction maps \citep[see e.g.][]{Lada_2010}.\footnote{Like \citet{Lada_2010}, we also subtract a constant background of $A_K=0.15$ mag from the Pipe nebula, due to its position close to the plane and towards the Galactic center. Furthermore, because the 2D dust extinction maps based on \citet{Leike_2020} are computed using only dust local to the cloud, rather than integrated along the line of sight, we limit our comparison to latitudes $|b| > 5^\circ$. In reality, this latitude cut only masks out a small section of the Pipe and Lupus clouds.} The masses are calculated inside the cloud boundaries by assuming the same mass surface density relation as \citet{Zari_2016}. This relation is equivalent to $\frac{\Sigma_{\rm gas}}{A_K} = \mu m_p \beta_K = 183 \; M_\sun \; \rm pc^{-2} \; mag^{-1}$, where $\mu = 1.37$ is the mean molecular weight, $m_p = 1.67 \times 10^{-24}$ g is the mass of the proton, and $\beta_K$ is the same gas-to-dust ratio that we adopt above, equivalent to $1.67 \times 10^{22}  \rm  \; cm^{-2} \; mag^{-1}$. To ensure a fair comparison, the NICEST maps from \S \ref{nicest} are convolved to the same resolution as the \citet{Leike_2020} based extinction maps. In both cases, the same distance $d_{cen}$ --- the average distance to a cloud's skeleton --- is adopted. 

\input{table1}

\begin{figure}[h!]
\begin{center}
\includegraphics[width=1.0\columnwidth]{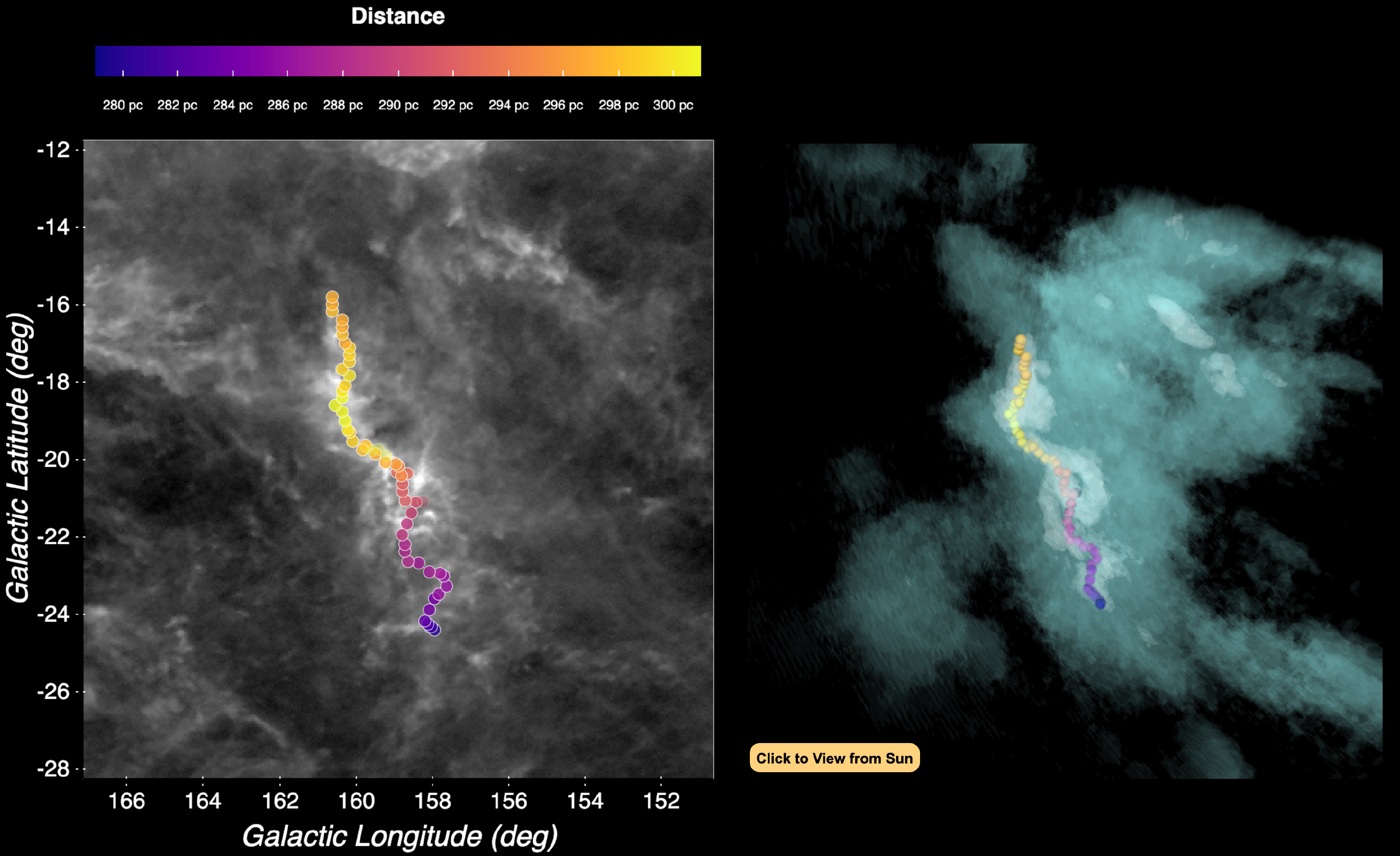}
\caption{Skeletonization results for the Perseus Molecular Cloud shown in 2D (left panel) and 3D (right panel). In both panels, the scatter points (colored by distance over the range $d=279-301$ pc) delineate the skeleton of the cloud, determined in 3D Cartesian space and projected back on the plane of the sky. In the 3D version of the figure we show the 3D volume density of the cloud $n$, where the cyan and white distributions indicate lower thresholds of $n = \rm 4 \; cm^{-3}$ and $n =  \rm 35 \; cm^{-3}$ respectively. In the plane of the sky version, we show the Planck reddening map \citep{Planck_2011}.  An interactive version of this figure is available at \href{https://faun.rc.fas.harvard.edu/czucker/Paper\_Figures/3D\_Cloud\_Topologies/perseus\_topology/perseus.html}{https://faun.rc.fas.harvard.edu/czucker/Paper\_Figures/3D\_Cloud\_Topologies/perseus\_topology/perseus.html} or in the online version of the published article. Comparable interactive figures for all clouds in the sample are available at \url{https://faun.rc.fas.harvard.edu/czucker/Paper_Figures/3D_Cloud_Topologies/gallery.html}. In the online version, the cyan and white boundaries (toggleable on/off) correspond to the FWHM of the inner and outer envelopes of the cloud, derived from the two-component Gaussian fitting results in \S \ref{results}. \label{fig:perseus_topology}}
\end{center}
\end{figure}

\begin{figure}[h!]
\begin{center}
\includegraphics[width=1.0\columnwidth]{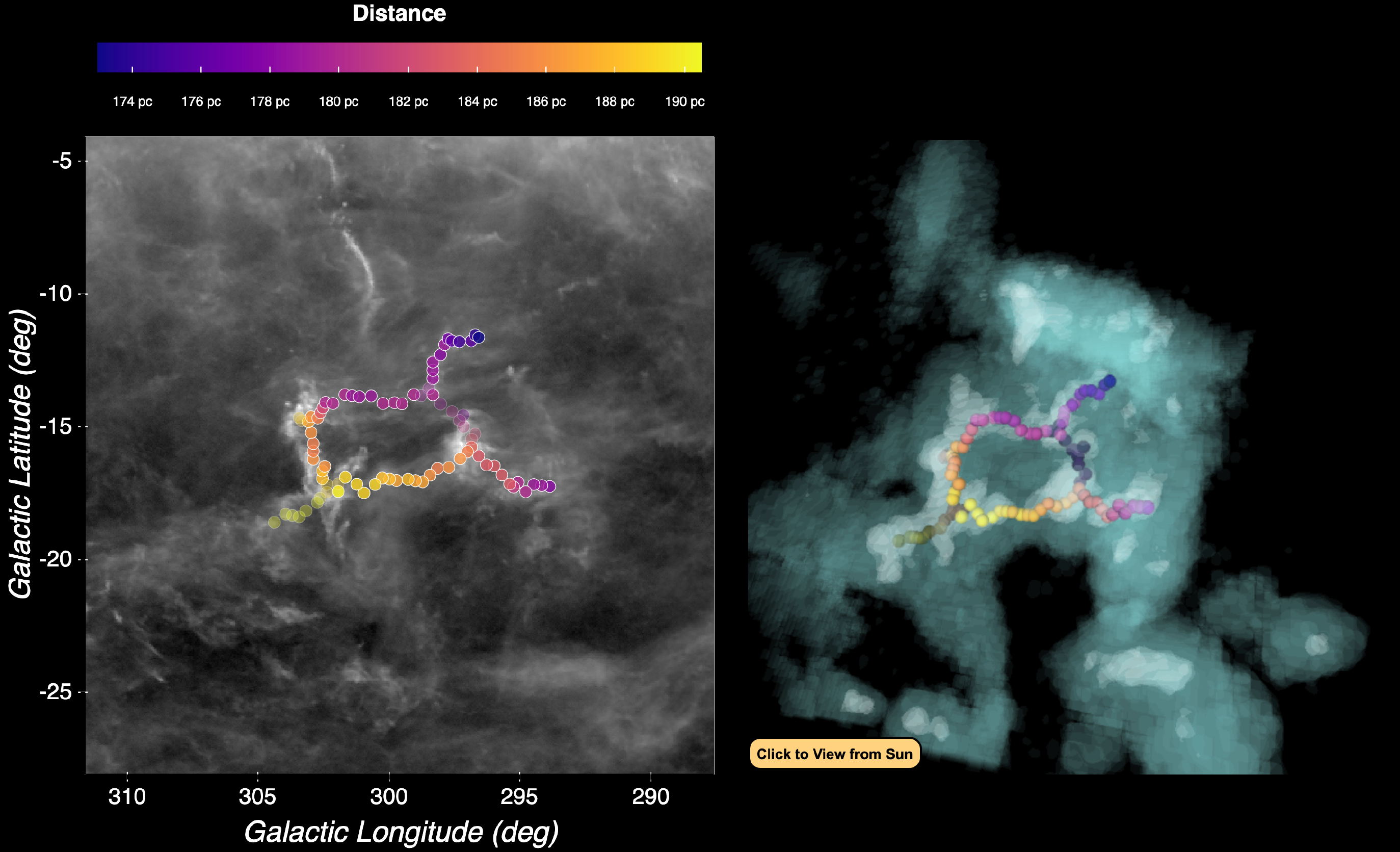}
\caption{Same as in Figure \ref{fig:perseus_topology} but for the Chamaeleon Molecular Cloud, with the spine (colored points) spanning a distance range of $d=173-190$ pc. Opaque spine points constitute the skeleton's main spine, while semi-transparent points are part of the full, unpruned skeleton. An interactive version of this figure is available at \href{https://faun.rc.fas.harvard.edu/czucker/Paper\_Figures/3D\_Cloud\_Topologies/chamaeleon\_topology/chamaeleon.html}{https://faun.rc.fas.harvard.edu/czucker/Paper\_Figures/3D\_Cloud\_Topologies/chamaeleon\_topology/chamaeleon.html} or in the online version of the published article. Comparable interactive figures for all clouds in the sample are available at \url{https://faun.rc.fas.harvard.edu/czucker/Paper_Figures/3D_Cloud_Topologies/gallery.html}.
\label{fig:chamaeleon_topology}}
\end{center}
\end{figure}

\section{Results} \label{results}

\subsection{Skeletonization} \label{skeletonization_results}
In Table \ref{tab:skeletonization} we summarize the properties of the skeletons, determined using \texttt{FilFinder3D's} longest path algorithm, for all clouds in our sample. This includes the minimum, median, and maximum distance to each cloud's skeleton, along with the extent of the cloud's skeleton along the Heliocentric Galactic cartesian $x$, $y$, and $z$ directions. We also include the number of skeletal components (skeletonized at a density threshold above $n = \rm 35 \; cm^{-3}$) and the total length of these components. The length is the total length of \textit{denser} features (computed using only the skeleton returned by the longest path algorithm) so it does not include features branching off the main trunk. The properties of both the longest path skeletons, and the full unpruned skeletons, are available online in machine readable format on the Harvard Dataverse (see \url{https://doi.org/10.7910/DVN/YQYBRD}).  The skeletonization results for two clouds (Perseus and Chamaeleon) are shown in Figures \ref{fig:perseus_topology} and \ref{fig:chamaeleon_topology}, respectively. These figures are interactive in their online version; it is possible to manipulate the skeleton in 3D, or hover over any point in the 2D skeleton to obtain its plane-of-the-sky coordinates and distance. Comparable interactive figures for the rest of the clouds in the sample are available at \url{https://faun.rc.fas.harvard.edu/czucker/Paper_Figures/3D_Cloud_Topologies/gallery.html} or in the online version of the published article. 

 We have separated Table \ref{tab:skeletonization} into two parts, representing ``complete" (top rows) and ``incomplete" (bottom rows) clouds. Incomplete clouds -- Orion A, Orion B, and $\lambda$ Orionis --- are those which lie at the very edge of the \citet{Leike_2020} 3D dust map, which extends to 370 pc along the cartesian $x$ and $y$ directions. While these three clouds are still capable of being skeletonized, it is likely that some fraction of each cloud lies beyond the map boundary, which would bias the skeletons to be mildly closer in distance than they actually are. This incompleteness will be further validated in \S \ref{mass_results}, where projected 2D extinction maps based on the \citet{Leike_2020} 3D dust distribution indicate that Orion A, Orion B, and $\lambda$ Orionis are the only three clouds in the sample whose masses are significantly discrepant from counterpart masses calculated out to ``infinity" \citep[that is, using the NICEST algorithm;][]{Lombardi_2009}. This discrepancy indicates that there is missing mass present beyond the boundaries of the map. 

\input{table2_excl2}

\begin{figure}[h!]
\begin{center}
\includegraphics[width=0.9\columnwidth]{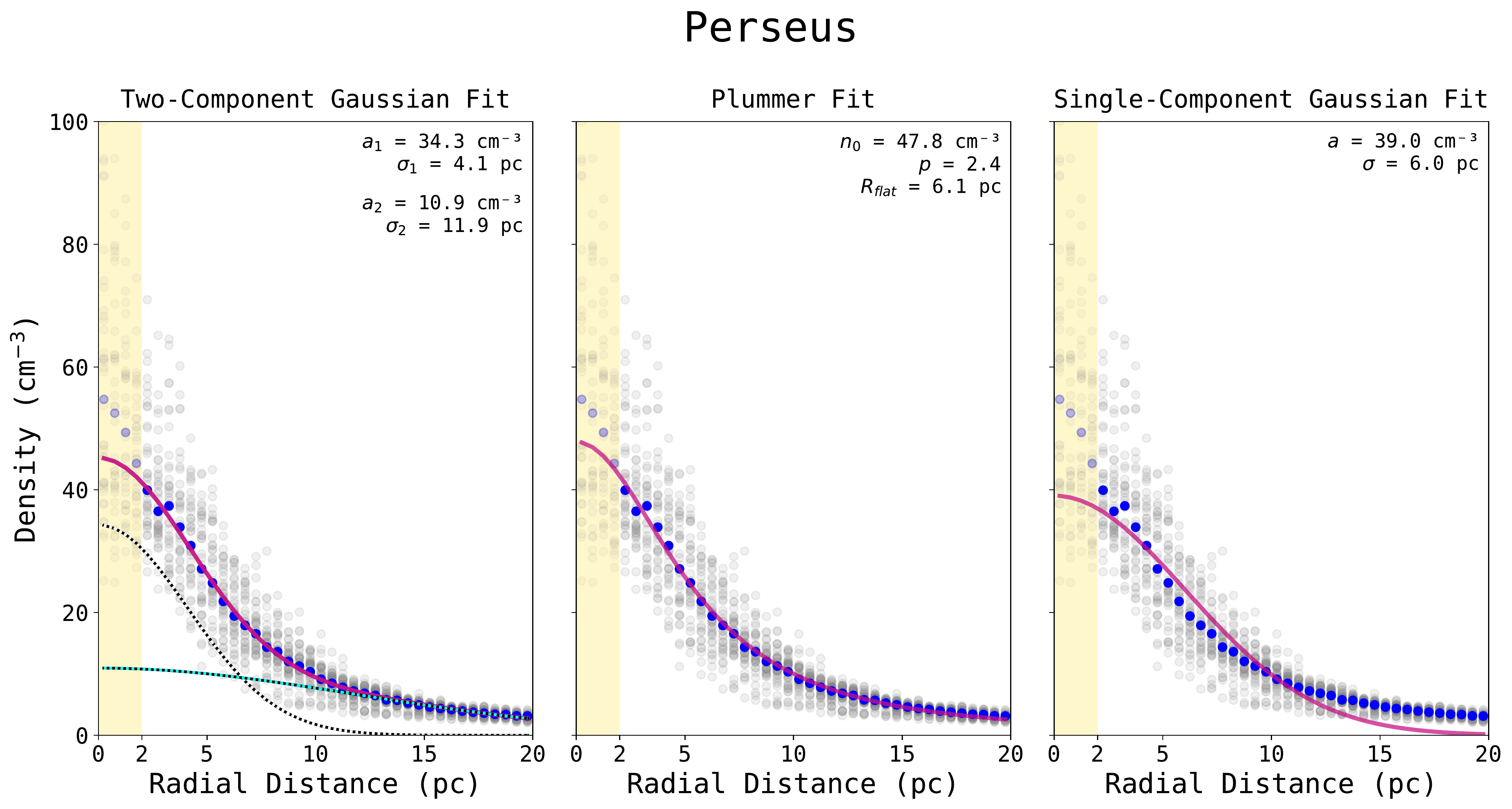}\vspace{0.5cm}
\includegraphics[width=0.9\columnwidth]{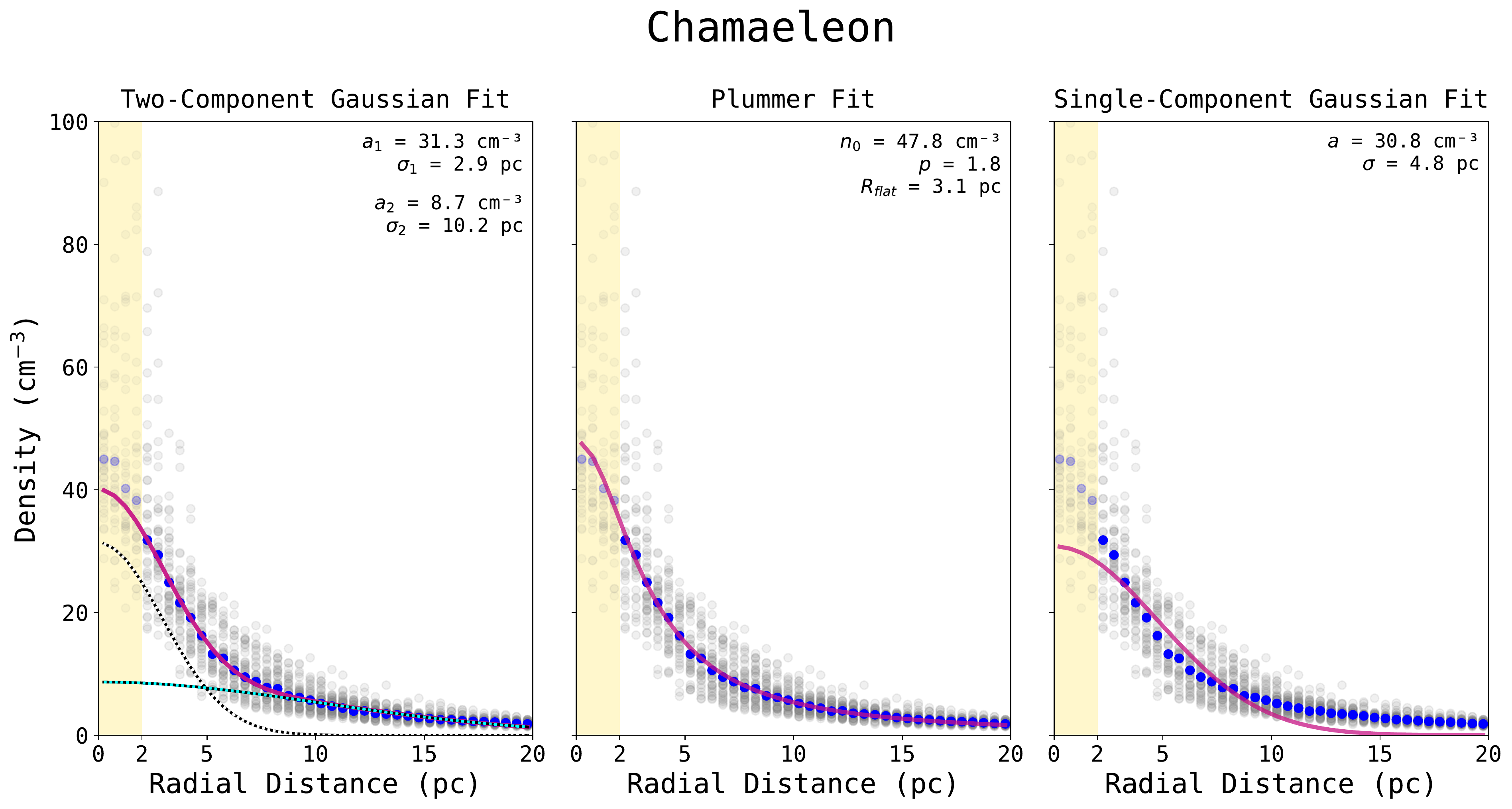}
\caption{Radial volume density profile for the Perseus Molecular Cloud (top) and the Chamaeleon Molecular Cloud (bottom). Because the \citet{Leike_2020} 3D dust map is more reliable at scales $>2$ pc and is not capable of probing the highest gas densities, we exclude the inner 2 pc (coincident with the yellow bar) from the fit.
In each panel the gray dots indicate the profiles for individual slices along the spine, while the blue dots indicate the median of all slices. We fit three functions to the averaged radial volume density profile, whose fits are shown in solid red: a two-component Gaussian function (left), a Plummer function (middle), and a single-component Gaussian function (right). In the left panel, we additionally overlay the decomposition of the two Gaussians via the dashed white and cyan lines, corresponding to the ``inner" (black-white dotted) and ``outer" (black-cyan dotted) components. Identical plots for the rest of the sample are available in the Appendix. \label{fig:profile_examples_excl2}}
\end{center}
\end{figure}

\begin{figure}[h!]
\begin{center}
\includegraphics[width=1\columnwidth]{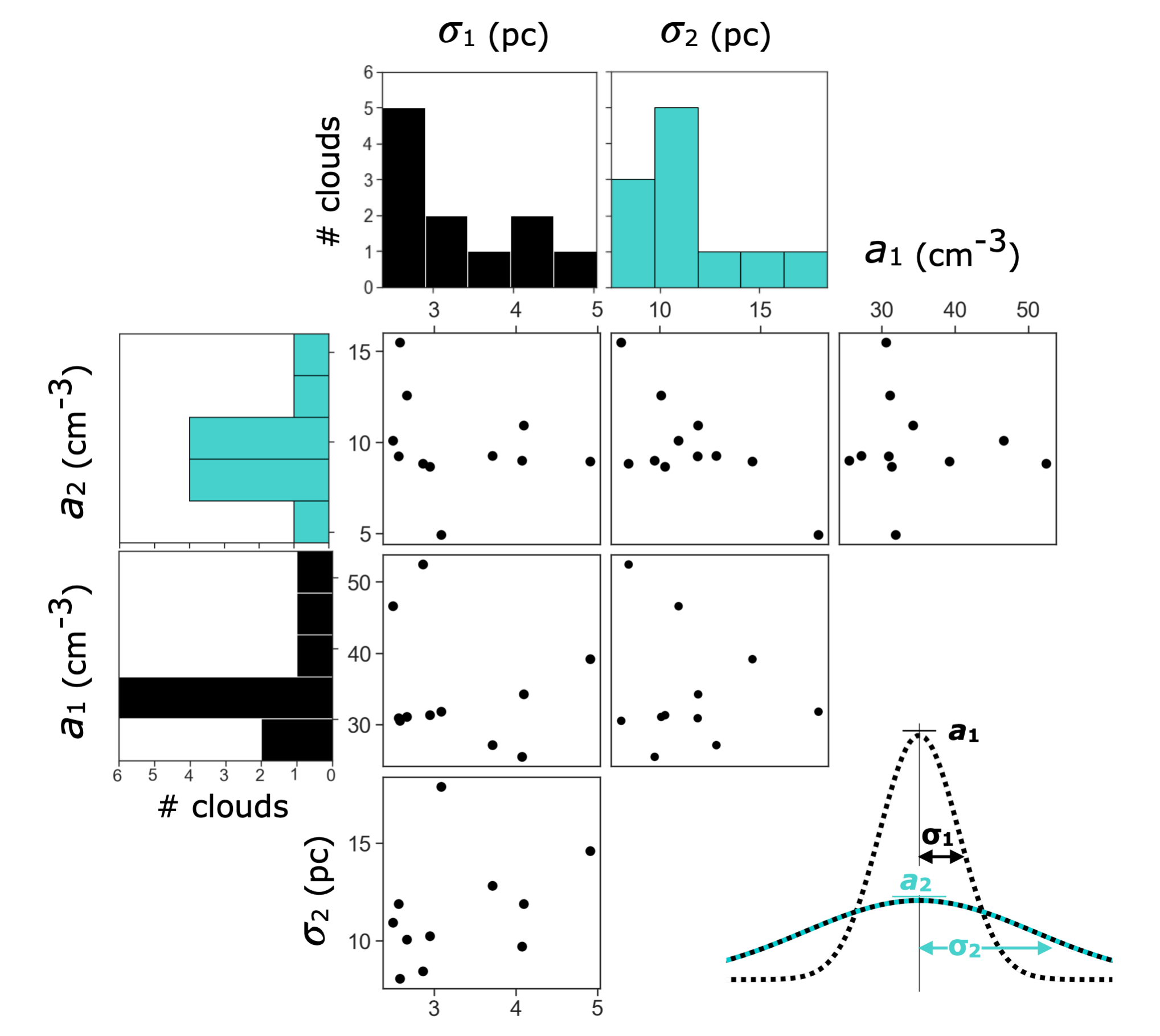}
\caption{Summary of the radial volume density profile fitting results for the two-component Gaussian function, including the standard deviation $\sigma_1$ of the inner Gaussian,  the standard deviation $\sigma_2$ of the outer Gaussian, the amplitude of the inner Gaussian $a_1$, and the amplitude of the outer Gaussian $a_2$. Each point represents a single local cloud in the sample. \label{fig:summary_twocomp}}
\end{center}
\end{figure}

\begin{figure}[h!]
\begin{center}
\includegraphics[width=1\columnwidth]{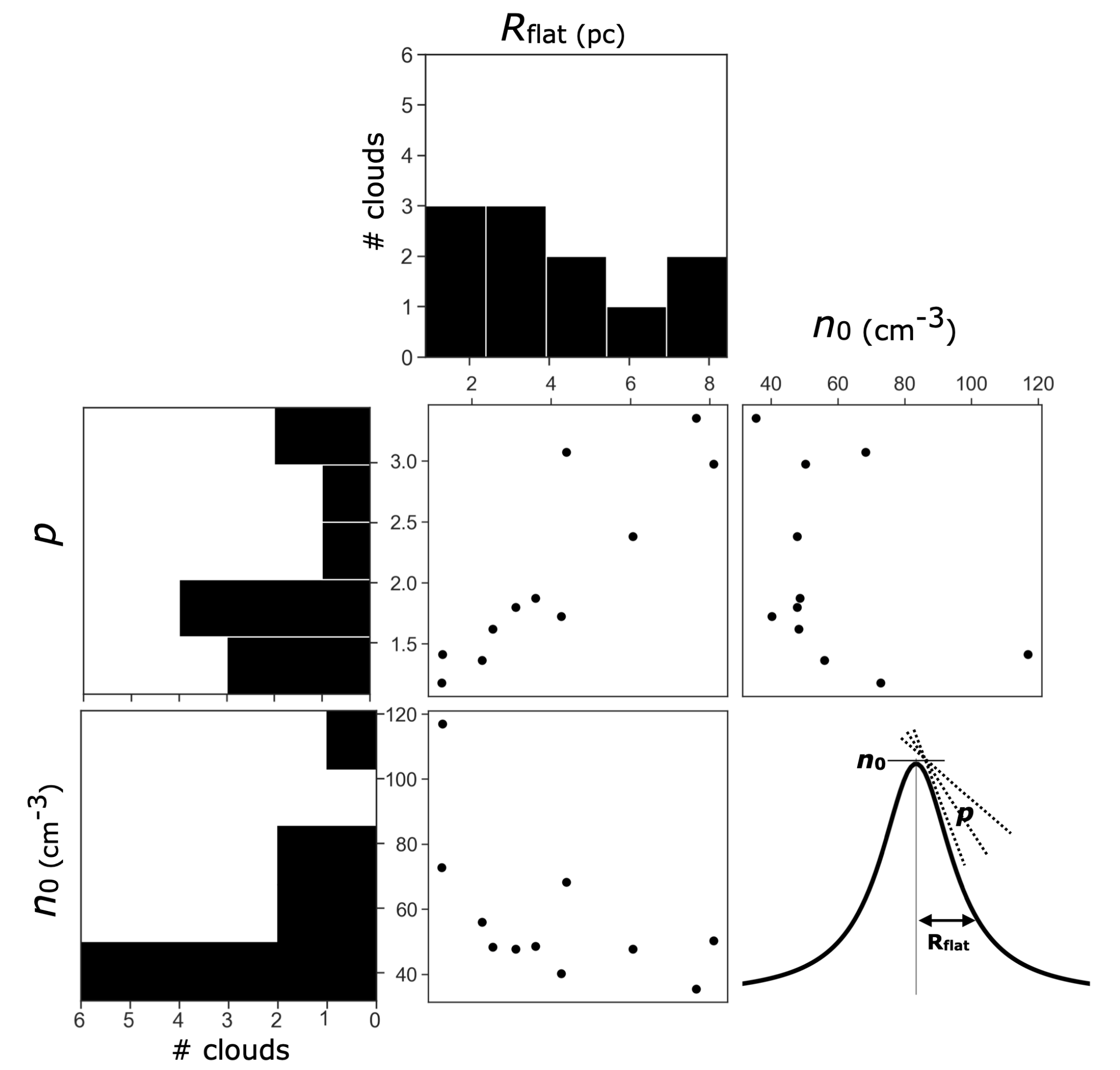}
\caption{Summary of the radial volume density profile fitting results for the Plummer function, including the distribution of the flattening radius $R_\mathrm{flat}$, the peak density $n_0$, and the index of the density profile $p$. Each point represents a single local cloud in the sample. We observe a strong degeneracy in the Plummer modeling, particularly between the $R_\mathrm{flat}$ and $p$ parameters. \label{fig:summary_plummer}}
\end{center}
\end{figure}

\clearpage
\subsection{Width Results} \label{width_results}

In Table \ref{tab:radial_profiles_excl2}, we summarize the results of our radial volume density fitting procedure. A machine readable version of Table \ref{tab:radial_profiles_excl2} is available at the Harvard Dataverse (\url{https://doi.org/10.7910/DVN/QKYR3G}). For each cloud and fit (two-component Gaussian, Plummer, single-component Gaussian) we list the average and 1$\sigma$ uncertainty on each parameter. In each case, we obtain a set of samples derived from the \texttt{dynesty} nested sampling procedure. The average value reported is the median (50th percentile) of the samples. The upper and lower bounds are the difference between the 50th and 84th percentiles of the samples, and the 16th and 50th percentiles of the samples, respectively. Cornerplots showing the marginalized posterior distributions for each cloud and type of fit are available online at the Harvard Dataverse (\url{https://doi.org/10.7910/DVN/YVQC7X}). The radial volume density profiles for two clouds --- Perseus and Chamaeleon --- are shown in Figure \ref{fig:profile_examples_excl2}, and similar figures are available for the rest of the clouds in the Appendix. The data behind these figures, including both the median and full ensemble of radial profiles computed for each cloud, are available at the Harvard Dataverse (see \url{https://doi.org/10.7910/DVN/JBRWHB}). 

We summarize the width results in Figure \ref{fig:summary_twocomp} and Figure \ref{fig:summary_plummer}. We exclude the single-component Gaussian from the summary due to its poor Bayes Factor relative to the two-component Gaussian and Plummer profiles (as justified in \S \ref{model_comp}). Figure \ref{fig:summary_twocomp} shows a cornerplot of the best-fit results for the two-component Gaussian model parameters, while Figure \ref{fig:summary_plummer} shows a cornerplot of the best-fit results for the Plummer model parameters. We find that for our two-component Gaussian results, inner widths $\sigma_1$ span 2.5 to 4.9 pc, with an average of $\approx$ 3.0 pc. The outer widths span 8 and 18 pc, with an average of $\approx$ 11 pc. We find a relatively tight spread in the ratio of the widths of the outer to inner envelopes ($\frac{\sigma_2}{\sigma_1}$), with an average value of 3.4 and a scatter of only 0.9. The amplitude $a_1$ of the inner Gaussian spans $n =$ 25 - 53 $\rm cm^{-3}$, about three times as large as the amplitude $a_2$ of the outer Gaussian, which spans $n =$ 5 to 15 $\rm cm^{-3}$. The widths of our Plummer fits $R_{\rm flat}$ span 1.2 - 8.1 pc with a typical value of $R_{\rm flat} = 3.6$ pc. The power-law indices $p$  of the Plummer fit range from 1.2 and 3.4, with a median value of 1.8. The peak density parameter $n_0$ ranges from $35 - 117 \; \rm cm^{-3}$, with a median of $\rm 49 \; cm^{-3}$. When the inner 2 pc of the profile is excluded, the Plummer fits often have a higher value of $n_0$ than measured, suggesting, as we will argue in \S \ref{fig:extinction_map}, that much of the densest structure in the 3D dust map is unresolved.  

\subsubsection{Model Comparison} \label{model_comp}
In addition to the best-fit parameters, we evaluate the performance of the models in light of the data. From Bayes' Theorem the posterior probability of a model $\mathcal{M}$ given the data $\mathcal{D}$ is proportional to:

\begin{equation}
p(\mathcal{M}\vert \mathcal{D}) \propto p(\mathcal{D} \vert \mathcal{M}) p(\mathcal{M})
\end{equation}

\noindent where $p(\mathcal{D}\vert \mathcal{M})$ is the likelihood of the data given the model and $p(\mathcal{M})$ is the prior probability of the model. To directly compare two models, $\mathcal{M}_1$ versus $\mathcal{M}_2$, we are interested in the ratio of their posterior probabilities (or posterior odds), given as:

\begin{equation}
\frac{p(\mathcal{M}_1 \vert \mathcal{D})}{p(\mathcal{M}_2\vert \mathcal{D})} = B_{12} \frac{p(\mathcal{M}_1)}{p(\mathcal{M}_2)}    
\end{equation}

\noindent with the Bayes Factor, $\mathcal{B}_{12}$, equal to the ratio of the marginal likelihood (also known as the evidence $\mathcal{Z}$) of the two models:

\begin{equation}
B_{12} = \frac{p(\mathcal{D}\vert \mathcal{M}_1)}{p(\mathcal{D} \vert \mathcal{M}_2)} = \frac{\mathcal{Z}_1}{\mathcal{Z}_2} 
\end{equation}

\noindent Assuming equal prior odds ($\frac{p(\mathcal{M}_1)}{p(\mathcal{M}_2)} = 1$), the posterior odds is simply the Bayes Factor. The logarithm of the evidence $\ln(\mathcal{Z})$ is conveniently returned for each model fit by the nested sampling package \texttt{dynesty} and is summarized in Table \ref{tab:radial_profiles_excl2}. The logarithm of the Bayes Factor is then given as: 

\begin{equation}
\ln(\mathcal{B}_{12}) = \ln(\mathcal{Z}_{1}) - \ln(\mathcal{Z}_{2})
\end{equation}

\noindent Computing $\ln(\mathcal{B}_{12})$ for the two-component Gaussian and Plummer Models, respectively, with regards to the single-component Gaussian model, we find that the two-component Gaussian has on average $\ln(\mathcal{B}_{12}) = 44$ over the single-component Gaussian, whereas the Plummer model has on average $\ln(\mathcal{B}_{12}) = 39$ over the single-component Gaussian. Based on the empirical ``Jeffrey's" scale for evaluating the relative strength of two models \citep{Jeffreys_1939}, $\ln(\mathcal{B}_{12}) > 5$ represents decisive evidence in favor of a given model. Because the Bayes Factor has a built in penalty for model complexity, the decisive evidence in favor of the two-component Gaussian and Plummer models is not due to the increased parameterization of the fit. Based on these Bayes Factor calculation, we can conclude that the single-component Gaussian fit is heavily disfavored with respect to both the two-component and Plummer fits. When comparing $\ln(\mathcal{B}_{12})$ for the two-component Gaussian and Plummer models, we find that the Plummer is decisively favored for two clouds, while the two-component Gaussian is decisively favored for four clouds, with the remaining clouds only having moderate evidence in favor of either model. Ruling out the single-component Gaussian model, we will discuss the theoretical implications of both the two-component Gaussian and Plummer profiles in \S \ref{discussion}. 

\subsection{Potential Additional Sources of Fitting Uncertainty} \label{caveats}

While the formal uncertainties on all parameters are reported in Table \ref{tab:radial_profiles_excl2}, there are potential additional systematic effects,  which we discuss here. 

\subsubsection{Degeneracy of the Plummer Profile}
While the Bayes Factor indicates that the Plummer function is sometimes preferred over the two-component Gaussian function, there are clear caveats to using a Plummer fit that should not go unsaid. In particular, it is well-known --- both from observations and simulations --- that two parameters in the Plummer function --- $p$ and $R_\mathrm{flat}$ are \textit{highly} degenerate. We see this trend clearly in Figure \ref{fig:summary_plummer}, where $p$ and $R_\mathrm{flat}$ are strongly correlated. In particular, as $R_\mathrm{flat}$ increases so does the value of $p$. This covariance has also been shown in synthetic filaments on much smaller scales (lengths $\approx$ a few parsecs, widths $\approx$ a few tenths of a parsec) in AREPO simulations from \citet{Smith_2014}, where the Plummer function was fit to the radial \textit{column} density distribution. We likewise see this in observations of the radial column density distributions of large-scale filaments (lengths $\approx$ tens of parsecs, widths $\approx$ 1 pc) \citep{Zucker_2018c}, where $p$ and $R_\mathrm{flat}$ show strong correlation. Accordingly, we urge caution when intepreting the Plummer profiles, as discussed in \S \ref{plummer_significance}.

\subsubsection{Intra-Cut Variation in the Radial Profiles}
In producing the averaged radial profile for fitting, we are not explicitly taking into account the variation in the radial profiles along the spine of each cloud. To summarize the variation, we have chosen to model the variable $\sigma_n^2$, which represents the scatter in the density profile values, capturing how much variation in the density we observe in a radial distance bin across all cuts. In general, for our better fitting models (Plummer and two-component Gaussian), we find that our fits prefer a small value for $\sigma_n^2$, equivalent to a few particles per cubic centimeter in density $n$. This trend is also evident in Figures \ref{fig:perseus_topology} and \ref{fig:chamaeleon_topology}. We do see a large scatter from cut to cut at small radial distances ($ r < 2$ pc), but these distances are excluded from the fit. For the remainder of the profile the scatter is significantly reduced, and is likely driving the low inferred value for $\sigma_n^2$. From this, we conclude that while inter-cut variation in the radial volume density profile may have some modest affect on our inferred value for the inner width $\sigma_1$ of the two-component Gaussian, it likely has less effect on the outerwidth $\sigma_2$ of the two-component Gaussian or the $R_\mathrm{flat}$ of our Plummer fits.

\subsubsection{Bias in Derived Widths as a Function of Cloud Distance} 

In order to confirm that we are probing intrinsic variation in the cloud widths, rather than the resolution of the underlying 3D dust map, it is necessary to verify that our derived cloud widths are \textit{not} dependent on the distance of the cloud. If this dependence exists, we  expect clouds at larger distances to have larger widths, due to degradation in the quality of \textit{Gaia} parallax measurements underpinning the \citet{Leike_2020} 3D dust map at larger distances. To examine this potential correlation, we plot all width parameters in Table \ref{tab:radial_profiles_excl2} as a function of cloud distance, which we show in Appendix \ref{distancebias} and Figure \ref{fig:distancebias}. We find no correlation between cloud distance and cloud width for any width parameter, suggesting that our properties are being driven by the intrinsic widths of clouds, rather than by distance biases.

\subsubsection{Including the Inner 2 pc of the Radial Profiles When Fitting}
In \S \ref{results}, we report results where radial distances between 0 and 2 pc (closest to the spine) are excluded from the radial profiles, due to combination of the 3D dust map's 1 pc voxel width and the caveat from \citet{Leike_2020} that scales less than 2 pc are unreliable. To ensure that our choice to exclude the inner 2 pc does not significantly affect our results, we refit the radial profiles using all radial distances between 0 and 20 pc, and those results are reported in Appendix \ref{appendix_full_profile}. Overall, we find that including the inner 2 pc has a modest effect on our results, changing the widths on average by $\lesssim 20\%$ across all models. However, the quality of the fit degrades, validating our choice to remove the inner 2 pc of each radial profile in our main analysis. 

\subsubsection{Uncertainties in the Underlying 3D Dust Distribution}

 All results reported in this section are derived from the mean 3D dust distribution reported in \citet{Leike_2020}. However, \citet{Leike_2020} also report twelve different realizations of their dust distribution, which are used to derive the mean map. Since the different realizations reflect the underlying uncertainty in the 3D dust reconstruction, it is necessary to quantify the effect this uncertainty has on our results. Accordingly, in Appendix \ref{appendix_samples}, we implement a more sophisticated fitting procedure, which simultaneously takes into account all twelve realizations of the 3D dust map when computing the model parameters. We find that overall, the uncertainty in the 3D dust map does not dominate our results, particularly for the two-component Gaussian fit, with best-fit parameters again typically changing $\lesssim 20\%$ across all clouds and all model fits. See Appendix \ref{appendix_samples} for full details.

\subsubsection{Relationship Between 3D Dust Correlation Kernel and Profile Shapes} \label{correlation_main}

Finally, to construct their 3D dust map, \citet{Leike_2020} model the dust as a log-normal Gaussian process, while simultaneously inferring the correlation kernel of the process, corresponding to the physical spatial correlation power spectrum of the dust. This correlation kernel acts as a prior on the dust extinction density\footnote{The dust extinction density in the context of the \citet{Leike_2020} is the optical depth in the Gaia $G$ band per one parsec.}. The adoption of a log-normal Gaussian process prior was primarily chosen for statistical reasons, to ensure that the dust density is always positive, and to allow the dust density to vary by orders of magnitude. While this kernel is inferred from the data, it is necessary to quantify the similarity between the shapes of the radial profiles and the shape of the inferred kernel. If the shapes of the kernel and the profiles differ, it is likely that the data in the vicinity of the cloud directly determines the profile shapes. However, if the shapes of the kernel and profiles look similar, it is possible that the kernel is being imprinted on the clouds' structure.

In Appendix \ref{appendix_correlation} and Figure \ref{fig:correlation} therein we compare the structure of our radial profiles to the structure of the \citet{Leike_2020} kernel inferred in the 3D dust map's reconstruction. We find that at radial distances between 0 and 10 pc, the cloud profiles deviate significantly from the kernel, indicating that their shapes are being driven by data local to the cloud. This deviation provides evidence that we are probing the \textit{intrinsic} structure of the cloud at smaller radial distances. However, at large radii ($r > 10$ pc), the shapes of the kernel and the cloud profiles become increasingly similar, particularly for the Musca cloud. This similarity indicates that the cloud structure at lower densities may be driven by the prior on the dust extinction density. However, since the kernel is inferred using data inclusive of the cloud’s structure (in addition to structure beyond each cloud of interest) the outer envelopes may still be physically meaningful.  Accordingly, while we will discuss the physical significance of the profiles at lower densities in \S \ref{discussion}, we caution against overinterpretation of the tails of the radial volume density until the lower density outer envelopes can be better validated in future work. 

\input{table3}

\subsection{2D Extinction Maps and Masses} \label{mass_results}
\begin{figure}[h!]
\begin{center}
\includegraphics[width=1.\columnwidth]{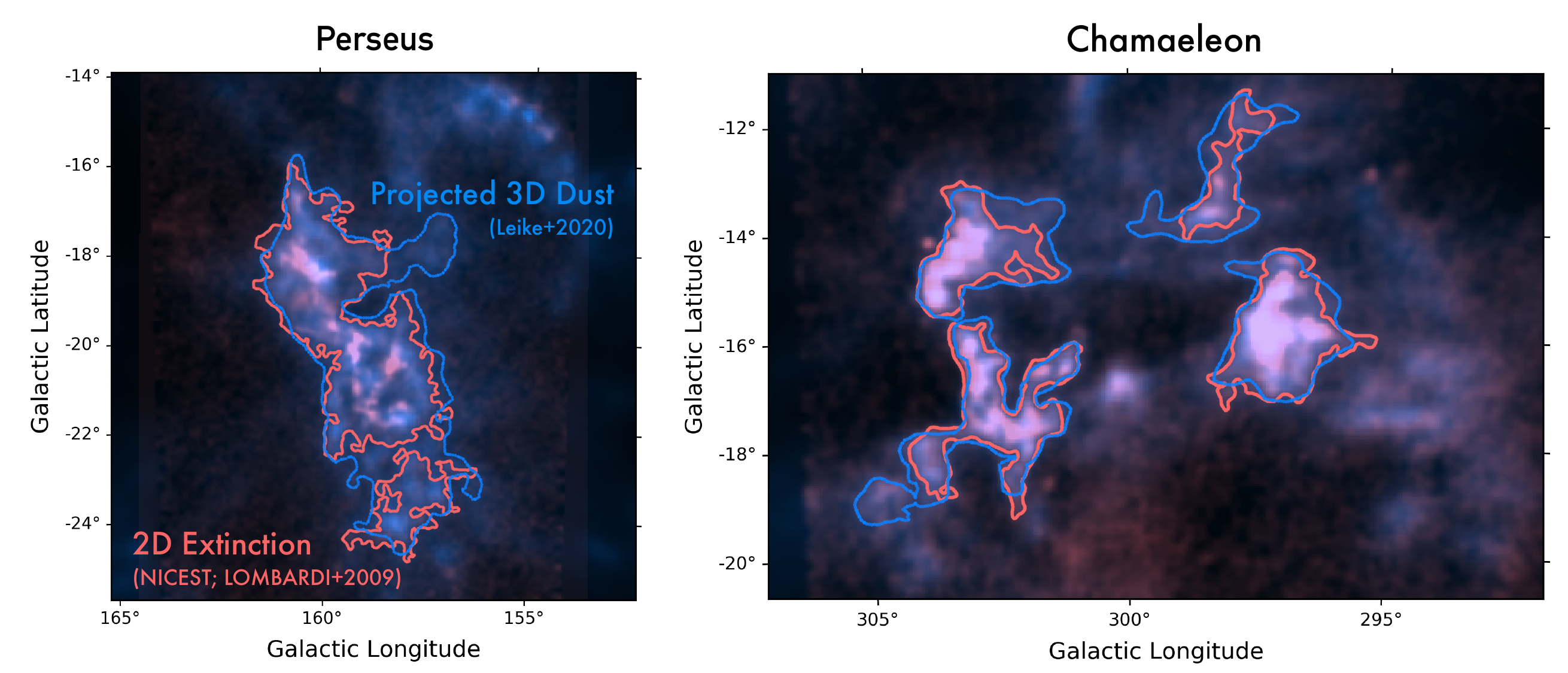}
\caption{A comparison of the $K$ band extinction ($A_K$) maps derived from the NICEST algorithm \citep{Lombardi_2009} (red) and from the \citet{Leike_2020} 3D dust map projected back on the plane of the sky (blue) for the Perseus Molecular Cloud (left) and Chamaeleon Molecular Cloud (right). The contours indicate the $A_K=0.1$ mag cloud boundaries which are used to derive the clouds masses listed in Table \ref{tab:mass}. \label{fig:extinction_map}}
\end{center}
\end{figure}

\begin{figure}[h!]
\begin{center}
\includegraphics[width=0.9\columnwidth]{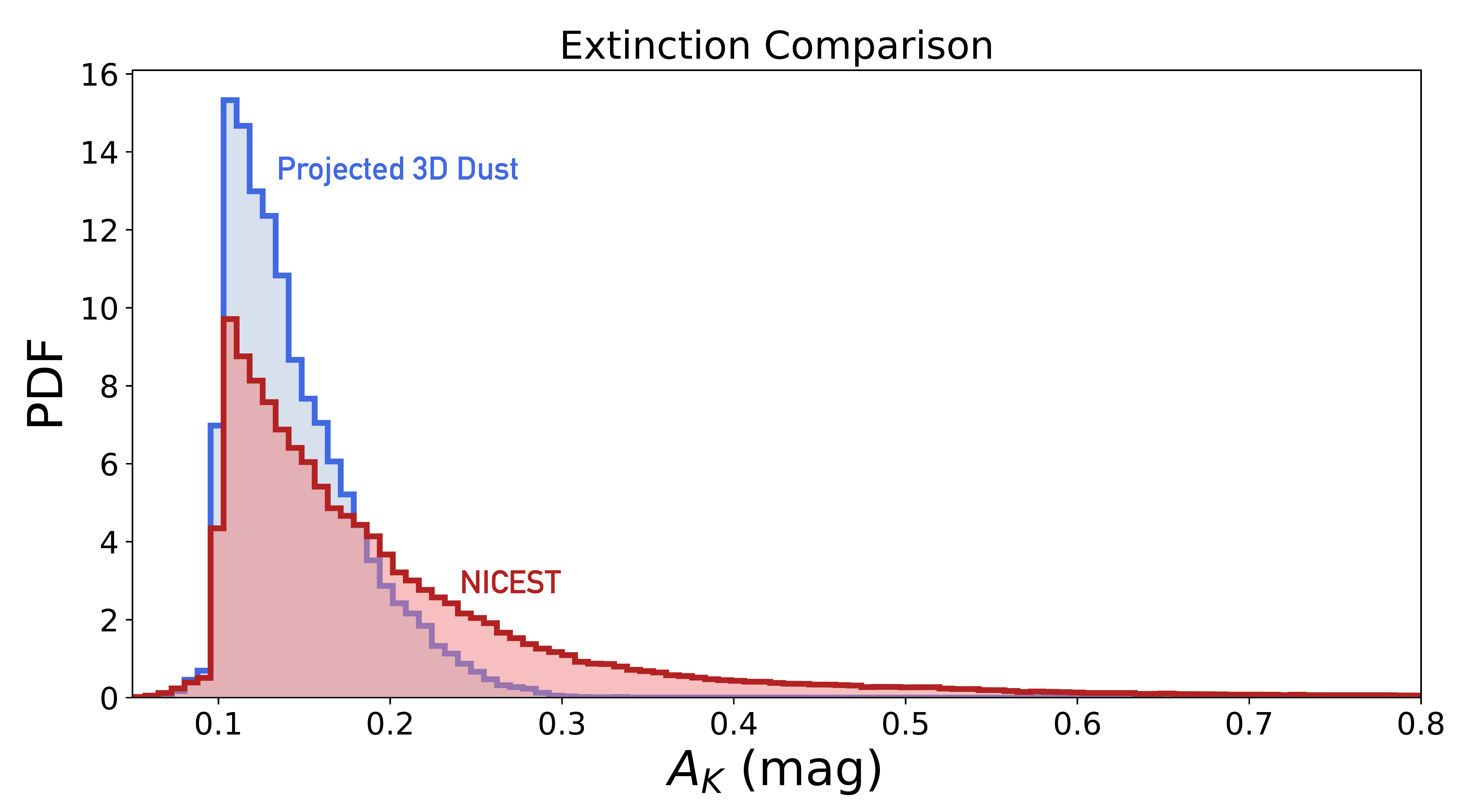}
\caption{A comparison of the PDFs between the $K$ band extinction ($A_K$) derived from the 2D NICEST algorithm \citep{Lombardi_2009} (red) and from the \citet{Leike_2020} 3D dust map projected back on the plane of the sky (blue). We only consider structure above the $A_K = 0.1$ mag threshold used to derive the cloud masses in Table \ref{tab:mass}.  We extract the extinction distribution above this $A_K = 0.1$ mag boundary for each cloud, and then stack the per-cloud results to obtain a representative histogram for each method. The histograms are normalized such that their total area is equal to one. While similar at the $A_K \approx 0.1$ mag level, traditional 2D dust extinction maps show a larger dynamic range, with increased sensitivity at higher extinction ($A_K \gtrsim 0.2-0.3$ mag). \label{fig:extinction_comp}}
\end{center}
\end{figure}

In this section, we compare the cloud masses derived from 3D dust to those from traditional 2D dust approaches, with the goal of quantifying any discrepancies. In Figure \ref{fig:extinction_map}, for the Perseus and Chamaeleon molecular clouds, we show a comparison between traditional 2D extinction maps \citep[from the NICEST algorithm;][see \S \ref{nicest}]{Lombardi_2009} and 2D extinction maps derived by projecting 3D dust data \citep[from][see \S \ref{leikenicest}]{Leike_2020} on the plane of the sky. Both maps have been convolved to the same (lower) angular resolution of the projected 3D dust maps, given the average distance to the cloud shown in Table \ref{tab:skeletonization}. In the same figures, we overlay for both NICEST (red) and 3D dust (blue) the $A_K = 0.1$ mag contours used to derive cloud masses in Table \ref{tab:mass}. As will become more clear in Figure \ref{fig:extinction_comp}, despite being morphologically very similar at the $A_K = 0.1$ mag level, the NICEST maps are able to probe a much larger dynamic range of extinction in comparison to the \citet{Leike_2020} derived results. In particular, the NICEST maps show much more structure at the highest extinction levels, revealing clumps, cores, and filaments that are invisible in the \citet{Leike_2020} projected version.

This lack of structure at high extinction is not surprising, given how the \citet{Leike_2020} 3D dust map is produced. In order to probe highly extinguished regions (see Figure \ref{fig:extinction_map}), any extinction-based dust mapping method must be able to see stars \textit{through} dust. In contrast to NICEST, which relies only on near-infrared photometry, the \citet{Leike_2020} analysis require \textit{Gaia} optical photometry with relatively high signal-to-noise, which precludes stars behind dense regions from being included in the analysis. Another possible source of discrepancy between the \citet{Leike_2020} projected 3D dust and traditional 2D dust maps might stem from the calibration of the likelihood function used to construct the \citet{Leike_2020} 3D dust map. Specifically, the likelihood used in the construction of the \citet{Leike_2020} 3D dust map is calibrated using stars in lines of sight with approximately zero extinction, which could introduce biases in regions of higher extinction. 

This missing high extinction information in the \citet{Leike_2020} 3D dust map is more apparent in Figure \ref{fig:extinction_comp}, which compares $A_K$ PDFs for the inherently 2D NICEST technique \citep{Lombardi_2009} and the 2D projection of the 3D maps \citep{Leike_2020}, across the full cloud sample. The histograms are computed by stacking every pixel above an $A_K = 0.1$ mag for every cloud. The $A_K$ PDF for the \citet{Leike_2020} method shows more extinction than NICEST at low values ($A_K \approx  0.1$ mag), and much less (essentially no) extinction above an $A_K \approx 0.3$ mag. While the shapes of the two PDFs are $\approx$ similar at the low extinction end ($A_K < 0.15$ mag), only the NICEST PDF  shows a tail extending to high extinction ($A_K \approx 1.0$ mag). This high extinction tail manifesting in the NICEST maps is consistent with the maximum $A_K$ we report on a cloud-by-cloud basis in Table \ref{tab:mass}. For the NICEST maps,  the maximum $A_K$ toward each cloud is $\approx 0.2 - 0.3$ mag, while for NICEST it is a factor of a few higher, with peak $A_K = 0.9 - 1.5$ mag inside each cloud boundary. So, these comparisons suggest that the \citet{Leike_2020} map is a faithful representation of projected dust column density up to $A_K = 0.2 - 0.3$ mag, equivalent to $A_V = 2 - 3$ mag.

How does reduced sensitivity at high extinction affect masses derived from the \citet{Leike_2020} map?  Table \ref{tab:mass} lists masses for clouds (above their $A_K = 0.1$ mag boundaries; see \S \ref{mass}) computed using both methods and the \textit{ratio} of those masses. In general, the ratio is near unity because mass at  high extinction makes up a small fraction of clouds' total mass \citep[see also e.g.][]{Lada_2010}. In all cases, NICEST masses are larger, but not by a significant amount: the ratio of the NICEST mass to the 3D dust mass ranges from 1.0 to 1.6, with an average of 1.2, for the ``complete clouds," listed above the line in Table \ref{tab:mass}. Discrepancies  are, as should be expected, significant for the ``incomplete" clouds (below the line, Orion A, Orion B, and $\lambda$ Orionis) which lie at the edge of the \citet{Leike_2020} 3D dust map. While the mass ratio for Orion B is only modestly higher than the rest of the sample ($\rm  \frac{Mass_{NICEST}}{Mass_{Leike}} = 1.7$), the NICEST mass for Orion A is $3.4\times$ larger than its 3D dust derived mass, and for $\lambda$ Orionis it is $13.7\times$ as large. This high ratio indicates that while Orion B is largely complete, significant amounts of mass are missing for Orion A and $\lambda$ Orionis, not only because the \citet{Leike_2020} method is not sensitive enough at high extinction, but also likely because some fraction of the mass lies beyond the \citet{Leike_2020} grid.

\section{Cloud-by-Cloud Exposition} \label{exposition}

In this section, we summarize the skeletonization and radial profile fitting results of \S \ref{results} on a cloud-by-cloud basis, and in light of complementary results for each cloud from the literature. A gallery of 3D interactive figures for the full sample is available \href{https://faun.rc.fas.harvard.edu/czucker/Paper\_Figures/3D\_Cloud_Topologies/gallery.html}{here}. 

\subsection{Chamaeleon}
As shown in Figure \ref{fig:chamaeleon_topology} (interactive), we find that the skeleton of the Chamaeleon Molecular Cloud is located at distances between 173 and 190 pc, with a median distance of 183 pc. Curiously, and as apparent in Figure \ref{fig:chamaeleon_topology}, we find that all the Chamaeleon clouds -- Chamaleon I, II, and III --- are connected via a hub-like structure, with a void in the center. The Chamaeleon complex is also physically connected to the Musca dark cloud at a distance of 172 pc. Musca is also shown in the 3D dust map in Figure \ref{fig:chamaeleon_topology} but skeletonized separately in this work. 

Based on  two-component Gaussian fitting results, Chamaeleon has an inner envelope full-width half max (FWHM$=2.35 \sigma$) of 6.8 pc, and an outer envelope FWHM of 24 pc, with a ratio of the outer to inner width equal to 3.5. 

\subsection{Ophiuchus}
As shown \href{https://faun.rc.fas.harvard.edu/czucker/Paper_Figures/3D_Cloud_Topologies/ophiuchus_topology/ophiuchus.html}{online} (interactive), we find that the skeleton of Ophiuchus is located at distances between 115 and 150 pc. The median distance is 136 pc, which is also the distance of L1688, consistent with previous distance estimates based on 3D dust mapping \citep[from][]{Zucker_2019, Zucker_2020} and from VLBI parallax measurements of YSOs \citep[from][]{Ortiz_Leon_2017}. There is a 10 pc distance gradient across the B44 streamer, which extends from $\rho$ Ophiuchus out to a distance of $d \approx 145$ pc. The farthest component is the ``arc" of Ophiuchus, lying at $d \approx 150$ pc, while the nearest component is the Ophiuchus north complex, lying between $d\approx 115 - 125$ pc. The effect of feedback from OB stars (e.g. $\sigma$ Sco, $\delta$ Sco) on the 3D distribution of dust, in particular the formation of the elongated streamers B44 and B45, will be discussed in detail in Alves et al. (2021, in prep.). While B44 appears as part of the main spine of Ophiuchus, B45 appears as a skeletal branch which is pruned prior to the width analysis. We find that the B45 streamer extends from B44, rather than from the core L1688 region. With a typical width of $\approx 1 \; \rm pc$ we believe this is an artifact of B45 being only marginally resolved in the \citet{Leike_2020} 3D dust, rather than a reflection of the true morphology of the B45 streamer. 

Based on the two-component Gaussian fitting results, Ophiuchus has an inner envelope FWHM of 6.3 pc, and an outer envelope FWHM of 24 pc, with a ratio of the outer to inner width equal to 3.8.

\subsection{Lupus}
As shown \href{https://faun.rc.fas.harvard.edu/czucker/Paper_Figures/3D_Cloud_Topologies/lupus_topology/lupus.html}{online} (interactive), the skeleton of the Lupus molecular cloud lies between 155 and 198 pc, with a median distance of 165 pc. On a region-by-region basis, we find Lupus 9 to be the farthest cloud in the complex, at a distance of $\approx 197$ pc. Lupus 3-8 are all connected within the same complex in 3D space (at the level of our thresholding algorithm, $n = \rm 35 \; cm^{-3}$) at a distance of $\approx 163$ pc. Lupus 1 lies the closest in distance, at $\approx 155$ pc, though the Lupus 1 line of sight suffers from confusion. By confusion, we mean that there is evidence of multiple components along the line of sight, in that there is a component immediately to the south of Lupus 1 (appearing to connect to Lupus 1 on the plane of the sky) but which lies at a much farther distance of 190 pc.  Our distances to Lupus 1, 3, 4, 5, and 6 are broadly consistent with those from \citet{Galli_2020_Lupus}, whose distances range from $150-160$ pc based on \textit{Gaia} DR2 astrometry towards young stars associated with the complex. Curiously, we find no prominent feature at higher gas densities ($n > \rm 35 \; cm^{-3}$) associated with Lupus 2. However, \citet{Galli_2020_Lupus} place Lupus 2 at a distance of 158 pc, suggesting it is indeed associated with the Lupus 1, 3, 4, 5, and 6 regions. 

Based on the two-component Gaussian fitting results, Lupus has an inner envelope FWHM of 6.1 pc, and an outer envelope FWHM of 19 pc, with a ratio of the outer to inner width equal to 3.1.

\subsection{Taurus}
As shown \href{https://faun.rc.fas.harvard.edu/czucker/Paper_Figures/3D_Cloud_Topologies/taurus_topology/taurus.html}{online} (interactive), the skeleton of the Taurus Molecular Cloud extends between 131 and 168 pc. When viewed top down, it is evident that Taurus is comprised of dust in two different layers, separated on average by about 10-15 pc. This layering is consistent with several previous studies, which find evidence of significant depth effects \citep[see e.g.][]{Galli_2018, Galli_2019, Zucker_2019} over this same distance range. The furthest component of Taurus, at a distance of 168 pc, is an elongated component of the cloud extending out from L1538 towards $(l,b) = (177.5^\circ, -8.5^\circ)$. 

Based on the two-component Gaussian fitting results, Taurus has an inner envelope FWHM of 6.1 pc, and an outer envelope FWHM of 28 pc, with a ratio of the outer to inner width equal to 4.6.

\subsection{Perseus}
As shown in Figure \ref{fig:perseus_topology} (interactive), the skeleton of the Perseus Molecular Cloud extends between 279 to 301 pc. It exhibits a prominent $\approx 20$ pc distance gradient. This distance gradient is very similar to that found independently in \citet{Zucker_2018a}, which also used 3D dust mapping to obtain distances to star-forming regions across Perseus, finding them to be located between 279 and 302 pc. This gradient is also broadly consistent with the VLBI parallax results towards young stars in Perseus from \citet{Ortiz_Leon_2018}, though \citet{Ortiz_Leon_2018} find IC348 to be at a distance of 320 pc, about 20 pc farther away than we find either in this work or in \citet{Zucker_2018a}. Upon further examination of the distribution of the 3D dust and young stars in this region, we find that neither the \citet{Zucker_2018a} nor \citet{Ortiz_Leon_2018} distance to IC348 is inaccurate. Rather, it appears that the IC348 stellar cluster has actually migrated beyond the distance of the dust cloud from which it formed, which results in the discrepancy between the dust- and star-based approaches. This migration is consistent with the estimated age of IC348 of $\approx 2-6$ Myr \citep{Muench_2007}. No such discrepancy is found for the younger NGC1333 cluster. When considering solely the 3D dust distribution, neglecting information from YSOs, no component of Perseus lies beyond 300 pc. 

Based on the two-component Gaussian fitting results, Perseus has an inner envelope FWHM of 9.6 pc and an outer envelope FWHM of 28 pc, with a ratio of the outer to inner width equal to 2.9.

\subsection{Musca}
As shown \href{https://faun.rc.fas.harvard.edu/czucker/Paper_Figures/3D_Cloud_Topologies/musca_topology/musca.html}{online} (interactive), the Musca dark cloud is located between distance of 171 and 173 pc. We find no evidence that Musca is a sheet viewed edge-on, but rather is a true filament in 3D space \citep[c.f.][]{Tritsis_2018}. It is physically connected to the Chamaeleon Molecular Cloud complex at a level of $n = 35 \; \rm cm^{-3}$ but is skeletonized separately in this work to obtain an independent width. 

Based on the two-component Gaussian fitting results, Musca has an inner envelope FWHM of 7.3 pc, and an outer envelope FWHM of 42 pc, with a ratio of the outer to inner width equal to 5.8. Musca is a clear outlier in the distribution of its outer Gaussian profile, which is significantly larger and more extended than the rest of the sample. At large radial distances, the shape of Musca's radial profile shows the highest correspondence to that of the underlying kernel inferred during the reconstruction of the \citet{Leike_2020} 3D dust map (see \S \ref{correlation_main} and Appendix \ref{appendix_correlation}), indicating that this highly extended envelope is likely prior-dominated. In addition to the outer envelope being prior-dominated, we also believe that the inner envelope of Musca -- manifesting as an infrared dark cloud -- is unresolved in our map, as 2D dust results targeting the infrared dark component of the cloud find widths as small as a few tenths of a parsec \citep{Cox_2016}. As a result, we urge extreme caution when interpreting these width results. 

\subsection{Pipe}
As shown \href{https://faun.rc.fas.harvard.edu/czucker/Paper_Figures/3D_Cloud_Topologies/pipe_topology/pipe.html}{online} (interactive), the Pipe dark cloud is located between a distance of 147 and 163 pc, with a 15 pc distance gradient. The stem of the Pipe nebula, near B59, is located at a distance of $150$ pc, and the bowl, extending down to latitudes of $b=2.4^\circ$, lies as far as 163 pc. About halfway down the stem, a thin tendril of gas connects the Pipe nebula with the Ophiuchus cloud complex, meeting an extension of Ophiuchus that reaches beyond the B44 streamer. However, the Pipe nebula is skeletonized separately from Ophiuchus in this work, in order to obtain an independent width. The distance to B59 in the stem of the Pipe nebula is smaller than, but still consistent with, independent \textit{Gaia} parallax measurements of YSOs in the complex \citep[163 pc, from][]{Dzib_2018}.  

Based on the two-component Gaussian fitting results, the Pipe nebula has an inner envelope FWHM of 9.6 pc and an outer envelope FWHM of 23 pc, with a ratio of the outer to inner width equal to 2.4. This ratio of the outer to inner width is the smallest of all clouds in the sample. 

\subsection{Cepheus}
As shown \href{https://faun.rc.fas.harvard.edu/czucker/Paper_Figures/3D_Cloud_Topologies/cepheus_topology/cepheus.html}{online} (interactive), the skeleton of the Cepheus molecular cloud is located between distances of 337 and 370 pc, with a median distance of 345 pc. Our skeletonization only includes the northern section of Cepheus, as the southern section (closer to the Galactic plane) actually lies at distances $> 900$ pc \citep{Grenier_1989, Zucker_2019}. The median skeleton distance is in excellent agreement with the average Cepheus ``near" distance of 352 pc, based on an independent 3D dust mapping pipeline from \citet{Zucker_2019}. Our Cepheus skeletonization includes  four isolated density features at a level above our chosen threshold  ($n = 35 \; \rm cm^{-3}$), the highest of any cloud in the sample. These features include the sub-regions L1148/L1552/L1155 ($d = 338$ pc), L1172 ($d = 340$ pc), L1228 ($d=368$ pc), L1241 ($d=349$ pc), and L1247/L1251 ($d=340$ pc). 

Based on the two-component Gaussian fitting results, Cepheus has an inner envelope FWHM of 5.9 pc, and an outer envelope FWHM of 26 pc, with a ratio of the outer to inner width equal to 4.4.

\subsection{Corona Australis}
As shown \href{https://faun.rc.fas.harvard.edu/czucker/Paper_Figures/3D\_Cloud\_Topologies/coraus\_topology/coraus.html}{online} (interactive), the skeleton of the Corona Australis cloud is located between distances of 136 and 179 pc, with a median distance of 161 pc. Corona Australis is one of the fluffiest local clouds in the \citet{Leike_2020} map and has no significant extended features above a density threshold of $n = 35 \; \rm cm^{-3}$ so it is skeletonized at a much lower level ($n = 5 \; \rm cm^{-3}$) to provide distance information. We find that the Corona Australis cloud exhibits a prominent distance gradient, with the head of the cloud about 20-30 pc closer than the tail. The most extinguished part of Corona Australis, which manifests as a dark cloud at optical wavelengths, lies at a distance of $d = 144$ pc, which is in good agreement with the results of \citet{Galli_2020} (149 pc), and \citet{Zucker_2019} (151 pc). The reason we determine a higher overall median spine distance is due to the existence of Corona Australis' diffuse tail, which extends as far as 179 pc at a latitude of $b=-26^\circ$.

\subsection{Orion A}
As shown \href{https://faun.rc.fas.harvard.edu/czucker/Paper_Figures/3D_Cloud_Topologies/oriona_topology/oriona.html}{online} (interactive), we determine that the skeleton of the Orion A cloud spans distances of 391 to 445 pc, with a median distance of 405 pc. The 3D dust distribution towards Orion A is peculiar, due to the fact that there is a large piece of the cloud missing towards the Orion Nebula Cluster --- the densest part of the entire complex. This lack of dust is \textit{not physical}, and represents the inability of the \citet{Leike_2020} to fully probe high gas densities. As a result, the head (toward the Orion Nebula Cluster) and tail of Orion A (toward L1647) appear not to connect as they should in 3D space.  Analysis of Orion's YSO distribution \citep{Grossschedl_2018} and other high resolution 3D dust maps of Orion A \citep{Rezaei_2020} clearly show that the head and tail \textit{are} physically connected. The missing ONC region could explain in part the significant discrepancy between the cloud mass derived from 2D dust extinction versus from the projected 3D dust distribution (see \S \ref{mass_results}). 

In general, we find good agreement with previous studies towards regions probed by the dust map, as well as a similar \textit{range} of distances. For example, \citet{Grossschedl_2018} find that the bulk of the Orion A cloud lies between distances of $d \approx 380 - 460$ pc, which is not inconsistent with our spread of $391 - 445 $ pc. However, there is no visible connection between the farthest point in Orion A (L1647-S at a distance of 445 pc) and the main Orion A cloud consisting of L1641, while this connection \textit{does} appear in other approaches \citep{Grossschedl_2018, Rezaei_2020}. This lack of connection at the tail could be physical, or it could be indicative of another missing mass situation, like toward the Orion Nebula Cluster.  In addition to mass missing \textit{within} the grid, it is also possible that there is mass missing \textit{beyond} the grid, since Orion A lies at the boundary of the \citet{Leike_2020} dust mass. Thus, our skeletonization and width analysis results for Orion A should be treated with more caution than for other ``complete" clouds. The most likely effect of Orion A lying at the edge of the grid is that we will slightly underestimate the distance to the complex. 

Based on the two-component Gaussian fitting results, we find that Orion A has an inner envelope FWHM of 6.8 pc and an outer envelope FWHM of 20 pc, with a ratio of the outer to inner width equal to 2.9. 

\subsection{Orion B}
As shown \href{https://faun.rc.fas.harvard.edu/czucker/Paper_Figures/3D_Cloud_Topologies/orionb_topology/orionb.html}{online} (interactive), we determine that the skeleton of the Orion B cloud spans distances of 397 to 406 pc, with a median distance of 404 pc. Our median distance is consistent with \citet{Kounkel_2018}, which leverages \textit{Gaia} DR2 parallax measurements toward young stars in the complex to obtain a distance of $407 \pm 4$ pc. Unlike Orion A, we find no evidence of a strong distance gradient, and also find that Orion B consists of a single density feature with a length of 69 pc. However, since Orion B again lies at the very edge of the \citet{Leike_2020} dust map, it is possible that parts of Orion B could still be missing, which would bias our distance estimates lower compared to a cloud that lies well within the map boundaries. Thus, like our skeletonization and width analysis results for Orion A, we caution that our results for Orion B should be treated with more caution than for other ``complete" clouds. 

Based on the two-component Gaussian fitting results, Orion B has an inner envelope FWHM equal of 11.4 pc and an outer envelope FWHM of 34 pc, with a ratio of the outer to inner width equal to 3.0.

\subsection{$\lambda$ Orionis}
Finally, as shown \href{https://faun.rc.fas.harvard.edu/czucker/Paper_Figures/3D_Cloud_Topologies/orionlam_topology/orionlam.html}{online} (interactive), we determine that the skeleton of the $\lambda$ Orionis cloud spans distances of 375 to 397 pc, with a median distance of 386 pc.  A clear ring-like morphology is observed in 2D dust extinction maps, which \citet{Dolan_2002} argue formed from ejected material driven by a supernovae blast at the center of the ring about 1 Myr ago. 

We observe the same ring-like morphology in 3D. However, the skeletonization, and resulting widths, should be considered with extreme caution, as the skeleton lies only a few parsecs from the boundary of the \citet{Leike_2020} 3D dust map, so the skeletonization is likely to underestimate the cloud's distance. There is clearly mass missing in the $\lambda$ Orionis region, given that the mass for the cloud based on 2D dust maps (from NICEST) is a factor of fourteen higher than what we calculate using just the 3D dust volume local to the cloud (see \S \ref{mass_results} and Table \ref{tab:mass}). Consistent with this missing mass scenario, average distances to the $\lambda$ Orionis region tend to be higher in other studies \citep[e.g. 402 pc from][]{Zucker_2020}. The average distance to $\lambda$ Orionis from \citet{Zucker_2020} is also consistent with independent distances of young stars, for which \citet{Kounkel_2018} obtain $d = 404 $ pc. 

Based on the two-component Gaussian fitting results, $\lambda$ Orionis has an inner envelope FWHM of 8.7 pc and an outer envelope FWHM equal of 30 pc, with a ratio of the outer to inner width equal to 3.4. Despite only measuring the cloud width using voxels in front of $\lambda$ Orionis (due to the \citet{Leike_2020} grid ending directly beyond the spine), these results are consistent with results for the ``complete" clouds. 

\section{Discussion} \label{discussion}
In Section \ref{results}, we have shown that the radial volume density profiles measured in 3D for local molecular clouds are well-described by two functions: a two-component Gaussian function and a Plummer function. A single-component Gaussian is universally a poor fit to the data. We have also shown that the \citet{Leike_2020} technique is capturing a majority of the cloud mass for all ``complete" clouds in the sample, deviating from 2D integrated dust extinction measurements \citep[from the NICEST methodology;][]{Lombardi_2009, Lombardi_2011} only at higher extinction ($A_K \gtrsim 0.3$ mag).  In this section, we discuss the potential implications of our results, and provide an initial comparison of the 2D and 3D shapes of the radial profiles for a subset of the sample, as a basis for future work. 

\subsection{Two-Component Gaussian --- Probing a Phase Transition?}
In terms of the two-component Gaussian model, one possible explanation for the shape of the radial density profiles is that we are probing gas in different phases, where the transition between the outer and inner Gaussian profiles represents a shift in temperature, density, and/or chemical composition of the gas. In the density range consistent with our radial profiles, possible transitions include a chemical transition between atomic and molecular hydrogen (the HI-to-$\rm H_2$ transition), or a thermal transition between the unstable neutral medium (UNM) and cold neutral medium (CNM), driven by the temperature dependence of the atomic cooling curve. While the data are suggestive, we do not yet have enough detailed evidence to prove the link between the shape of the radial volume density profiles and a phase transition. Thus, in what follows, we only seek to present a plausibility argument.

\subsubsection{Chemical Transition Between Atomic and Molecular Gas} \label{HI_to_H2}

The first explanation for the bimodality in the radial profiles is that the broader, lower amplitude ``outer" Gaussian probes the shape of the atomic HI gas envelope, while the narrower higher amplitude ``inner" Gaussian probes the molecular phase. If so, the noticeable break in volume density coincident with the start of the inner Gaussian would represent a chemical transition of the gas, delineating the radial distance in 3D space at which point molecular hydrogen becomes self-shielded. Models suggest that the transition to self-shielding --- at which point molecular hydrogen can persist --- can be either sharp or gradual, depending on the strength of the UV radiation field relative to the gas density. However, once a critical column density of $\rm H_2$ has been produced, the fraction of hydrogen in molecular form should rise rapidly \citep{Bialy_2016, Sternberg_2014, Krumholz_2008}. If this transition is occurring in local molecular clouds --- as predicted by analytic theory --- it might explain why we find the two-component Gaussian model to be so stongly favored over a single-component Gaussian model.

We have known for a few decades that HI envelopes around molecular clouds are common \citep[see e.g.][]{Moriarty_Schieven_1997, Williams_1996, Williams_1995}. These HI envelopes are significantly extended with respect to the molecular gas and may be either the remnants of atomic clouds which condensed to form GMCs, or photodissociated gas on the outskirts of the molecular cloud. More likely, the atomic envelope arises from a combination of both phenomena. In addition to being observed in solar neighborhood clouds \citep[e.g. in Perseus and Taurus; see][]{Stanimirovic_2014, Pascucci_2015, Lee_2015} and toward the inner galaxy \citep{Beuther_2020, Bialy_2017}, atomic envelopes have also been observed at much lower resolution in the nearby Large Magellanic Cloud \citep{Fukui_2009}. Recent numerical simulations characterizing the properties of highly resolved synthetic molecular clouds \citep[e.g.][]{Duarte_Cabral_2016,Seifried_2017} also note the presence of large extended HI envelopes. 

We find that the typical amplitudes of the outer Gaussian profiles are consistent with the typical densities of the atomic envelopes seen in observations, mainly via HI emission on the plane of the sky. In this study, the peak densities of the outer envelopes --- characterized by the amplitude parameter $a_2$ --- span $n = 5 - 15 \; \rm cm^{-3}$, with an average amplitude of $n = 9 \; \rm cm^{-3}$, falling off to only a few particles per cubic centimeter at large radii. This range is consistent with HI densities inferred for atomic envelopes around molecular clouds in both observations and simulations. Looking at a sample of 123 HI envelopes in the Large Magellanic Cloud, \citet{Fukui_2009} derive a typical envelope density of $\rm 10 \; cm^{-3}$, calculated by dividing each cloud's HI column density by its approximate size on the plane of the sky.\footnote{We note, however, that the HI-to-H2 transition strongly depends on the dust-to-gas ratio, and the Large Magellanic Cloud has lower metallicity than Milky Way \citep{LMC_2013}, so any direct comparison between the two systems should be treated with caution.} Using GALFA HI and IRAS data, \citet{Bialy_2015} also calculate the atomic hydrogen volume density towards the Perseus molecular cloud, finding that the atomic densities vary between $n  = 2 - 11 \; \rm cm^{-3}$. This range of densities is similar to the range of densities captured by the outer Gaussian in the radial volume density profile we measure for Perseus, spanning $n = 12 \; \rm cm^{-3}$ toward the spine, down to $n \approx 2 \; \rm cm^{-3}$ toward the outskirts of the cloud.

In terms of the inner Gaussian, we find typical peak densities --- parameterized by $a_1$ --- to lie between $n \approx 25 - 52 \; \rm cm^{-3}$. While the amplitude for the averaged profile is of this order, a significant fraction of cuts show peak densities with $n = 100 \; \rm cm^{-3}$, with some extending as high as $n = 200 \; \rm cm^{-3}$.  While these average volume densities are not high enough to be observationally probed by CO-bright molecular gas manifesting at $n \gtrsim 200 \; \rm cm^{-3}$ \citep[see e.g.][]{Busch_2019}, they are consistent with densities of so-called ``CO-dark" gas, a phase associated with warmer, more diffuse molecular hydrogen which is not detectable in CO or HI. While invisible in both CO and HI, $\gamma$ ray observations --- capable of tracing a cloud's gas mass independent of its chemical state --- show that diffuse molecular gas surrounds all nearby CO-detected molecular clouds targeted in this study. This CO-dark phase effectively acts as an observational bridge between clouds' extended atomic envelopes and their embedded CO-bright cores \citep{Grenier_2005}, and would mark the true chemical transition between atomic and molecular gas. \citet{Tang_2016} use \textsc{[CII]} emission to trace CO-dark diffuse molecular gas in the Galactic plane, finding that this gas manifests at a range of densities (from $n \approx 60 - 1000 \; \rm cm^{-3}$), with an average density of $n \approx 200 \; \rm cm^{-3}$.\footnote{While \textsc{[CII]} emission is commonly used as a tracer of CO dark gas, we note that a recent study by \citet{Hall_2020} towards the local Perseus Molecular Cloud concludes that a \textsc{[CII]} envelope can also arise from the CNM phase, without a need to invoke CO-dark molecular gas.} The average density from \citet{Tang_2016} is a factor of a few larger than our average peak density $a_1$, but the lower estimates from \citet{Tang_2016} remain consistent with our results. Also consistent with these observations, analysis of the CO-dark diffuse molecular gas for synthetic clouds from the SILCC-Zoom project \citep{Seifried_2019} find number densities between $10 - 1000 \; \rm cm^{-3}$, with an average density of CO-dark gas on the order of $100 \; \rm cm^{-3}$ --- only a factor of two higher than what we estimate here for the inner Gaussian. Given that the volume densities derived from the 3D dust data --- particularly for the inner Gaussian component --- should be lower limits (see also \S \ref{systematics}), our results are generally in agreement with literature estimates for CO-dark molecular gas from both observations and simulations.

Finally, there is a straightforward test to disprove this HI-to-$\rm H_2$ transition hypothesis, involving a comparison between the radial extent of CO emission projected on the plane of the sky and radial extent of the inner Gaussian component derived here. The detection of CO emission significantly beyond the boundaries of the inner Gaussian profile would be in direct contradiction to the proposed origin, since no molecular gas should lie beyond the inner Gaussian. To test this scenario, we build CO radial profiles for a subset of the clouds in the sample, as described in detailed in Appendix \ref{co_profiles}. We find that the CO profiles shows widths equal to or marginally smaller than that of the inner Gaussian profiles, consistent with the scenario that the inner Gaussian profiles could be tracing CO dark gas.  

\subsubsection{Thermal Transition Between Unstable and Cold Neutral Medium}
While a diffuse molecular gas component offers one explanation for the shape of the inner Gaussian profile, no $\rm H_2$ component necessarily needs to be invoked to explain the clouds' density structure. It is well established that the atomic neutral medium consists of two thermally stable states at significantly different densities and temperatures --- the warm neutral medium (WNM) at $n \approx 0.2 - 0.9 \; \rm cm^{-3}$ and the cold neutral medium (CNM) at $n \approx 7-70 \; \rm cm^{-3}$ \citep{Field_1969, Wolfire_2003, Murray_2018_Sponge}. This two-phase structure is a natural consequence of the density and temperature dependence of the cooling curve at play in the diffuse interstellar medium. While the peak densities we find for the inner Gaussian ($a_1 = 25 - 52 \; \rm cm^{-3}$) are consistent with lower-density estimates of CO-dark diffuse molecular gas, they are also in strong agreement with the densities of the CNM. However, the peak densities of the outer envelopes are much larger than expected for the WNM.  In traditional two-phase models, HI at intermediate temperatures should be expected to quickly evolve into one of the stable phases. However, ample evidence has recently suggested that a significant fraction of Galactic HI lies in the thermally unstable phase, driven by dynamical processes in the interstellar medium \citep[e.g. turbulence; c.f.][]{Heiles_2003, Roy_2013, Murray_2018_Sponge}. Thus, if the inner envelope is probing the CNM, then the outer envelopes (with peak densities $a_2 = 5-15 \; \rm cm^{-3}$) may probe the thermally unstable HI lying at densities between the WNM and CNM with $n \approx 1 - 10 \; \rm cm^{-3}$. The transition observed in the two-component Gaussian profiles would then be driven not by a change in the chemical state of the gas, but by a change in the thermal phase of the atomic neutral medium.

Consistent with this scenario, \citet{Murray_2018_Sponge} analyzed HI spectra toward 57 lines of sight taken from the 21-SPONGE survey to infer the fraction of HI mass in the thermally unstable regime, finding that 20\% is in the UNM phase, as opposed to  28\% in the CNM phase (with the remaining mass in the WNM phase). \citet{Murray_2018_Sponge} further find no significant difference in the UNM fraction as a function of latitude, with both high ($|b| > 10^\circ$) and low latitude lines of sight showing similar mass distributions. Given its ubiquity, both from \citet{Murray_2018} and complementary studies \citep{Roy_2013, Heiles_2003}, the UNM-to-CNM transition represents a viable alternative to the HI-to-$\rm H_2$ transition for explaining the two-component Gaussian shape of the radial profiles. We note, however, that while the \citet{Murray_2018_Sponge} lines of sight span a wide range in latitudes, none of them intersect the clouds targeted in this work. 

Nevertheless, to broadly verify the consistency of our results with those of \citet{Murray_2018_Sponge}, we can compare for each cloud the relative amount of mass contained in the outer and inner envelope. If our results are consistent with \citet{Murray_2018_Sponge}, we would expect the ratio of the outer and inner envelope masses to be $\frac{M_2}{M_1} = \frac{M_{\rm UNM}}{M_{\rm CNM}} = \frac{20\%_{\rm UNM}}{28\%_{\rm CNM}} = 0.71$. We compute this ratio in two ways. First, we compute the mass of the inner envelope, $M_1$, and outer envelope $M_2$ by taking the definite integral of the Gaussian density profiles (shown via the black-white dotted and cyan-black dotted curves in e.g. Figure \ref{fig:perseus_topology}) over the volume, assuming a cylindrical geometry:

$$M_1 = 2 \pi l \, \mu m_p \int^{+\infty}_{-\infty} a_1 e^{\frac{-r^2}{2\sigma_1^2}} \, r  \, dr $$

$$M_2 =  2 \pi l \, \mu m_p \int^{+\infty}_{-\infty} a_2 e^{\frac{-r^2}{2\sigma_2^2}} \, r \, dr $$

\noindent where the parameters $a_1$, $\sigma_1$, $a_2$, $\sigma_2$ are the same as those reported in Table \ref{tab:radial_profiles_excl2}, $\mu = 1.37$ is the mean molecular weight, corrected for the helium abundance, $m_p$ is the mass of a proton, and $l$ is the length of the cloud reported in Table \ref{tab:skeletonization}. Computing $M_1$ and $M_2$ numerically and taking the ratio across the sample we obtain a median ratio for $\frac{M_2}{M_1} = 2.9$, with a spread of 1.3, which is higher than the predicted \citet{Murray_2018_Sponge} results of 0.72. Given that we are likely underestimating the densities in potentially CNM-associated regions, one possible explanation is that while the typical extents of the inner and outer components could be correct, their amplitudes, particularly for the inner Gaussian, may not be. Thus, a second, more extreme way to test this is to assume that the inner envelope is entirely composed of CNM, and UNM only exists beyond the edge of the inner envelope. To compute this, we create cylindrical representations of the inner and outer envelopes, where the width of each 3D cylinder is equivalent to the FWHM derived from the $\sigma_1$ and $\sigma_2$ values reported in Table \ref{tab:radial_profiles_excl2} (FWHM $= 2.355 \times \sigma$). These cylinders are toggle-able on and off in the \href{https://faun.rc.fas.harvard.edu/czucker/Paper\_Figures/3D\_Cloud\_Topologies/gallery.html}{gallery of interactive cloud figures}. The cylinders define a mask in 3D volume density space over which the masses are computed:

$$M_{\rm cylinder_{inner}} = \Sigma_i \; \mu m_p \times n_{i, \rm inner} \; dV$$

$$M_{\rm cylinder_{outer}} = \Sigma_i \; \mu m_p \times n_{i, \rm outer} \; dV$$

\noindent Here, $dV$ is the volume element of an individual voxel $i$ in the 3D volume density map, equal to 1 pc$^{3}$. The $\Sigma_i \; n_{i, \rm inner}$ and $\Sigma_i \; n_{i, \rm outer}$ terms represent the sum over the set of voxels (in gas density space $n$) which lie within the 3D inner and outer cylindrical regions, respectively. \textit{Under the assumption that no UNM lies within the bounds of the inner envelope}, the resulting mass ratio would be:

$$\frac{M_2}{M_1} = \frac{M_{\rm cylinder_{outer}} - M_{\rm cylinder_{inner}}}{M_{\rm cylinder_{inner}}}$$

\noindent As expected, we calculate a lower value of $\frac{M_2}{M_1}$, with a median of 2.1 and a spread of 1.0. In reality the true mass ratio is likely somewhere between the two examples given here, and additional follow-up work will be needed to rule out the possibility of a UNM-to-CNM transition.

\subsubsection{Constant Ratio Between Outer and Inner Cloud Envelopes}
Regardless of whether the transition between the outer and inner envelopes is due to a a) chemical transition between HI and $\rm H_2$ gas b) a thermal transition between the UNM and the CNM; or neither, we find that the \textbf{ratio between the widths of the outer and inner envelopes across the sample is roughly 3.4,} with a relatively small scatter of $\approx 1$.  This ratio could be physically significant, with the potential to constrain the formation timescale of $\rm H_2$, or the rate at which gas flows between different thermal phases of the diffuse interstellar medium (e.g. between the UNM and CNM). More targeted analytic modeling \citep[e.g. in the style of][]{Sternberg_2014, Bialy_2016, Krumholz_2009}, further contextualization with CO and HI spectral-line maps, and comparisons to simulations that track HI, $\rm H_2$ and CO formation and destruction in a time-dependent chemical network \citep[e.g. the Cloud Factory Simulations;][]{Smith_2020} should shed more light on this behavior.

\begin{figure}[h!]
\begin{center}
\includegraphics[width=1\columnwidth]{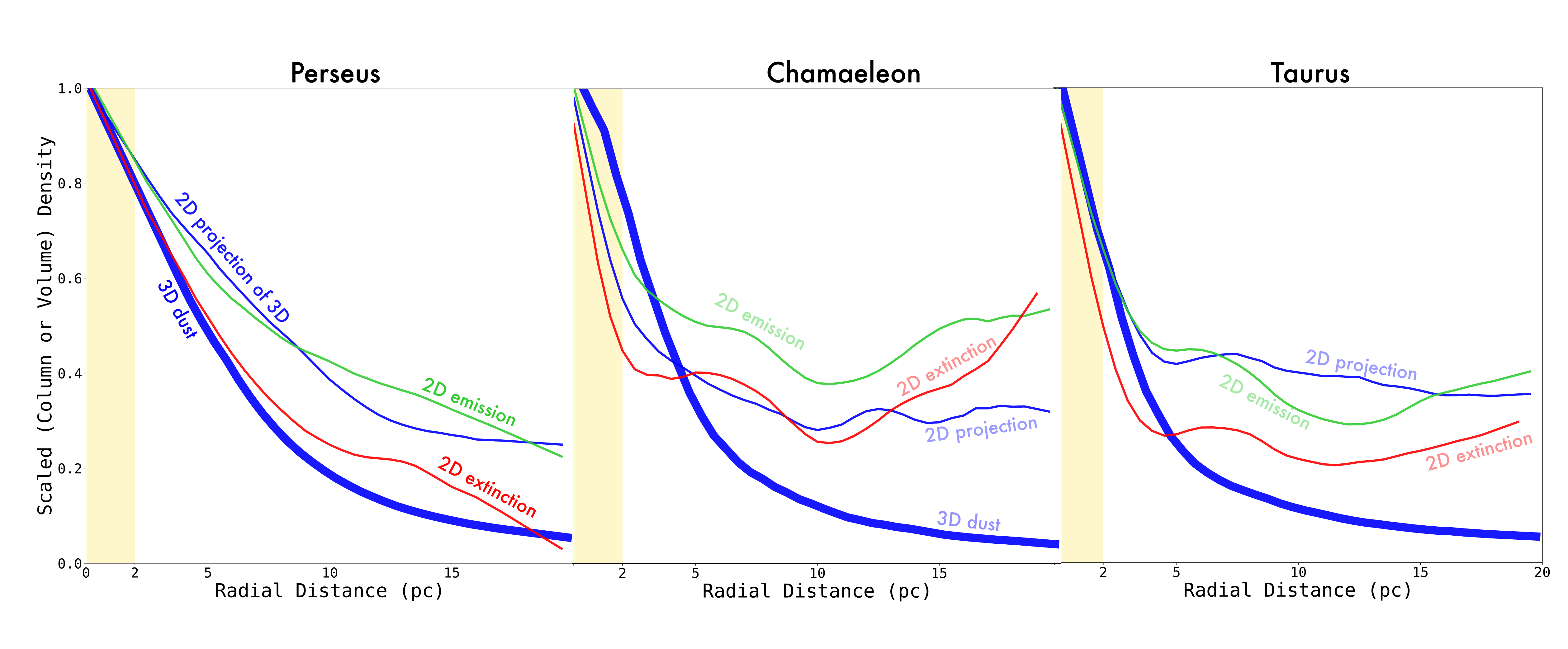}
\caption{Comparison between 2D radial column density profiles and 3D radial volume density profiles for the Perseus (left), Chamaeleon (middle) and Taurus (right) molecular clouds. The \textbf{thick blue lines} show  \textbf{3D volume density }derived from the \citet{Leike_2020} 3D dust map. The \textbf{thin blue lines} show  \textbf{2D column density} profiles derived by \textbf{projecting} the   \citet{Leike_2020} 3D dust map on to the plane of the sky (\S \ref{leikenicest}). The \textbf{red lines} show \textbf{2D} column density from the \textbf{NICEST} map  \citep{Lombardi_2009}. The \textbf{green lines} show profiles derived from \textbf{2D dust emission} maps \citep{Schlegel_1998}. All maps have been scaled to a peak value of 1.0, to facilitate a comparison of the shapes of the volume and column density profiles. While the 2D and 3D results largely agree at radial distances $< 5$ pc, significant deviations are observed at larger radial distances, particularly for the Chamaeleon and Taurus clouds. \label{fig:2d_vs_3d}}
\end{center}
\end{figure}

\subsection{An Isothermal Self-Gravitating Cylinder?} \label{plummer_significance}
For the past ten years, Plummer models \citep[originally describing the stellar surface density of globular clusters;][]{Plummer_1911} have often been used to characterize the radial profiles of filaments extracted from $\rm H_{2}$ column density maps as part of the Herschel Gould Belt Survey \citep{Andre_2010}. These studies \citep{Arzoumanian_2011, Arzoumanian_2019, Palmeirim_2013, Cox_2016} adopt an idealized Plummer-like analytic model for the (column) density profile of a cylindrical filament, featuring a denser, flat inner portion and an outer portion that falls off like a power-law at larger radii: 
\begin{equation} \label{eq:plummer_2d_3d}
n(r) = \frac{n_{0}}{\left[1+\left({r/R_{\rm flat}}\right)^{2}\right]^{p/2}}\  \longrightarrow 
 \Sigma(r) = A_p \frac{N_{\rm 0}}{\left[1+\left({r/R_{\rm flat}}\right)^{2}\right]^{\frac{p-1}{2}}}, \ \ 
\end{equation}

where $n(r)$ describes the volume density profile adopted throughout this work, and $\Sigma (r)$ is the corresponding column density profile employed in the literature, with $N_0$ the peak column density, $R_\mathrm{flat}$ the flattening radius, $p$ the index of the density profile, and $A_p$ is a constant factor that accounts for the filament’s inclination angle to the plane of the sky. In column density studies, the inclination of the filament to the plane of the sky is assumed to be zero, which is not the case in our volume density analysis. Following the formalism outlined in \citet{Ostriker_1964} and Equation \ref{eq:plummer_2d_3d}, the  density structure of an isothermal, self-gravitating gaseous cylinder in hydrostatic equilibrium corresponds to a special case with a power-law index $p = 4$ or $n \propto r^{-4}$ at large radii ($r >> R_{\rm flat}$). Analyzing the \textit{Herschel} column density distribution of filaments, various works \citep{Arzoumanian_2011, Arzoumanian_2019, Palmeirim_2013, Cox_2016} have found that  these filaments are inconsistent with the $p=4$ scenario, instead preferring much shallower power-law indices ($p = 1.5 - 2.5$).  

The \textit{Herschel} filaments analyzed in previous work lie \textit{within} the local molecular clouds we analyze here, persisting at much higher column densities ($N_{\rm H_2} > 10^{22} \; \rm cm^{-2}$), and on much smaller scales (widths $\approx$ a few tenths of a parsec, lengths $\approx$ a few parsecs). As such, they are unresolved in the \citet{Leike_2020} 3D dust maps we analyze. Nevertheless, the same formalism can be applied to the much larger molecular clouds these dense filaments inhabit, since the analytic models are scale-free, and we have shown here that local molecular clouds are well-described as cylindrical filaments, even at lower densities ($n < 100 \;  \rm cm^{-3}$), where their aspect ratios are often in excess of $3:1$ (see lengths and widths in Tables \ref{tab:skeletonization} and \ref{tab:radial_profiles_excl2}). Thus, while we target a very different regime -- about two orders of magnitude lower in density, and an order of magnitude larger spatial scale -- the comparison to the classical $p=4$ \citet{Ostriker_1964} model for an isothermal, self-gravitating filament in hydrostatic equilibrium, is still relevant. 

We find that the local, large-scale, molecular clouds studied here have  shallower radial density profiles, with typical power-law indices $p$ between $1.5-2.5$, than would an isothermal  self-gravitating cylinder in hydrostatic equilibrium. This shallower power-law, while inconsistent with isothermal self-gravitating filaments in the absence of a magnetic field, could be consistent with some models of magnetized filaments \citep{Fiege_2000}.  Previous studies of the structure of the magnetic field in solar neighborhood clouds find a strong coupling between the gas column density and the magnetic field orientation, which is consistent with the idea that magnetic fields play an important role in structuring the interstellar medium in and around local molecular clouds \citep[see e.g.][]{Soler_2019}. 

However, given that the parameters of the Plummer profile are \textit{highly} degenerate (see discussion in \S \ref{width_results}), and driven by behavior at the tail of the distribution (at very low densities), we are hesitant to assign any physical significance to the shape of the Plummer profile, other than broadly noting the inconsistency with the \citet{Ostriker_1964} result. Similar to the two-component Gaussian results, future comparison with numerical simulations should allow for more robust interpretation of the physical significance, if any, of the shape of the Plummer profiles.

\subsection{Comparison between 3D Volume Density and 2D Column Density Radial Profiles}
\textbf{The ultimate goal of this work is to lay the foundation for a robust characterization and interpretation of the structure of molecular clouds.} To do so will require the contextualization of these newly presented 3D results with extant 2D approaches. Traditionally, there have been three ways of measuring the column density structure of molecular clouds: near-infrared extinction mapping \citep[e.g. the NICEST method, \S \ref{nicest};][]{Lombardi_2009}, thermal dust emission mapping in the far-infrared \citep[e.g. the SFD or Planck dust maps;][]{Schlegel_1998, Planck_2016}, and spectral-line mapping of CO isotopologues. A systematic analysis of and comparison to the 2D radial profiles of molecular clouds is beyond the scope of this work. So, as a precursor to a  more expansive study, we offer in Figure \ref{fig:2d_vs_3d} an initial comparison of our 3D radial volume density profile results with traditional 2D column density results for three clouds in the sample (Perseus, Chamaeleon, and Taurus). We compare the 3D radial profiles with three measurements of the 2D radial profile: one obtained from near-infrared extinction using the NICEST maps (see \S \ref{nicest}), one obtained from far-infrared dust emission maps \citep[][]{Schlegel_1998}, and one obtained by projecting the 3D dust map from \citet{Leike_2020} on to the plane of the sky, using the same column density maps from \S \ref{leikenicest}. Each curve has been smoothed with a kernel roughly equal to 2 pc (the effective spatial resolution of the \citet{Leike_2020} dust map), to alleviate noise in the profiles at large radial distances and facilitate comparison between the different tracers.\footnote{In all three cases, the radial profiles are obtained identically to our 3D results presented in \S \ref{profile_build}, except the threshold for mask (and thus skeleton) creation was defined in 2D column density space as opposed to 3D volume density space. All 2D maps are convolved to the same resolution. The difference in ``zero point" emission and extinction levels in traditional 2D approaches precludes the application of an identical column density threshold for all three column density maps. To avoid biases in our analysis, we define thresholds using the total area of the cloud masks, rather than absolute column density. That is, we adopt the same $A_K = 0.1$ mag threshold (converted to $N_{\rm H}$) used in the \S \ref{mass} NICEST results to define the optimal mask, and set the thresholds for the 2D dust emission \citep{Schlegel_1998} and projected \citet{Leike_2020} 2D results so they occupy approximately the same area. By doing so, all masks are defined using a narrow range in column density, spanning $\rm N = 1 - 3 \times 10^{21} \; cm^{-2}$. A tutorial for how to build 2D radial column density profiles with the \texttt{RadFil} package is available on Zenodo \citep{zucker_zenodo}.}  We largely defer comparison to the 2D structure of clouds obtained from velocity-integrated spectral-line maps to future work, given the need for careful accounting of opacity effects and the presence of CO dark gas. In order to facilitate a comparison of the \textit{shapes} of the profiles, we scale the radial volume density profile and the three radial column density profiles to a peak value of one. 

In contrast to the 3D radial profiles, the 2D radial profiles show a much larger variation in structure both within and between clouds. In general, we find good agreement between 2D and 3D results at small radial distances ($r < 5$ pc), particularly for the Perseus and Taurus clouds. However, at radial distances $>5$ pc, we observe a flattening of the 2D radial profiles that does not manifest in 3D. \textit{These larger radial distances are beyond the $A_K = 0.1$ mag boundary used to validate the 3D-dust-based cloud masses in \S \ref{mass_results}.} We observe  flattening at large radii even in the projected dust map from \citet{Leike_2020} (thin blue trace), which makes use of the same data as the 3D radial profile (thick blue trace), but simply integrates along the line of sight. The flattening seen (only) in 2D suggests two underlying causes. First, diffuse dust unassociated with (but potentially in the same general vicinity of) the cloud could be contaminating the inferred radial profiles at large radial distances. Second, projection effects intrinsic to the cloud itself (e.g. due to the cloud's orientation with respect to the observer) could be driving this variation. In either or both scenarios, discrepancy between 2D and 3D cloud structure will have profound implications for interpreting the shapes of the column density PDFs of molecular clouds \citep[c.f.][]{Alves_2017,Chen_2019}. These deviations also underline the difficulty of validating the atomic outer envelopes of clouds and the tails of the Plummer profiles (\S \ref{plummer_significance}) using traditional 2D dust maps. Disentangling these effects will require careful calibration of 2D dust extinction and emission maps; careful supplementation of the 2D dust information with opacity-corrected spectral line maps of CO, HI, and \textsc{[CII]}, and contextualization of both types of data in light of synthetic observations of local cloud analogs.  

\section{Conclusion} \label{conclusion}
The rise of \textit{Gaia} has transformed our understanding not just of the 3D distribution of stars, but also of the 3D distribution of dust that reddens their colors. The recent release of the \citet{Leike_2020} 3D dust map, which provides 1 pc spatial resolution out to a distance of several hundred parsecs, offers an unprecedented opportunity to measure and describe the shape and internal structure of local molecular clouds on scales previously obtainable only in simulations. Converting from differential dust extinction in the \textit{Gaia} G-band to volume density using a wavelength-dependent extinction curve, we characterize the shapes and thicknesses of famous local clouds in 3D gas density space for the first time. Local clouds appear filamentary, even at gas densities $n < 100 \;  \rm cm^{-3}$. Thus, we start by extending the \texttt{FilFinder} algorithm -- originally applied to 2D maps of dust emission \citep{Koch_2015} --- into 3D, to determine the skeletons of nearby clouds, equivalent to a one-voxel-wide representation of their 3D structure. We then extend the \texttt{RadFil} algorithm \citep{Zucker_2018b} to measure the radial volume density profiles with respect to  the cloud spines. Our results are as follows:

\begin{itemize}
\item By projecting each cloud's skeleton -- determined in 3D $(x,y,z)$ cartesian space --- back to sky coordinates $(l,b,d)$, we constrain the distances to famous local molecular clouds. The 1 pc uncertainty on the distances allows for complete characterization of distance gradients and the presence of multiple components. 

\item We model each cloud's radial volume density profile with three functions: a two-component Gaussian, a single-component Gaussian, and a Plummer-like function. Comparing the relative quality of models, we find that a single-component Gaussian function is universally disfavored, with significantly worse odds compared with the two-component Gaussian and Plummer models.

\item For the two-component Gaussian model, we find widths $\sigma_2$ of the ``outer" Gaussian between $8 - 18$ pc, with amplitudes between $n = 5 - 15$ $\rm cm^{-3}$. For the ``inner" Gaussian, we find widths $\sigma_1$ of $2.5 - 4.9$ pc, with amplitudes between $n = 25 - 52$ $\rm cm^{-3}$. The typical ratio of the outer to inner widths is $3.4:1$, with relatively small scatter ($< 1$). We hypothesize that the boundary might represent a chemical transition between HI and $\rm H_{2}$ gas in local clouds, with the inner regions probing warm, diffuse molecular gas, and the outer regions tracing the extended atomic envelopes. Another plausible scenario includes a thermal transition between the unstable neutral medium (UNM) and the cold neutral medium (CNM). Future comparison with numerical simulations \citep[e.g.][]{Smith_2020} will be needed to verify these hypotheses.

\item We produce synthetic 2D dust extinction maps ($A_K$) by projecting the \citet{Leike_2020} 3D dust distribution on the plane of the sky. We compare the 2D extinction maps derived from 3D dust with traditional integrated 2D dust extinction maps \citep[from the NICEST algorithm;][]{Lombardi_2009}. Computing cloud masses, we determine that the \citet{Leike_2020} map is recovering a majority of each cloud's mass. However, the 3D dust is insensitive to structure above $A_K = 0.3$ mag ($A_V = 3$ mag)

\item All data underpinning this work are publicly available online at the Harvard Dataverse (\url{https://dataverse.harvard.edu/dataverse/3D\_cloud\_structure/}). An online gallery of interactive figures illustrating our cloud results is also available online \href{https://faun.rc.fas.harvard.edu/czucker/Paper\_Figures/3D\_Cloud\_Topologies/gallery.html}{here}. 

\end{itemize}

Ultimately, these results are only the beginning of what is possible through the combination of \textit{Gaia} astrometry, large photometric surveys, innovative computational and statistical techniques, and exploratory 3D data visualization. The quality of the \citet{Leike_2020} 3D dust map, and of similar maps extending out many more kiloparsecs \citep{Green_2019, Zucker_2020}, relies on accurate measurements of the distance and integrated extinction to individual stars. Not only will future \textit{Gaia} data releases produce increasingly better estimates of stellar distance, but upcoming photometric surveys like LSST will allow for improved constraints on the integrated extinction. Future 3D dust maps, built on much larger catalogs of stellar properties (e.g. the \texttt{brutus} algorithm; Speagle et al. 2021, in prep.) should allow for improved models of the 3D dust distribution, probing not only clouds at larger distances, but also with improved sensitivity to higher densities within molecular clouds. 

\acknowledgements
We would like to thank our anonymous referee, whose thoughtful reading of our manuscript improved the quality of this work, particularly with regards to the physical interpretation of the radial profile results. 

We would like to thank Charles Lada, Thomas Dame, Robert Benjamin, and Gina Panopoulou for helpful feedback and discussions which improved the quality of this manuscript. 

The visualization, exploration, and interpretation of data presented in this work was made possible using the \texttt{glue} visualization software, supported under NSF grant numbers OAC-1739657 and CDS\&E:AAG-1908419. \\

D.P.F. and C.Z. acknowledge support by NSF grant AST-1614941, ``Exploring the Galaxy: 3-Dimensional Structure and Stellar Streams.'' \\

We thank the developers of the X-Toolkit \citep{XTK}, which enabled the fast volume rendering of our results in the browser, in Figures \ref{fig:perseus_topology} and \ref{fig:chamaeleon_topology}.

\software{\texttt{astropy} \citep{Astropy_2018},  \texttt{bokeh} \citep{Bokeh}, \texttt{dynesty} \citep{Speagle_2020}, \texttt{glue} \citep{glueviz_2017}, \texttt{numpy} \citep{numpy}, \texttt{PyVista} \citep{Sullivan_2019}, \texttt{scipy} \citep{scipy}, \texttt{FilFinder} \citep{Koch_2015}, \texttt{scikit-image} \citep{scikit_image}, \texttt{skan} \citep{NunezIglesias_2018_skan}, \texttt{RadFil} \citep{Zucker_2018b}}

\bibliographystyle{aasjournal}
\bibliography{topology_revised.bib}{}

\appendix
\restartappendixnumbering

\section{Radial Profiles for Full Cloud Sample}
\begin{figure}[h!]
\gridline{\fig{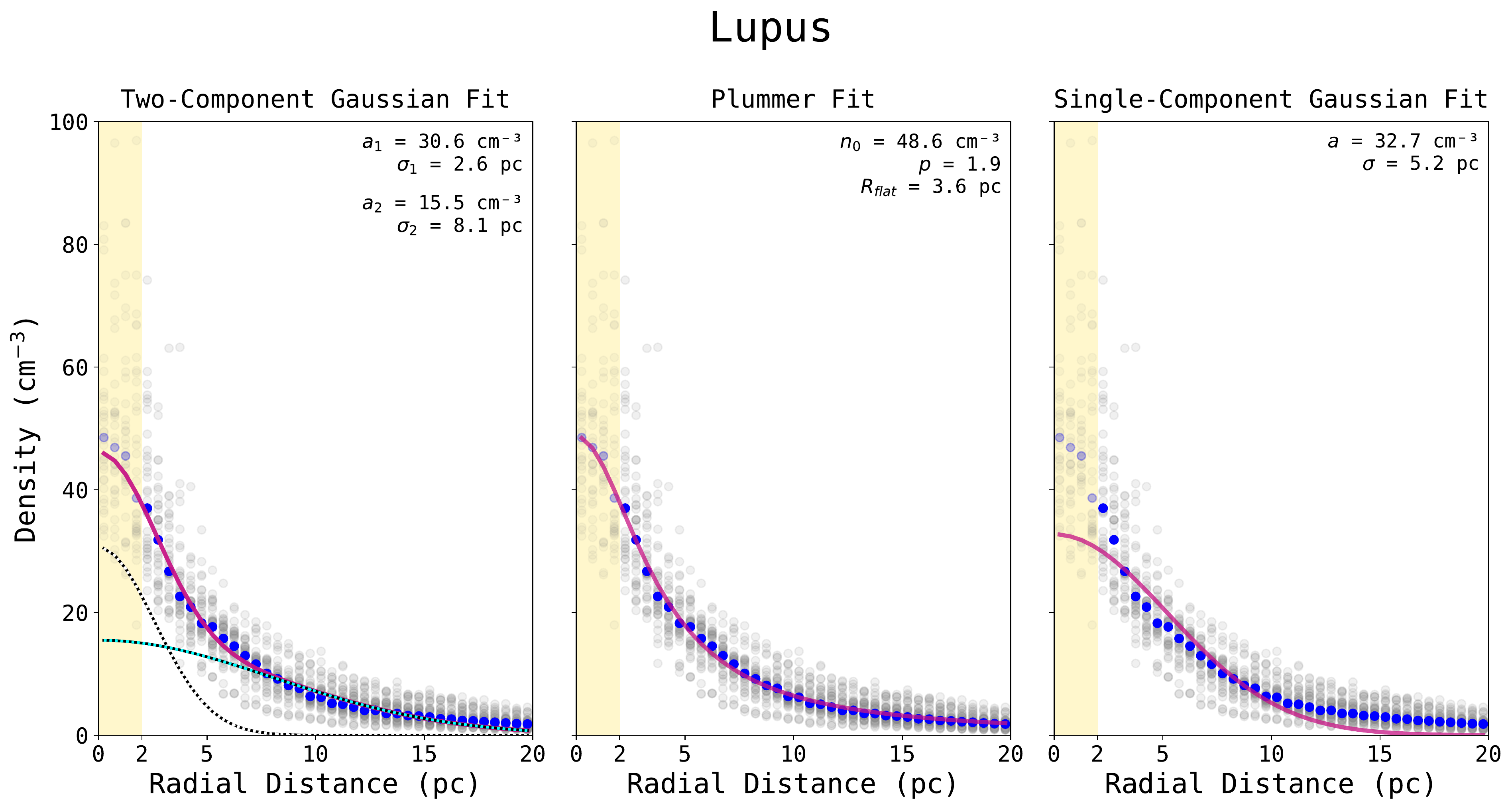}{0.6\textwidth}{}\vspace{-0.8cm}}
\gridline{\fig{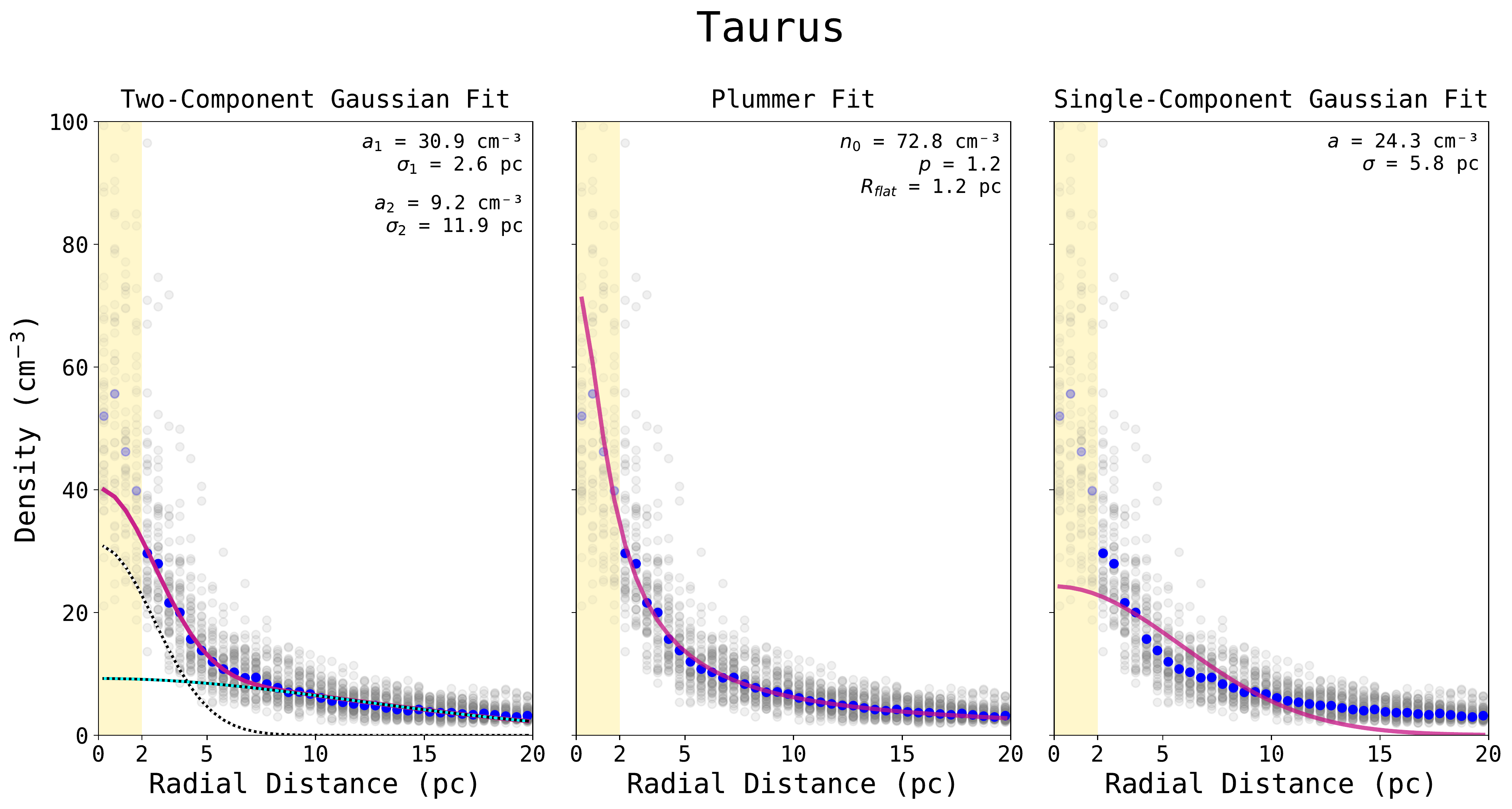}{0.6\textwidth}{}\vspace{-0.8cm}}
\gridline{\fig{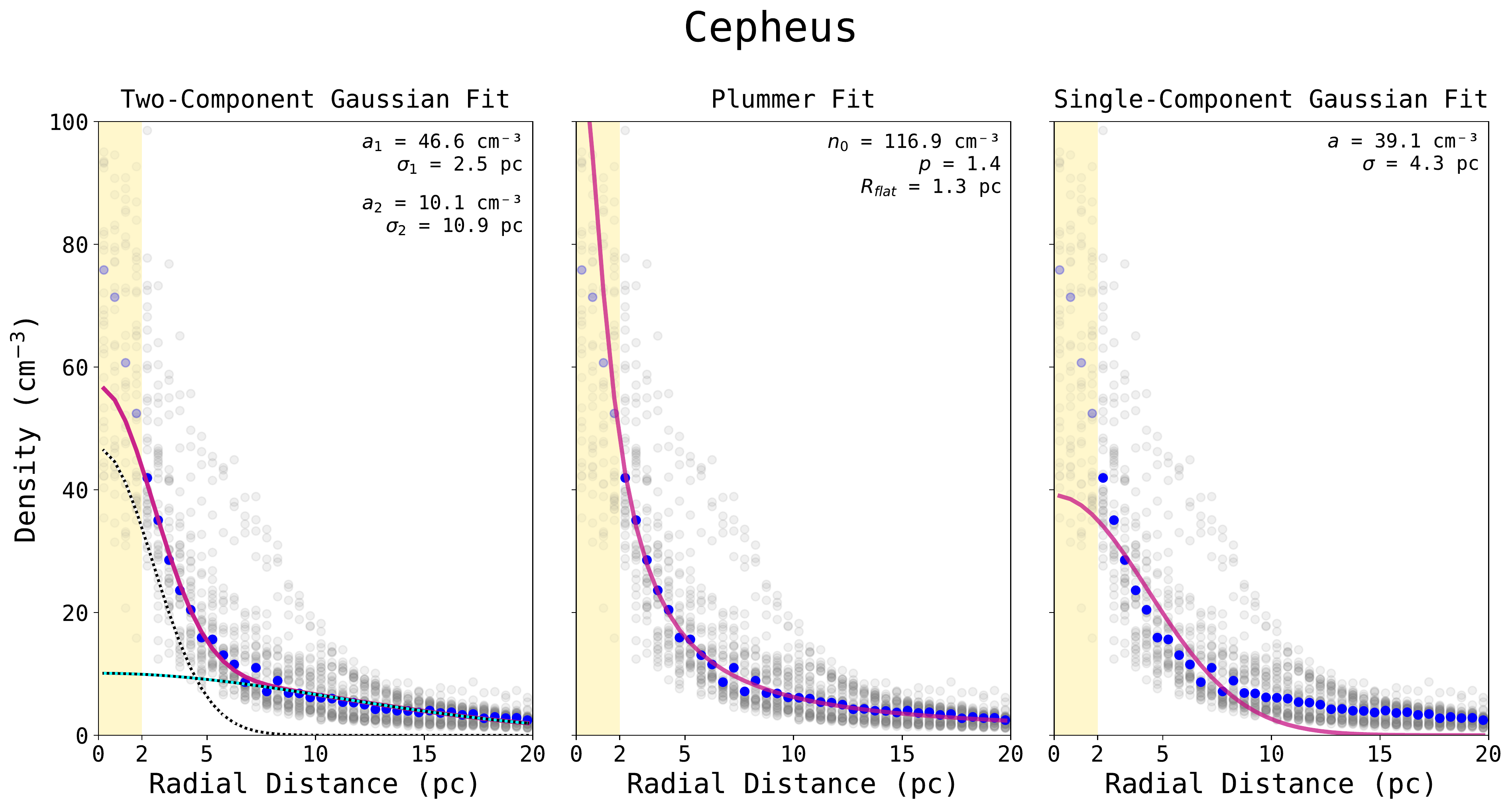}{0.6\textwidth}{}\vspace{-0.8cm}}
\caption{Same as in Figure \ref{fig:profile_examples_excl2} except for the Lupus molecular cloud (top), the Taurus molecular cloud (middle), and the Cepheus molecular cloud (bottom) \label{fig:profile_batch1}}
\end{figure}

\begin{figure}[h!]
\gridline{\fig{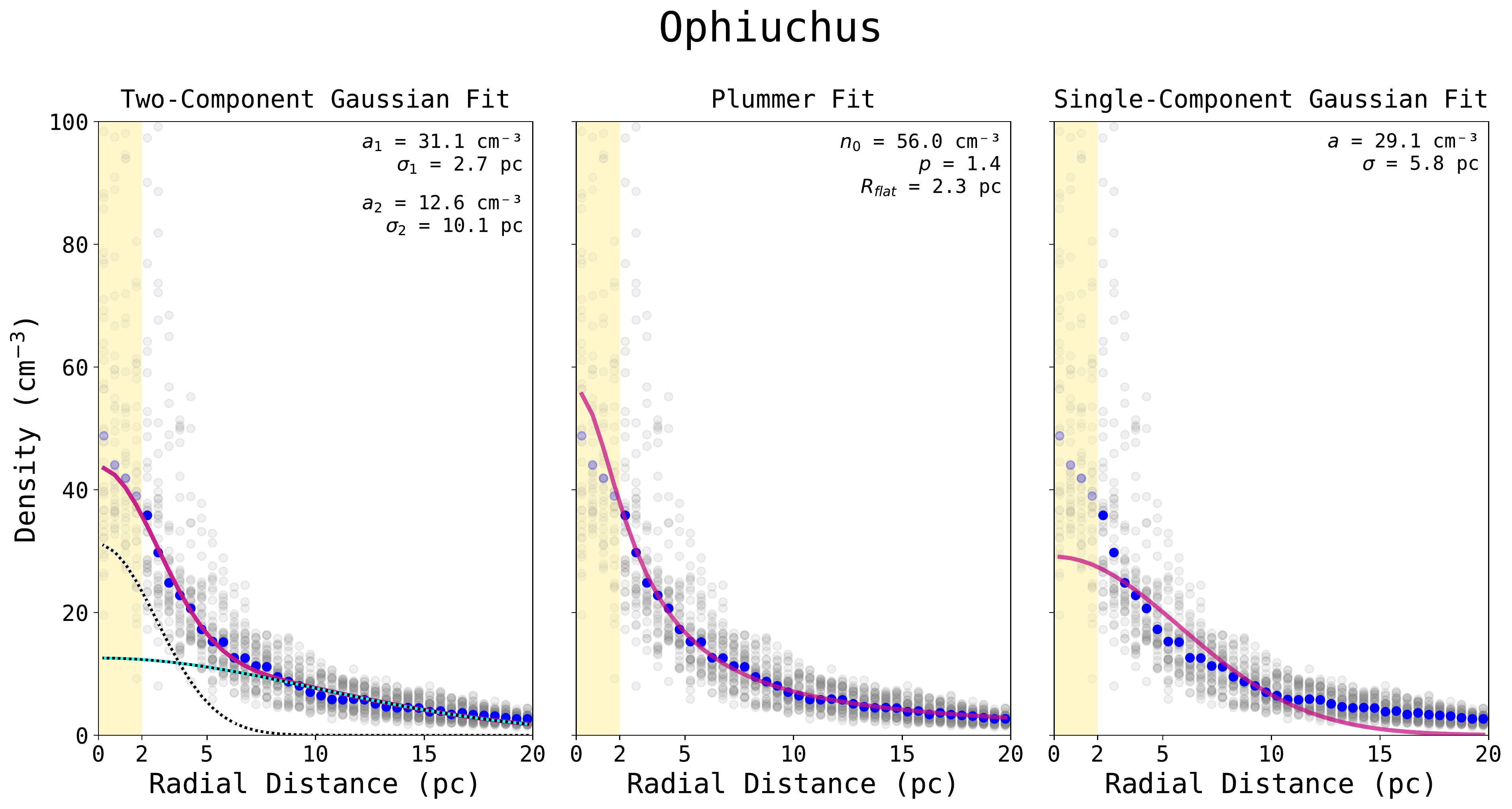}{0.6\textwidth}{}\vspace{-0.8cm}}
\gridline{\fig{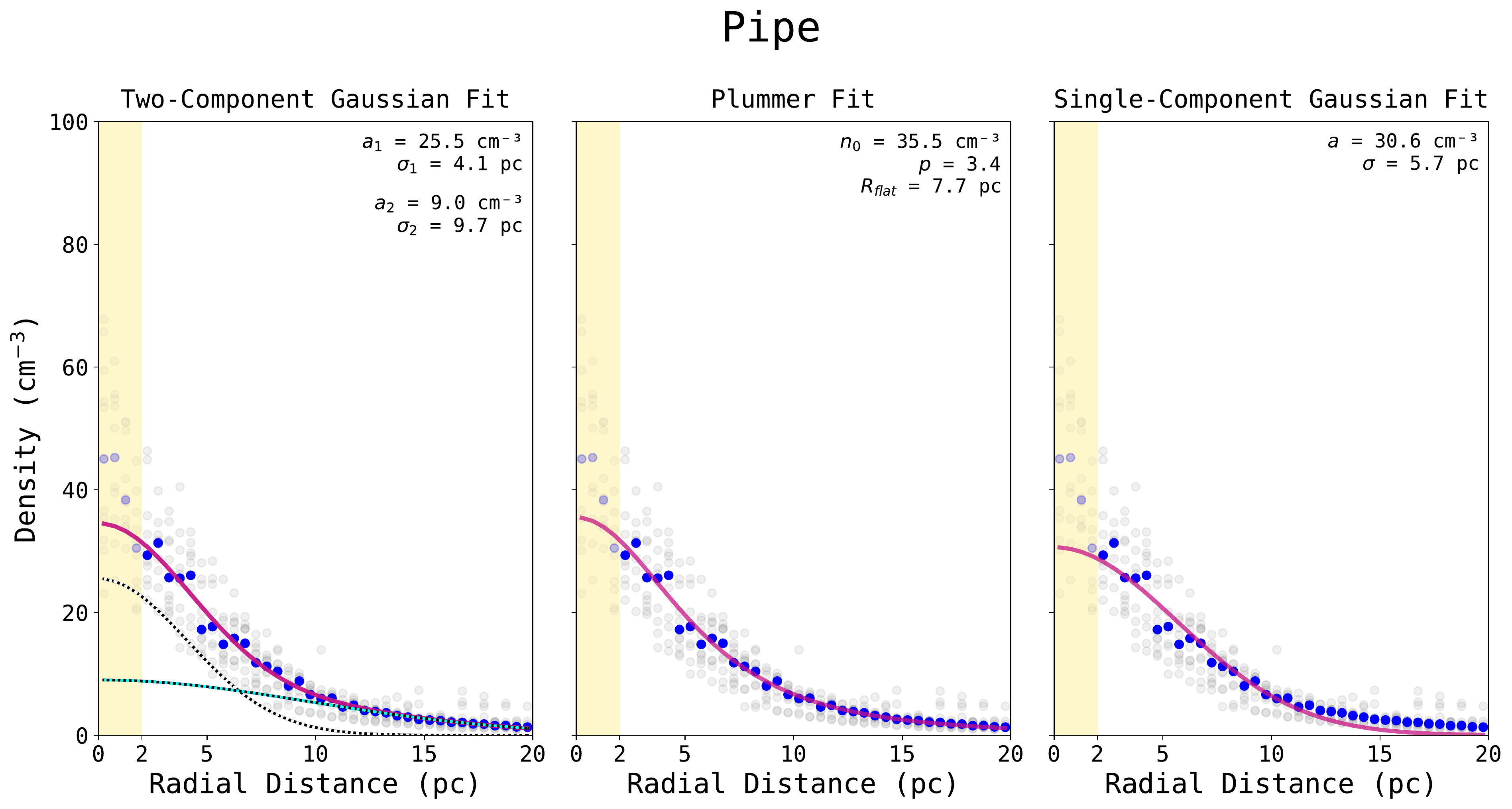}{0.6\textwidth}{}\vspace{-0.8cm}}
\gridline{\fig{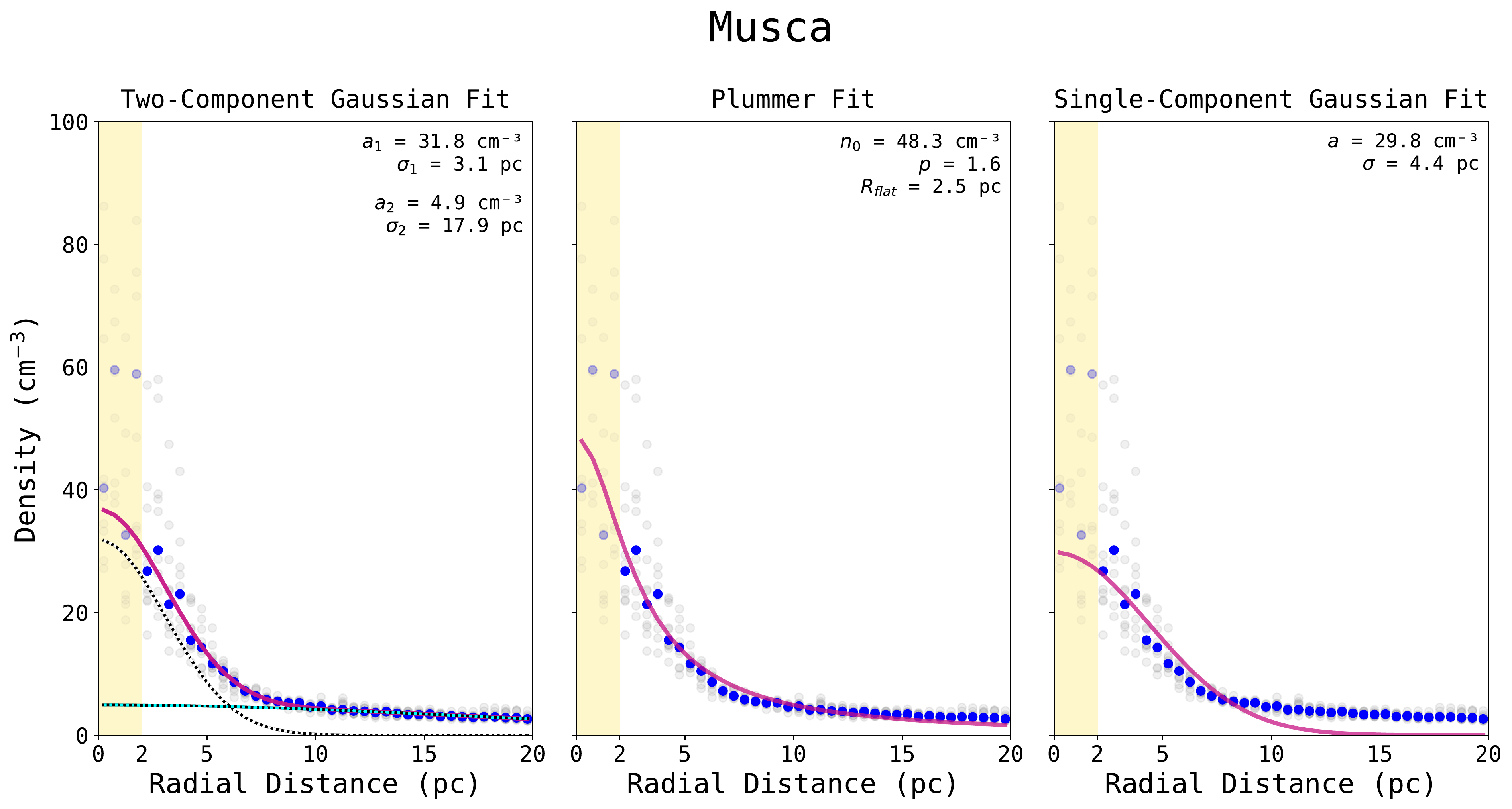}{0.6\textwidth}{}\vspace{-0.8cm}}
\caption{Same as in Figure \ref{fig:profile_examples_excl2} except for the Ophiuchus molecular cloud (top), the Pipe nebula (middle), and the Musca dark cloud (bottom)\label{fig:profile_batch2}}
\end{figure}

\begin{figure}[h!]
\gridline{\fig{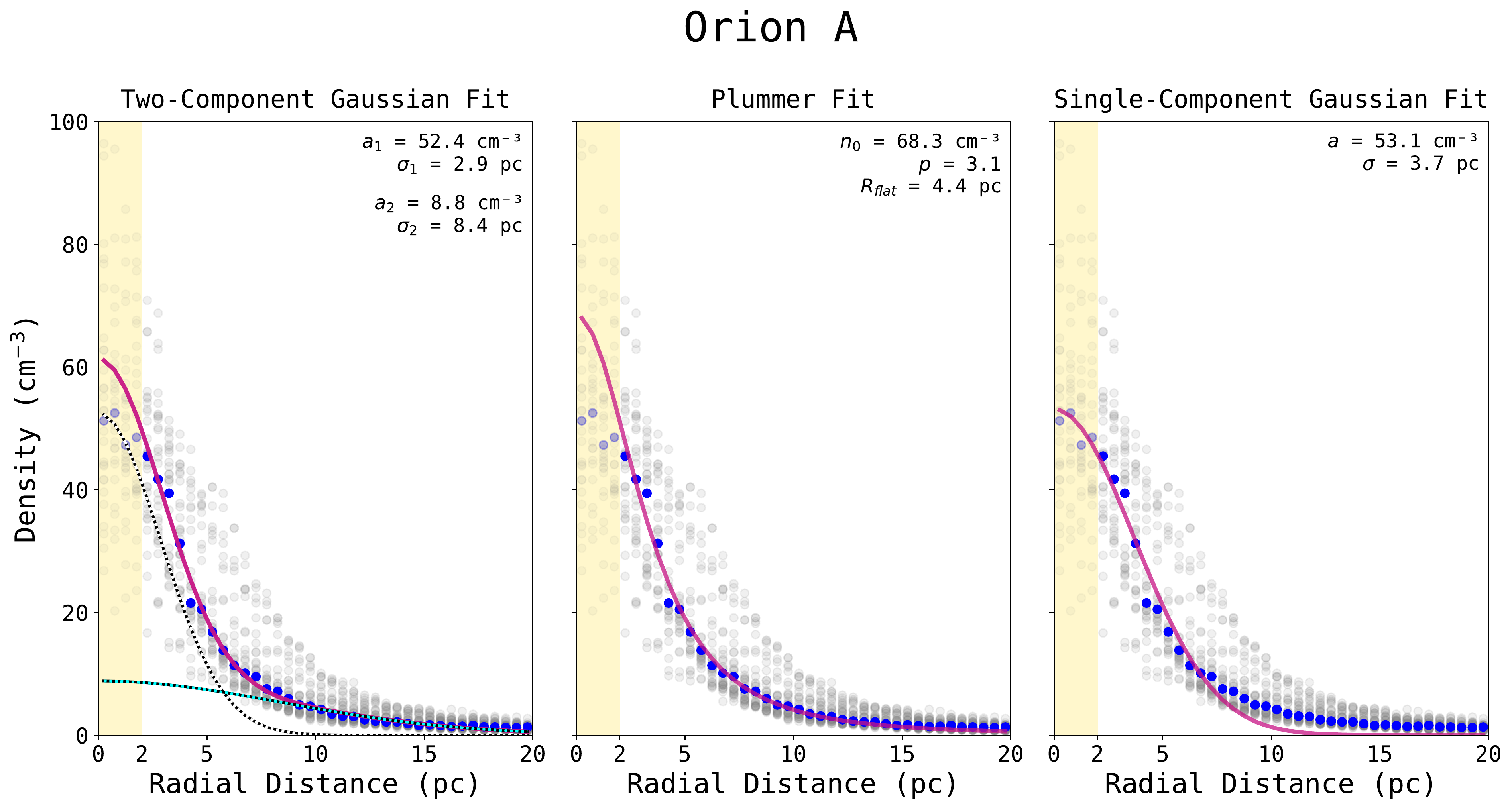}{0.6\textwidth}{}\vspace{-0.8cm}}
\gridline{\fig{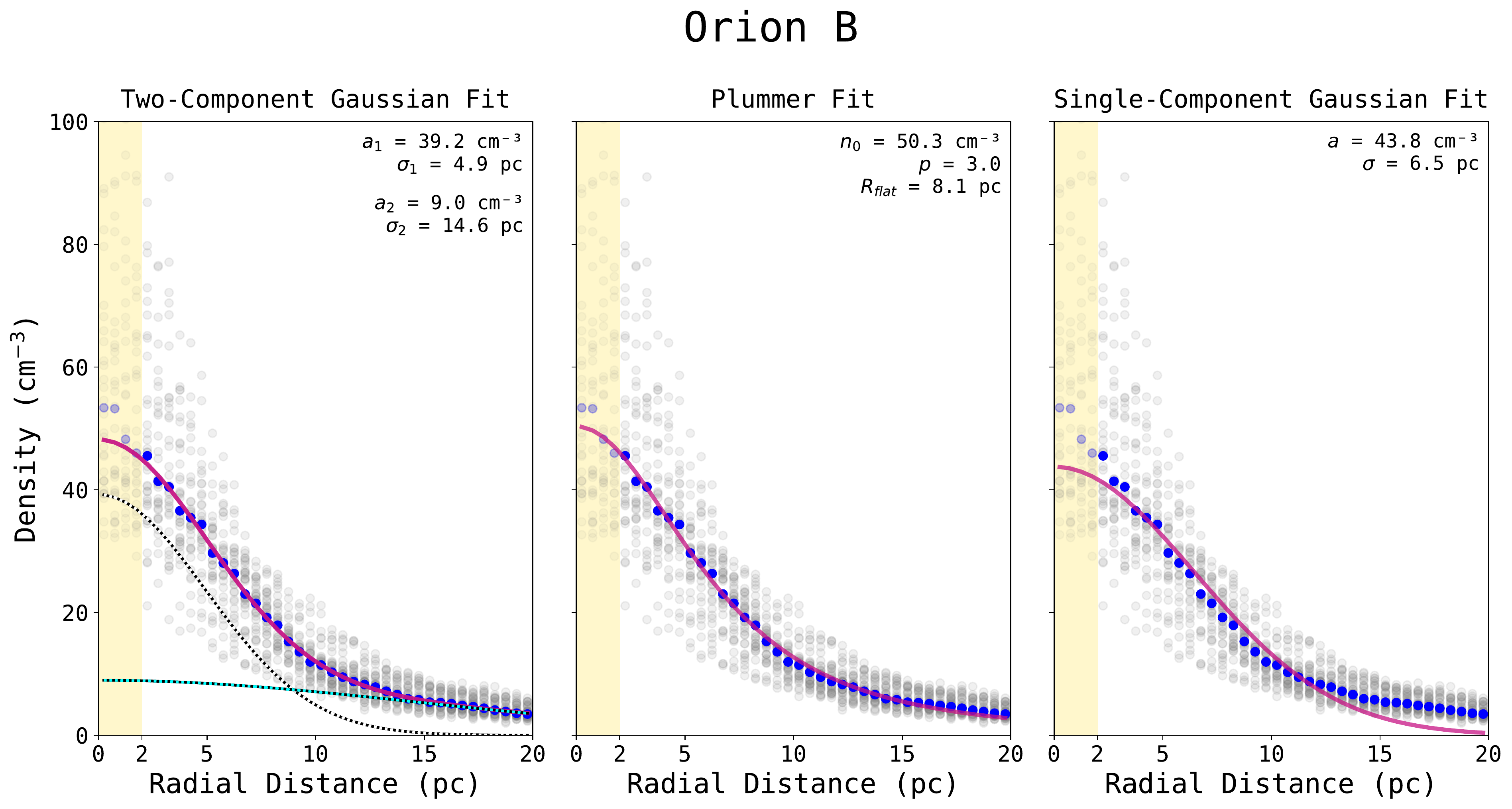}{0.6\textwidth}{}\vspace{-0.8cm}}
\gridline{\fig{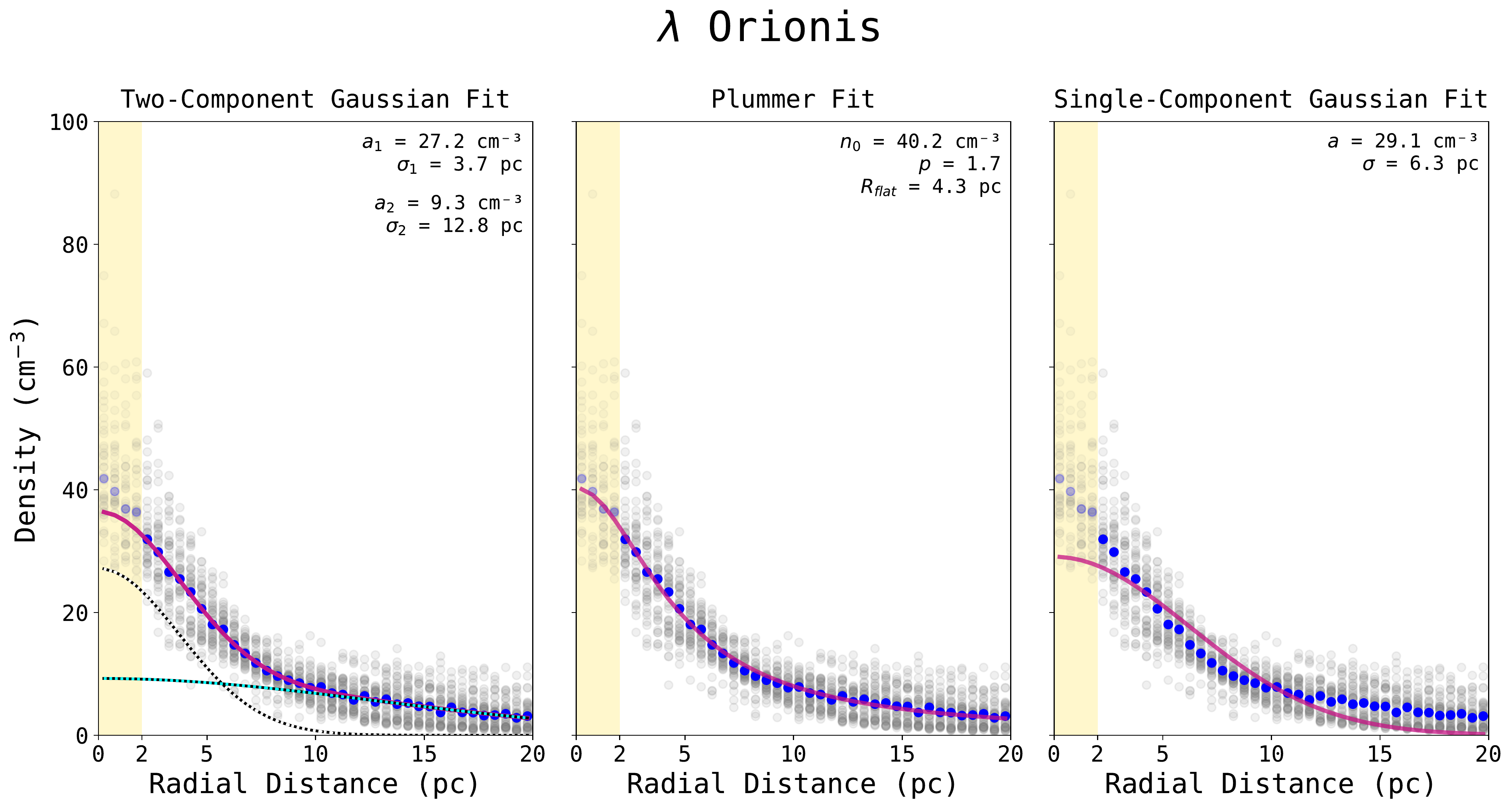}{0.6\textwidth}{}\vspace{-0.8cm}}
\caption{Same as in Figure \ref{fig:profile_examples_excl2} except for the Orion A molecular cloud (top), the Orion B molecular cloud (middle), and the $\lambda$ Orionis molecular cloud (bottom). These clouds are at the edge of the \citet{Leike_2020} 3D dust grid, so the results should be treated with more caution than our ``complete" cloud sample.\label{fig:profile_batch3} }
\end{figure}

In Figure \ref{fig:profile_examples_excl2}, we present the radial density profiles for two clouds in the sample --- Perseus and Chamaeleon. In Figures \ref{fig:profile_batch1}, \ref{fig:profile_batch2}, and \ref{fig:profile_batch3}, we show the same radial density profiles for the remaining nine clouds.

\newpage

\restartappendixnumbering

\section{Cloud Widths as a Function of Distance} \label{distancebias}
\begin{figure}[h!]
\begin{center}
\includegraphics[width=0.5\columnwidth]{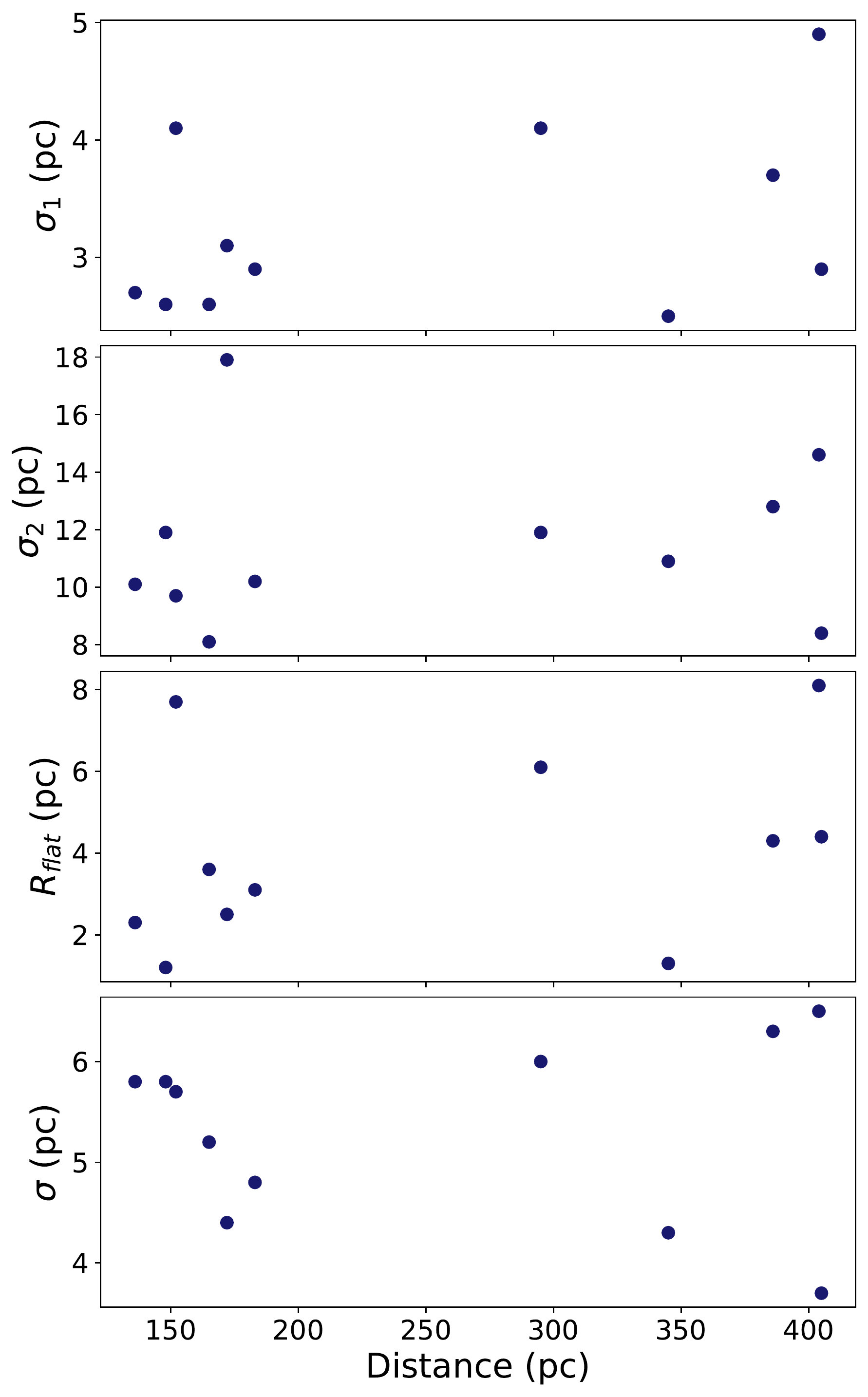}
\caption{The derived width parameters as a function of cloud distance, with the ``inner" width $\sigma_1$ of the two-component Gaussian fit, the ``outer" width $\sigma_2$ of the two-component Gaussian fit, the flattening radius $R_\mathrm{flat}$ of the Plummer fit, and the standard deviation $\sigma$ of the single-component Gaussian fit shown from top to bottom.  \label{fig:distancebias}}
\end{center}
\end{figure}

To determine whether the distance of the cloud influences the widths we derive, we plot the four width parameters fitted for in this study ($\sigma_1$ and $\sigma_2$ from the two-component Gaussian function, $R_\mathrm{flat}$ from the Plummer function, and $\sigma$ from the single-component Gaussian) as a function of distance. The results are shown in Figure \ref{fig:distancebias}. If distance was influencing the cloud widths, we would expect the cloud widths to increase as a function of distance. As we show in Figure \ref{fig:distancebias}, we find no such correlation, indicating that the cloud distance is not strongly biasing our results. 

\restartappendixnumbering
\clearpage
\newpage
\section{Comparison between 3D Dust Correlation Kernel and Measured Profiles} \label{appendix_correlation}

In this section, we compare the shape of the correlation kernel of the \citet{Leike_2020} 3D dust reconstruction with the shape of our cloud profiles. Our goal is to determine the extent to which our cloud profiles are driven by data local to the cloud, versus the prior on the dust extinction density. \citet{Leike_2020} assume the prior on the dust extinction density $s_x$ to be positive and spatially correlated, enforced via a log-normal Gaussian process prior of the form: 

\begin{align*}
    s_x &= \rho_0\,\text{exp}(\tau_x) \\
    \tau & \curvearrowleft \mathcal{G}(\tau \vert 0,T) 
\end{align*}

where $\rho_0$ is the prior median extinction density and $T$ is the correlation kernel of the Gaussian Process $\tau_x$. The $\rho_0$ term is a hyperparameter of the model, and is set to $\rho_0 = \frac{1}{\rm 1000 \; pc^{-1}}$. The correlation kernel $T$ is inferred during their reconstruction. To parallelize the reconstruction, \citet{Leike_2020} split the local volume into eight octants, with the correlation kernel inferred on an octant by octant basis. We take samples of the correlation kernel inferred for the octant in which each cloud resides, which is equivalent to obtaining samples of the two-point correlation matrix of the logarithmic dust density. The two-point correlation matrix directly indicates how the density at one point is correlated with another point in 3D space.  

\begin{figure}
\begin{center}
\includegraphics[width=0.9\columnwidth]{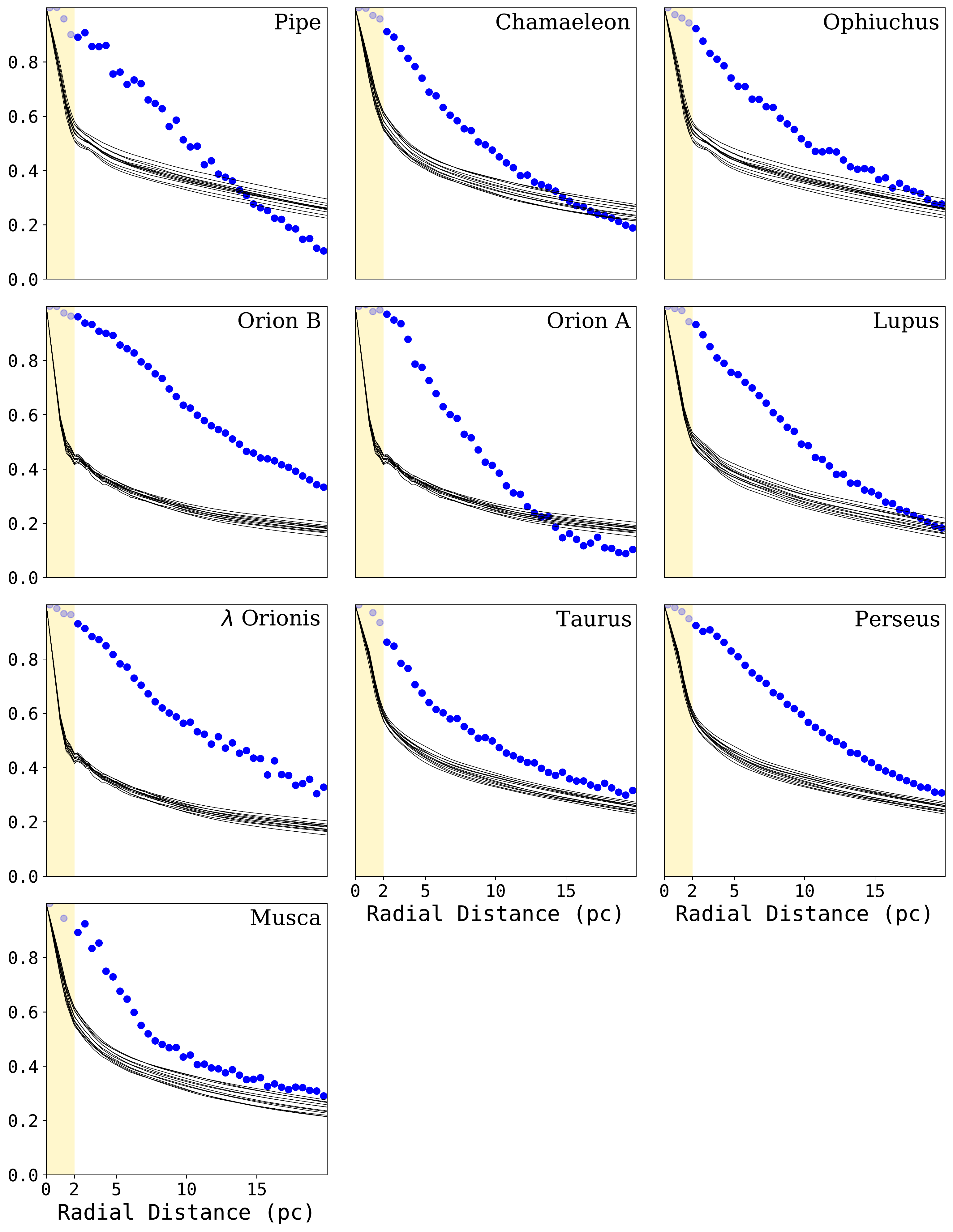}
\caption{Comparison between samples of the two-point correlation function derived from the \citet{Leike_2020} 3D dust kernel (black lines) and the derived correlation of each cloud's radial profile obtained in this work (blue points). Both the 3D dust map's correlation kernel and correlation of the cloud radial profiles have been scaled to a peak value of one. At smaller radial distances, our measured profiles deviate significantly from the underlying correlation kernel used to derive the map, indicating that the inner widths we obtain for the clouds are robust. However, at larger radial distances,  the shape of the kernel and profiles become similar, indicating the dust at larger radial distances may be prior-dominated. However, since the prior is inferred using real cloud structure, the outer envelopes could still be physically meaningful. See Appendix \ref{appendix_correlation} for full details. \label{fig:correlation}}
\end{center}
\end{figure}

To facilitate a comparison between the two-point correlation matrix and our profiles, we compute the correlation of each measured profile $C_p (r)$. Before doing so, we convert the radial profiles as a function of gas density $n(r)$, utilized throughout this study, back to the native units of the \citet{Leike_2020} 3D dust map, by dividing each profile by a factor of 880 (see \S \ref{data}) to obtain $s_x(r)$. Since the two-point correlation matrix describes the logarthmic dust density, rather than the dust density, we further transform $s_x (r)$ into logarithmic density space, hereafter $p(r)$:

\begin{equation} \label{eq:logdensity}
p(r) = \text{log} (s_x (r) \times 1000) 
\end{equation}

\noindent where the factor of 1000 accounts for the scaling of the hyperparameter $\rho_0$. To calculate the correlation of the profile $C_p (r)$, we compute:

\begin{equation}
C_p (r) = p(r) \times p(0)
\end{equation}

where $p(0)$ is the logarithmic density of the radial profile in Equation \ref{eq:logdensity} at a radial distance of 0 pc. Given the overdensity of the dust at $r = 0\,\text{pc}$, this comparison describes how much of the dust at larger radial distances can be explained by the correlation kernel of the dust map. In Figure \ref{fig:correlation} we plot $T$ versus $C_p(r)$ for each cloud, where $T$ constitutes samples of two-point correlation matrix inferred in the cloud's octant, and $C_p(r)$ is the correlation of the profile we construct in \S \ref{profile_build}. We normalize both $T$ and $C_p(r)$ to a peak value of 1.0. 

We observe that the shape of $C_p(r)$ and $T$ are qualitatively different between radial distances of 2 and 10 pc, indicating that the ``inner" widths of our two-component Gaussian results are driven by data local to the cloud, rather than the kernel used in the 3D dust's reconstruction. However, for radial distances $>$ 10 pc, we see the slopes of $C_p(r)$ and $T$ become increasingly similar, particularly for the Musca, Taurus, Ophiuchus, and Orion A clouds. The similarity between the shape of $C_p(r)$ and $T$ at larger radial distances could indicate that the kernel is driving the structure of clouds at lower densities. However, since the kernel is inferred using data inclusive of the cloud's structure (in addition to structure beyond each cloud of interest) the outer envelopes could still be physically meaningful. More validation in future work will be needed to fully verify the relationship, if any, between the outer envelopes we observe in our radial profiles and the atomic gas envelopes of local molecular clouds, as discussed in \S \ref{discussion}. 

\clearpage

\restartappendixnumbering

\section{Radial Profile Fitting Results for the Full Profile} \label{appendix_full_profile}
In \S \ref{results}, we discuss the width results where the inner profiles (radial distances $<$ 2 pc) are excluded from the fit, due to the \citet{Leike_2020} map's 1 pc grid and its caveat that features less than 2 pc in size should be treated with caution. For completeness, in Table \ref{tab:radial_profiles_incl2},  we report the results of the full profile fits, including radial distance bins between 0 and 20 pc.

In general, we find that including the inner 2 pc of each radial profile in the fit has a modest effect on the model parameters.  When including the inner 2 pc for the two-component Gaussian-fit (in comparison to the results from \S \ref{results}), the inner width $\sigma_1$ changes on average by 15\%, while the outer width typically differs by 17\%. For the single-component Gaussian, $\sigma$ differs on average by 20\%. We see the largest difference in the Plummer model, with $R_{flat}$ changing by 28\% on average. Overall, however, the quality of all model fits decreases when including the inner 2 pc, which is why we report the results excluding the inner 2 pc, over the full profile results, in the main text. The decline in the quality of the fit, as determined using the same procedure outlined in \S \ref{model_comp}, is consistent with our understanding that the \citet{Leike_2020} dust map is unreliable at scales $< 2$ pc, and is also less accurate at reconstructing the highest gas densities, which are closest to the spine. Thus, it is not surprising that the fit improves when only higher fidelity measurements, farther from the spine, are included. 

\input{table2_incl2}

\newpage 
\clearpage

\restartappendixnumbering
\section{Radial Profile Fitting Results using Samples of the 3D Dust Map} \label{appendix_samples}
In deriving their 3D dust map, \citet{Leike_2020} generate twelve posterior samples, or realizations, of the extinction density, and the 3D dust map derived from the mean of these posterior samples is used in the fitting results reported in \S \ref{results}. However, because the uncertainty in the underlying 3D dust reconstruction is not accounted for when characterizing the radial profiles, it is necessary to quantify how much our results change when different realizations of the 3D dust distribution are used, rather than just the mean. To do so, we create a new likelihood function, building on the formalism described in \S \ref{fitting}. Recall that when fitting 3D dust determined by the mean of the samples, we adopt a Gaussian log-likelihood that takes the following form:

\begin{equation}
\mathcal{L}_\mathrm{mean} = \ln\, p(n \big| \theta,\sigma_n^{2} ) = - \frac{1}{2} \sum_i \left[ \frac{(n_i - n_{\theta,i})^{2}}{\sigma_n^2} + \ln(2\pi \sigma_n^{2}) \right]
\end{equation}

where $n_i$ is the measured profile in the $ith$ radial distance bin characterized by the mean 3D dust map, $n_{\theta,i}$ is the model density profile in the $ith$ distance bin (for the single Gaussian, two-component Gaussian, or Plummer fit), $\theta$ constitutes the parameters characterizing the model density profile for each fit (e.g. $\sigma$, $a$ for the single Gaussian fit), and $\sigma_n^{2}$ is the square of the scatter in the density (across all bins) that we infer as part of our modeling. 

Our new likelihood function, taking into account all twelve realizations of the 3D dust map,  is of the form:

\begin{equation}
\label{eq:samples_likelihood}
\mathcal{L}_\mathrm{samples} = \ln[\exp(\mathcal{L}_{{\rm sample}_1}) + \exp(\mathcal{L}_{{\rm sample}_2}) + ... + \exp(\mathcal{L}_{{\rm sample}_{12}})] - \ln(12)
\end{equation}

where $L_{\mathrm{sample}_1}$  is calculated equivalently to  ${L}_\mathrm{mean}$, except the measured radial density profile $n$ is derived from the 1st sample of the 3D dust map. There are twelve samples, so we compute the radial profile for each, assuming the same filament spine as determined for the mean dust map. We exclude the inner 2 pc of the radial profiles when fitting, as we do for the results reported in Table \ref{tab:radial_profiles_excl2}. Each of the twelve samples is a possible reality with a probability of $\frac{1}{12}$, and therefore the combined probability is their average. Thus, the total log-likelihood is the average of the individual likelihoods derived from each realization. We sample for the best-fit model parameters for all three functions (single-component Gaussian, two-component Gaussian, Plummer) for all clouds, using the same priors and the same \texttt{dynesty} nested sampling framework we employ in \S \ref{fitting}. Our results are summarized in Table \ref{tab:samples_results}. 

Overall, we find that that uncertainty in the profile fitting derived from the uncertainty in the 3D dust map has only a modest effect on our results. For the two-component Gaussian, the typical percent difference between the sample-based inference and the mean-based inference is 2\% for $\sigma_2$, 10\% for $\sigma_1$, 21\% for $a_2$ and 11\% for $a_1$. For the Plummer fit, $p$ changes by an average of 8\%, $n_0$ by 17\%, and $R_\mathrm{flat}$ by 22\%. However, we do note that for three clouds --- Pipe, $\lambda$ Orionis, and Musca, the $R_\mathrm{flat}$ parameter changes more drastically, increasing by a factor of two when using the 3D dust samples rather than the mean. This larger change could be tied to the degeneracy in the Plummer modeling described in \S \ref{caveats}, which provides another reason not to overinterpret the Plummer models, as we caution in \S \ref{plummer_significance}. The two-component Gaussian fit is much more robust, and the sample-based fitting results reported in Table \ref{tab:samples_results} do not affect any physical interpretation of the profile shapes described in \S \ref{discussion}.

\input{table2_samples}

\newpage 
\clearpage

\restartappendixnumbering
\section{Comparison to CO Radial Profiles} \label{co_profiles}

We build radial profiles of the CO emission in order to test the HI-to-$\rm H_2$ transition hypothesis presented in \S \ref{HI_to_H2}. Specifically, if CO emission is detected significantly beyond the boundaries of the inner Gaussian profile, the two component Gaussian would not represent a phase transition between atomic and molecular gas, disproving the hypothesis discussed in \S \ref{HI_to_H2}. For this test, we leverage data from the 1.2 m CfA CO survey \citep{Dame_2001}, selecting all clouds which have broad enough CO coverage to probe potential emission at large radial distances (seven of eleven clouds in the sample). For each cloud, we produce integrated intensity maps of the $\rm ^{12}CO$ emission over the velocity range of $\rm -15$ to $15 \;  \rm km \; s^{-1}$. We then smooth each map to the resolution of the \citet{Leike_2020} data at the distance of the cloud (see Table \ref{tab:skeletonization}) and define 2D masks by thresholding the integrated intensity maps at a level of $10 \rm \; K \; km \; s^{-1}$. We then follow the procedure for building radial profiles in 2D space, as presented in \citet{Zucker_2018b}, which involves skeletonizing the mask and taking cuts across the spine. We employ the ``shift" option to account for CO emission offset from the main spine, ensuring the emission peaks at a radial distance of zero. 

\begin{figure}
\begin{center}
\includegraphics[width=0.9\columnwidth]{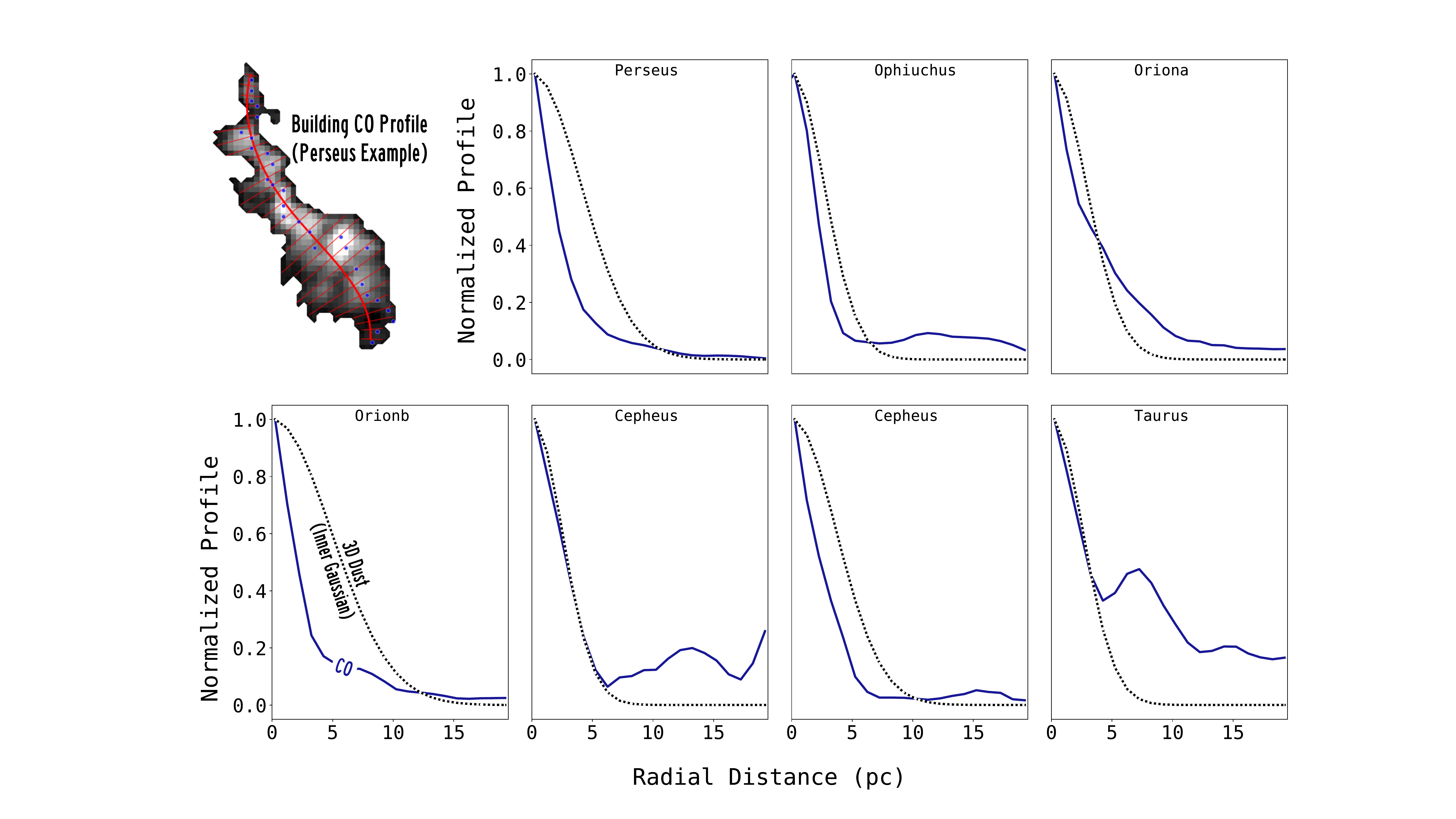} 
\caption{Comparison to CO radial profiles for seven clouds in the sample. The top left hand panel shows a schematic for building 2D radial profiles from the $\rm ^{12}CO$ integrated intensity maps for an example cloud (the Perseus Molecular Cloud), including the spine (thick red line), the distribution of integrated CO emission inside the filament mask (background grayscale), the perpendicular cuts across the spine (thin red lines) and the shifted peak integrated emission inside the filament mask (blue points). Note that the radial profiles are measured well beyond the boundaries of the mask, and the mask extent is only used for defining the spine and the optional shifting procedure. The remaining panels show normalized CO integrated intensity radial profiles (navy blues curves) alongside the inner component of the 3D dust derived profiles (black dotted curves) for the two-component Gaussian fits presented in the main text. Both components have been normalized to a peak value of one. We find no significant CO emission beyond the boundary of the 3D-dust-derived inner Gaussian profile, finding no direct evidence against the HI-to-$\rm H_2$ transition scenario presented in \S \ref{HI_to_H2} \label{fig:co_profiles}}
\end{center}
\end{figure}

The results are presented in Figure \ref{fig:co_profiles}. In the top left hand corner of Figure \ref{fig:co_profiles} we show an example of what the 2D skeletonization of the CO integrated intensity masks looks like for a single cloud in the sample --- the Perseus Molecular Cloud. In the subsequent panels of Figure \ref{fig:co_profiles}, we show normalized CO integrated intensity profiles (blue) alongside the normalized density profiles (black) for the inner Gaussian component based on the 3D dust mapping. Both profiles have been scaled to a peak value of one for comparison purposes. We find no extended CO emission beyond the bounds of the inner Gaussian, with most CO profiles showing smaller widths. The one exception is the Taurus Molecular Cloud, whose CO profile is almost identical to the 3D dust profile out to a radial distance of 5 pc, but also shows a secondary peak at larger radial distances. We find that the secondary peak is entirely due to projection effects, stemming from the two components of Taurus (separated by about 10-15 pc along the line of sight) which project to similar areas on the plane of the sky. Thus, we are unable to disprove the HI-to-$\rm H_2$ transition hypothesis given the results of the CO radial profile analysis.

\end{document}

%% file: table1.tex
\tabletypesize{\small}
\setlength{\tabcolsep}{7pt}
\begin{deluxetable*}{cccccccccccccc}
\tablecaption{Skeletonization Results \label{tab:skeletonization} }
\colnumbers
\tablehead{\colhead{Cloud} & \colhead{l} & \colhead{b} & \colhead{$d_{min}$} & \colhead{$d_{med}$} & \colhead{$d_{max}$} & \colhead{$x_{min}$} & \colhead{$x_{max}$} & \colhead{$y_{min}$} & \colhead{$y_{max}$} & \colhead{$z_{min}$} & \colhead{$z_{max}$}  & \colhead{$N_{components}$} &  \colhead{$l_{skeleton}$}\\
\colhead{} & \colhead{\scriptsize $^\circ$} & \colhead{\scriptsize $^\circ$} & \colhead{\scriptsize pc} & \colhead{\scriptsize pc} & \colhead{\scriptsize pc} & \colhead{\scriptsize pc} & \colhead{\scriptsize pc} & \colhead{\scriptsize pc} & \colhead{\scriptsize pc} & \colhead{\scriptsize pc} & \colhead{\scriptsize pc} &  & \colhead{\scriptsize pc} }
\startdata
Chamaeleon & 299.0 & -15.9 & 173 & 183 & 190 & 69 & 99 & -159 & -148 & -57 & -35 & 1 & 68 \\
Ophiuchus & 355.4 & 17.0 & 115 & 136 & 150 & 104 & 144 & -26 & 5 & 24 & 48 & 3 & 74 \\
Lupus & 341.7 & 5.1 & 155 & 165 & 198 & 139 & 196 & -73 & -28 & 4 & 47 & 3 & 80 \\
Taurus & 172.6 & -15.6 & 131 & 148 & 168 & -166 & -122 & 6 & 44 & -45 & -26 & 3 & 79 \\
Perseus & 158.9 & -20.2 & 279 & 295 & 301 & -270 & -236 & 95 & 100 & -116 & -81 & 1 & 57 \\
Musca & 301.0 & -9.7 & 171 & 172 & 173 & 86 & 91 & -146 & -145 & -36 & -22 & 1 & 16 \\
Pipe & 2.1 & 4.1 & 147 & 152 & 163 & 146 & 163 & -8 & 7 & 6 & 18 & 1 & 27 \\
Cepheus & 112.1 & 15.9 & 337 & 345 & 370 & -133 & -65 & 300 & 323 & 81 & 133 & 4 & 52 \\
Corona Australis\tablenotemark{\tiny a} & 6.5 & -24.6 & 136 & 161 & 179 & 132 & 162 & -2 & 24 & -77 & -33 & 1 & 86 \\
\cline{1-14}
Orion A* & 212.5 & -19.6 & 391 & 405 & 445 & -349 & -315 & -241 & -177 & -149 & -118 & 3 & 52 \\
Orion B* & 205.5 & -14.0 & 397 & 404 & 406 & -362 & -332 & -190 & -157 & -114 & -76 & 1 & 69 \\
$\lambda$ Orionis* & 194.3 & -14.6 & 375 & 386 & 397 & -367 & -356 & -129 & -73 & -109 & -73 & 3 & 62 \\
\enddata
\tablecomments{Properties of the 3D skeletons for local molecular clouds. (1) Name of the cloud. (2-3) Median Galactic longitude $l$ and latitude $b$ of the points defining the cloud's skeleton. (4 - 6) Minimum, median, and maximum distance of the points defining the skeleton. (7-8) Minimum and maximum extent of the cloud's skeleton in the Heliocentric Galactic cartesian $x$ direction. (9-10) Minimum and maximum extent of the cloud's skeleton in the Heliocentric Galactic cartesian $y$ direction. (11-12) Minimum and maximum extent of the cloud's skeleton in the Heliocentric Galactic cartesian $z$ direction.  (13) Number of skeletons per cloud, determined by the number of significant features above a density threshold of $n = \rm 35 \; cm^{-3}$. (14) Total length of the set of skeletons in (13), computed using only each skeleton's main trunk via \texttt{FilFinder's} longest path algorithm. At the \href{https://dataverse.harvard.edu/dataverse/cloud\_topologies}{Harvard Dataverse}, we provide tables in FITS format with the ($x,y,z,l,b,d$) values of each point in the spine (for both the longest path skeleton summarized here, and the full, unpruned skeleton).  We also provide a machine readable version of this table (see \url{https://doi.org/10.7910/DVN/CAVMAQ}). Interactive figures highlighting these skeletonization results are available at \url{https://faun.rc.fas.harvard.edu/czucker/Paper_Figures/3D\_Cloud\_Topologies/gallery.html}.}
\tablenotetext{a}{Corona Australis is too diffuse to meet the minimum density threshold to be included in the width analysis for this study. We skeletonize the cloud at a much lower threshold ($n = 5 \; \rm cm^{-3}$) to provide distance information, but do not include it in subsequent analysis in Tables \ref{tab:radial_profiles_excl2} and \ref{tab:mass}.}
\tablenotetext{*}{Clouds below the horizontal line -- Orion A, Orion B, and $\lambda$ Orionis -- should be treated with caution, as they lie at the very edge of the \citet{Leike_2020} 3D dust grid and are subjected to additional biases. Part of Orion A, toward the Orion Nebula Cluster, is also devoid of dust in the \citet{Leike_2020} 3D dust map despite being highly extinguished. See \S \ref{exposition} for full details.}

\end{deluxetable*}

%% file: table2_excl2.tex
\movetabledown=1.5in
\begin{rotatetable}
\begin{deluxetable}{@{\extracolsep{2pt}}ccccccccccccccccc}
\tablecaption{Radial Volume Density Fitting Results \label{tab:radial_profiles_excl2}}
\tabletypesize{\scriptsize}
\setlength{\tabcolsep}{3pt}
\def\arraystretch{2.0}
\colnumbers
\tablehead
{
\colhead{} &
  \multicolumn{7}{c}{Two-Component Gaussian}&
  \multicolumn{5}{c}{Plummer}&
  \multicolumn{4}{c}{Single Gaussian}\\
\cline{2-8} \cline{9-13} \cline{14-17}\\[-0.5em]
\small Cloud & \small $\sigma_1$ &  \small $a_1$ &  \small $\sigma_2$ &  \small $a_2$ &  \small $\sigma_{n, \rm G2}^2$  & \small $\frac{\sigma_2}{\sigma_1}$  & $lnZ_{\rm G2}$ &  \small $n_0$ &  \small $R_{\rm flat}$ &  \small $p$ &  \small $\sigma_{n, \rm P}^2$ &  $lnZ_{P}$ & \small $\sigma$ &  \small $a$ &  \small $\sigma_{n, G1}^2$  & $lnZ_{\rm G1}$ \\[0.1in]  &  \scriptsize pc & \scriptsize cm$^{-3}$ & \scriptsize pc &\scriptsize cm$^{-3}$ & \scriptsize cm$^{-3}$ & & & \scriptsize cm$^{-3}$ &  \scriptsize pc & &
 \scriptsize cm$^{-3}$ & &  \scriptsize pc & \scriptsize cm$^{-3}$ & \scriptsize cm$^{-3}$ &}
\startdata
Chamaeleon & $2.9_{-0.1}^{+0.1}$ & $31.3_{-0.6}^{+0.5}$ & $10.2_{-0.3}^{+0.4}$ & $8.7_{-0.5}^{+0.5}$ & $0.2_{-0.0}^{+0.0}$ & {3.5} & {-2.9}& $47.8_{-1.2}^{+1.5}$ & $3.1_{-0.2}^{+0.1}$ & $1.8_{-0.1}^{+0.1}$ & $0.1_{-0.0}^{+0.0}$ & {-1.5} & $4.8_{-0.3}^{+0.3}$ & $30.8_{-2.0}^{+2.1}$ & $6.1_{-1.3}^{+1.9}$ & {-58.0}\\
Ophiuchus & $2.7_{-0.2}^{+0.2}$ & $31.1_{-1.5}^{+1.6}$ & $10.1_{-0.5}^{+0.6}$ & $12.6_{-1.1}^{+1.1}$ & $0.8_{-0.2}^{+0.2}$ & {3.8} & {-27.2}& $56.0_{-2.9}^{+3.5}$ & $2.3_{-0.2}^{+0.2}$ & $1.4_{-0.0}^{+0.0}$ & $0.3_{-0.1}^{+0.1}$ & {-11.9} & $5.8_{-0.4}^{+0.5}$ & $29.1_{-2.0}^{+2.0}$ & $9.5_{-2.0}^{+2.6}$ & {-64.7}\\
Lupus & $2.6_{-0.2}^{+0.3}$ & $30.6_{-1.8}^{+1.8}$ & $8.1_{-0.5}^{+0.6}$ & $15.5_{-2.0}^{+1.9}$ & $0.9_{-0.2}^{+0.3}$ & {3.1} & {-29.2}& $48.6_{-1.9}^{+2.3}$ & $3.6_{-0.3}^{+0.3}$ & $1.9_{-0.1}^{+0.1}$ & $0.5_{-0.1}^{+0.1}$ & {-19.3} & $5.2_{-0.3}^{+0.3}$ & $32.7_{-1.7}^{+1.8}$ & $6.2_{-1.3}^{+1.7}$ & {-58.1}\\
Taurus & $2.6_{-0.1}^{+0.1}$ & $30.9_{-1.1}^{+1.2}$ & $11.9_{-0.5}^{+0.6}$ & $9.2_{-0.5}^{+0.5}$ & $0.4_{-0.1}^{+0.1}$ & {4.6} & {-17.4}& $72.8_{-11.1}^{+22.7}$ & $1.2_{-0.3}^{+0.3}$ & $1.2_{-0.0}^{+0.0}$ & $0.4_{-0.1}^{+0.1}$ & {-17.3} & $5.8_{-0.6}^{+0.6}$ & $24.3_{-2.2}^{+2.4}$ & $10.3_{-2.3}^{+3.3}$ & {-65.7}\\
Perseus & $4.1_{-0.1}^{+0.1}$ & $34.3_{-0.9}^{+0.9}$ & $11.9_{-0.6}^{+0.8}$ & $10.9_{-1.0}^{+1.0}$ & $0.3_{-0.1}^{+0.1}$ & {2.9} & {-15.0}& $47.8_{-0.9}^{+0.9}$ & $6.1_{-0.3}^{+0.3}$ & $2.4_{-0.1}^{+0.1}$ & $0.5_{-0.1}^{+0.1}$ & {-20.7} & $6.0_{-0.2}^{+0.2}$ & $39.0_{-1.5}^{+1.6}$ & $6.9_{-1.4}^{+2.1}$ & {-60.4}\\
Musca\tablenotemark{\tiny{a}} & $3.1_{-0.1}^{+0.1}$ & $31.8_{-1.5}^{+1.5}$ & $17.9_{-3.0}^{+4.1}$ & $4.9_{-0.6}^{+0.7}$ & $1.3_{-0.3}^{+0.4}$ & {5.8} & {-34.4}& $48.3_{-5.5}^{+8.4}$ & $2.5_{-0.5}^{+0.6}$ & $1.6_{-0.2}^{+0.2}$ & $2.4_{-0.5}^{+0.7}$ & {-44.5} & $4.4_{-0.4}^{+0.4}$ & $29.8_{-2.9}^{+3.0}$ & $8.9_{-2.0}^{+2.7}$ & {-63.8}\\
Pipe & $4.1_{-0.4}^{+0.4}$ & $25.5_{-2.8}^{+2.7}$ & $9.7_{-1.2}^{+2.5}$ & $9.0_{-3.4}^{+3.3}$ & $1.5_{-0.3}^{+0.5}$ & {2.4} & {-36.7}& $35.5_{-1.4}^{+1.5}$ & $7.7_{-1.0}^{+1.2}$ & $3.4_{-0.5}^{+0.6}$ & $1.4_{-0.3}^{+0.4}$ & {-35.5} & $5.7_{-0.2}^{+0.2}$ & $30.6_{-1.1}^{+1.2}$ & $3.5_{-0.7}^{+1.1}$ & {-49.3}\\
Cepheus & $2.5_{-0.1}^{+0.1}$ & $46.6_{-1.6}^{+1.7}$ & $10.9_{-0.6}^{+0.7}$ & $10.1_{-0.7}^{+0.7}$ & $0.7_{-0.2}^{+0.2}$ & {4.4} & {-26.4}& $116.9_{-17.5}^{+23.7}$ & $1.3_{-0.2}^{+0.2}$ & $1.4_{-0.0}^{+0.0}$ & $0.5_{-0.1}^{+0.2}$ & {-21.6} & $4.3_{-0.4}^{+0.4}$ & $39.1_{-3.7}^{+3.9}$ & $14.3_{-3.0}^{+4.2}$ & {-71.4}\\ \cline{1-17}
Orion A* & $2.9_{-0.1}^{+0.1}$ & $52.4_{-1.8}^{+1.8}$ & $8.4_{-0.8}^{+1.0}$ & $8.8_{-1.7}^{+1.9}$ & $1.3_{-0.3}^{+0.4}$ & {2.9} & {-35.0}& $68.3_{-3.0}^{+3.4}$ & $4.4_{-0.4}^{+0.5}$ & $3.1_{-0.3}^{+0.3}$ & $1.5_{-0.3}^{+0.5}$ & {-37.2} & $3.7_{-0.1}^{+0.1}$ & $53.1_{-2.4}^{+2.5}$ & $5.8_{-1.3}^{+1.8}$ & {-57.5}\\
Orion B* & $4.9_{-0.1}^{+0.1}$ & $39.2_{-1.1}^{+1.1}$ & $14.6_{-1.3}^{+1.9}$ & $9.0_{-1.2}^{+1.3}$ & $0.3_{-0.1}^{+0.1}$ & {3.0} & {-13.6}& $50.3_{-0.8}^{+0.8}$ & $8.1_{-0.4}^{+0.5}$ & $3.0_{-0.2}^{+0.2}$ & $0.6_{-0.1}^{+0.2}$ & {-23.6} & $6.5_{-0.2}^{+0.2}$ & $43.8_{-1.3}^{+1.4}$ & $5.9_{-1.2}^{+1.8}$ & {-58.0}\\
$\lambda$ Orionis* & $3.7_{-0.1}^{+0.1}$ & $27.2_{-0.5}^{+0.5}$ & $12.8_{-0.4}^{+0.5}$ & $9.3_{-0.4}^{+0.4}$ & $0.1_{-0.0}^{+0.0}$ & {3.4} & {-0.7}& $40.2_{-1.0}^{+1.1}$ & $4.3_{-0.2}^{+0.3}$ & $1.7_{-0.1}^{+0.1}$ & $0.3_{-0.1}^{+0.1}$ & {-14.0} & $6.3_{-0.4}^{+0.3}$ & $29.1_{-1.6}^{+1.6}$ & $7.2_{-1.5}^{+2.0}$ & {-60.2}\\
\enddata
\tablecomments{Cloud thickness results, obtained via radial volume density profile fitting of each cloud's 3D density distribution. (1) Name of the cloud. (2 - 6) Best-fit parameters for the two-component Gaussian fit, including the standard deviation and amplitude of the narrower Gaussian ($\sigma_1$, $a_1$), the standard deviation and amplitude of the broader Gaussian ($\sigma_2$, $a_2$), and the square of the modeled scatter in the density measurements ($\sigma_{n, \rm G2}^2$). (7) Ratio of the outer to inner Gaussian standard deviations.  (8) Logarithm of the evidence of the two-component Gaussian fit. (9-12) Best-fit parameters of the Plummer fit, including the peak amplitude $n_0$, the flattening radius $R_{\rm flat}$, the power index of the density profile $p$, and the square of the modeled scatter in the density measurements ($\sigma_{n,\rm P}^2$). (13) Logarithm of the evidence of the Plummer fit. (14-16) Standard deviation $\sigma$, amplitude $a$, and the square of the modeled scatter in the density measurements ($\sigma_{n,\rm G1}^2$) for the single component Gaussian fit. (17) Logarithm of the evidence of the single-component Gaussian fit. We report the single-component Gaussian results here for completeness only, as it is a poor fit to the data particularly at large radial distances. A machine readable version of this table is available at \url{https://doi.org/10.7910/DVN/QKYR3G}.}
\tablenotetext{a}{At small radial distances, Musca is largely unresolved in the \citet{Leike_2020} map, and at large radial distances, its structure is consistent with being prior-dominated, so we urge caution when interpreting the results.} 
\tablenotetext{*}{Clouds below the horizontal line -- Orion A, Orion B, and $\lambda$ Orionis -- should be treated with caution, as they lie at the very edge of the \citet{Leike_2020} 3D dust grid and are subjected to additional biases. Part of Orion A, toward the Orion Nebula Cluster, is also devoid of dust in the \citet{Leike_2020} 3D dust map despite being highly extinguished. See \S \ref{exposition} for full details.}

\end{deluxetable}
\end{rotatetable}

%% file: table3.tex
\setlength{\tabcolsep}{7pt}
\begin{deluxetable*}{cccccc}
\tablecaption{Mass and Extinction Results \label{tab:mass} }
\tabletypesize{\small}
\colnumbers
\tablehead{\colhead{Cloud} & \colhead{${\rm Mass}_{\rm NICEST}$} & \colhead{${\rm Mass}_{\rm Leike}$} & \colhead{$\rm  \frac{Mass_{\rm NICEST}}{Mass_{Leike}}$} & \colhead{Maximum $A_{K\rm, NICEST}$} & \colhead{Maximum $A_{K\rm,Leike}$} \\
\colhead{} &  \colhead{$M_\sun$} &  \colhead{$M_\sun$} & \colhead{} &  \colhead{mag} & \colhead{mag}}
\startdata
Chamaeleon & 4.9e+03 & 4.4e+03 & 1.1 & 1.19 & 0.29 \\
Ophiuchus & 9.4e+03 & 7.9e+03 & 1.2 & 1.99 & 0.31 \\
Lupus & 1.0e+04 & 8.9e+03 & 1.2 & 1.55 & 0.28 \\
Taurus & 1.6e+04 & 1.5e+04 & 1.0 & 0.90 & 0.38 \\
Perseus & 1.6e+04 & 1.3e+04 & 1.2 & 1.40 & 0.28 \\
Musca & 5.9e+02 & 4.9e+02 & 1.2 & 0.47 & 0.19 \\
Pipe & 8.5e+02 & 5.2e+02 & 1.6 & 1.29 & 0.23 \\
Cepheus & 1.2e+04 & 1.0e+04 & 1.2 & 1.15 & 0.33 \\ \cline{1-6}
Orion A* & 3.2e+04 & 9.4e+03 & 3.4 & 1.48 & 0.34 \\
Orion B* & 2.9e+04 & 1.7e+04 & 1.7 & 1.34 & 0.33 \\
$\lambda$ Orionis* & 1.2e+04 & 8.6e+02 & 13.7 & 1.36 & 0.22 \\
\enddata
\tablecomments{Summary of mass and extinction results for local molecular clouds. (1) Cloud Name  (2) Mass derived from a traditional 2D extinction map computed using the NICEST algorithm from \citet{Lombardi_2009}. (3) Mass derived for the cloud from the \citet{Leike_2020} 3D dust map, which has been projected back on the plane of the sky. In both cases, the mass is computed above an $A_K = 0.1$ mag cloud boundary. (4) Ratio of the mass from columns (2) and (3). In (5) Maximum $A_K$ inside the cloud boundary based on the NICEST algorithm. (6) Maximum $A_K$ inside the cloud boundary for the 2D extinction map derived from the 3D dust data.  We provide a machine readable version of this table at the \href{https://dataverse.harvard.edu/dataverse/cloud\_topologies}{Harvard Dataverse} (\url{https://doi.org/10.7910/DVN/EIPHPR}).}
\tablenotetext{*}{Clouds below the horizontal line -- Orion A, Orion B, and $\lambda$ Orionis -- should be treated with caution, as they lie at the very edge of the \citet{Leike_2020} 3D dust grid and are subjected to additional biases. Part of Orion A, toward the Orion Nebula Cluster, is also devoid of dust in the \citet{Leike_2020} 3D dust map despite being highly extinguished. See \S \ref{exposition} for full details.}
 
\end{deluxetable*}

%% file: table2_incl2.tex
\movetabledown=1.5in
\begin{rotatetable}
\begin{deluxetable}{@{\extracolsep{2pt}}ccccccccccccccccc}
\tablecaption{Radial Volume Density Fitting (Full Profile) \label{tab:radial_profiles_incl2}}
\tabletypesize{\scriptsize}
\setlength{\tabcolsep}{3pt}
\def\arraystretch{2.0}
\colnumbers
\tablehead
{
\colhead{} &
  \multicolumn{7}{c}{Two-Component Gaussian}&
  \multicolumn{5}{c}{Plummer}&
  \multicolumn{4}{c}{Single Gaussian}\\
\cline{2-8} \cline{9-13} \cline{14-17}\\[-0.5em]
\small Cloud & \small $\sigma_1$ &  \small $a_1$ &  \small $\sigma_2$ &  \small $a_2$ &  \small $\sigma_{n, \rm G2}^2$  & \small $\frac{\sigma_2}{\sigma_1}$  & $lnZ_{G2}$ &  \small $n_0$ &  \small $R_{\rm flat}$ &  \small $p$ &  \small $\sigma_{n, \rm P}^2$ &  $lnZ_{\rm P}$ & \small $\sigma$ &  \small $a$ &  \small $\sigma_{n, \rm G1}^2$  & $lnZ_{\rm G1}$ \\[0.1in]  &  \scriptsize pc & \scriptsize cm$^{-3}$ & \scriptsize pc &\scriptsize cm$^{-3}$ & \scriptsize cm$^{-3}$ & & & \scriptsize cm$^{-3}$ &  \scriptsize pc & &
 \scriptsize cm$^{-3}$ & &  \scriptsize pc & \scriptsize cm$^{-3}$ & \scriptsize cm$^{-3}$ &}
\startdata
Chamaeleon & $2.6_{-0.1}^{+0.1}$ & $34.5_{-0.7}^{+0.7}$ & $9.4_{-0.4}^{+0.5}$ & $10.2_{-0.7}^{+0.7}$ & $0.4_{-0.1}^{+0.1}$ & {3.6} & {-19.7}& $46.1_{-0.3}^{+0.3}$ & $3.4_{-0.1}^{+0.1}$ & $1.9_{-0.1}^{+0.1}$ & $0.3_{-0.1}^{+0.1}$ & {-10.8} & $3.8_{-0.2}^{+0.2}$ & $41.1_{-1.6}^{+1.6}$ & $10.0_{-2.0}^{+2.8}$ & {-73.1}\\
Ophiuchus & $2.4_{-0.1}^{+0.1}$ & $33.1_{-1.0}^{+0.9}$ & $9.6_{-0.5}^{+0.5}$ & $13.6_{-0.9}^{+1.0}$ & $0.9_{-0.2}^{+0.3}$ & {3.9} & {-33.8}& $48.1_{-0.5}^{+0.5}$ & $2.9_{-0.1}^{+0.1}$ & $1.5_{-0.0}^{+0.0}$ & $0.5_{-0.1}^{+0.1}$ & {-22.6} & $4.3_{-0.3}^{+0.3}$ & $40.6_{-2.0}^{+2.2}$ & $16.9_{-3.5}^{+4.9}$ & {-82.3}\\
Lupus & $2.3_{-0.1}^{+0.1}$ & $31.4_{-1.3}^{+1.3}$ & $7.7_{-0.3}^{+0.4}$ & $17.1_{-1.4}^{+1.3}$ & $1.0_{-0.2}^{+0.3}$ & {3.3} & {-34.3}& $49.0_{-0.5}^{+0.5}$ & $3.6_{-0.1}^{+0.2}$ & $1.9_{-0.1}^{+0.1}$ & $0.5_{-0.1}^{+0.2}$ & {-24.6} & $4.2_{-0.2}^{+0.2}$ & $42.9_{-1.7}^{+1.8}$ & $12.1_{-2.5}^{+3.4}$ & {-76.6}\\
Taurus & $2.0_{-0.1}^{+0.1}$ & $43.1_{-1.1}^{+1.0}$ & $10.4_{-0.7}^{+0.8}$ & $11.2_{-0.9}^{+0.9}$ & $1.8_{-0.3}^{+0.5}$ & {5.2} & {-45.3}& $56.8_{-1.0}^{+1.0}$ & $1.9_{-0.1}^{+0.1}$ & $1.4_{-0.1}^{+0.1}$ & $1.8_{-0.4}^{+0.5}$ & {-46.0} & $3.1_{-0.2}^{+0.3}$ & $48.2_{-3.0}^{+3.2}$ & $24.4_{-5.0}^{+7.1}$ & {-89.3}\\
Perseus & $3.3_{-0.2}^{+0.2}$ & $35.0_{-1.9}^{+1.9}$ & $9.7_{-0.6}^{+0.8}$ & $16.5_{-2.0}^{+2.0}$ & $1.8_{-0.4}^{+0.5}$ & {3.0} & {-45.0}& $52.5_{-0.6}^{+0.6}$ & $4.8_{-0.3}^{+0.3}$ & $2.0_{-0.1}^{+0.1}$ & $1.1_{-0.2}^{+0.3}$ & {-37.3} & $5.2_{-0.2}^{+0.2}$ & $46.8_{-1.5}^{+1.5}$ & $13.6_{-2.8}^{+3.7}$ & {-79.0}\\
Musca\tablenotemark{\tiny{a}} & $2.6_{-0.2}^{+0.3}$ & $42.8_{-3.5}^{+3.1}$ & $14.7_{-4.9}^{+6.1}$ & $6.1_{-1.7}^{+2.6}$ & $23.6_{-4.5}^{+6.8}$ & {5.7} & {-89.0}& $49.9_{-3.3}^{+3.2}$ & $3.9_{-1.0}^{+1.6}$ & $2.6_{-0.7}^{+1.4}$ & $26.2_{-5.4}^{+7.5}$ & {-91.4} & $3.2_{-0.3}^{+0.3}$ & $46.7_{-3.3}^{+3.3}$ & $33.6_{-6.7}^{+10.0}$ & {-95.7}\\
Pipe & $2.6_{-0.2}^{+0.3}$ & $24.4_{-1.8}^{+2.4}$ & $7.6_{-0.4}^{+0.6}$ & $18.1_{-2.8}^{+1.4}$ & $4.0_{-0.9}^{+1.1}$ & {2.9} & {-58.3}& $42.0_{-1.2}^{+1.2}$ & $4.8_{-0.6}^{+0.7}$ & $2.2_{-0.3}^{+0.4}$ & $3.8_{-0.8}^{+1.1}$ & {-57.2} & $4.9_{-0.2}^{+0.2}$ & $37.4_{-1.4}^{+1.4}$ & $10.4_{-2.2}^{+2.9}$ & {-73.6}\\
Cepheus & $2.0_{-0.1}^{+0.1}$ & $60.6_{-1.3}^{+1.3}$ & $9.1_{-0.6}^{+0.7}$ & $13.5_{-1.1}^{+1.2}$ & $2.1_{-0.4}^{+0.6}$ & {4.6} & {-48.2}& $77.9_{-0.7}^{+0.8}$ & $2.1_{-0.1}^{+0.1}$ & $1.6_{-0.0}^{+0.0}$ & $0.8_{-0.2}^{+0.2}$ & {-32.9} & $2.8_{-0.2}^{+0.2}$ & $67.9_{-3.1}^{+3.1}$ & $28.3_{-5.9}^{+8.2}$ & {-92.3}\\\cline{1-17}
Orion A* & $3.3_{-0.1}^{+0.1}$ & $48.6_{-2.0}^{+1.4}$ & $11.2_{-2.2}^{+3.9}$ & $5.0_{-1.5}^{+2.0}$ & $2.4_{-0.5}^{+0.7}$ & {3.4} & {-48.6}& $54.2_{-1.0}^{+1.0}$ & $7.0_{-0.7}^{+0.7}$ & $4.8_{-0.7}^{+0.7}$ & $3.3_{-0.7}^{+0.9}$ & {-54.3} & $3.7_{-0.1}^{+0.1}$ & $52.4_{-1.1}^{+1.1}$ & $5.4_{-1.1}^{+1.7}$ & {-62.8}\\
Orion B* & $4.5_{-0.2}^{+0.2}$ & $38.1_{-2.3}^{+2.1}$ & $12.0_{-1.1}^{+1.6}$ & $12.6_{-2.3}^{+2.4}$ & $1.2_{-0.2}^{+0.3}$ & {2.7} & {-36.5}& $51.6_{-0.5}^{+0.5}$ & $7.5_{-0.4}^{+0.4}$ & $2.8_{-0.2}^{+0.2}$ & $1.0_{-0.2}^{+0.3}$ & {-34.7} & $6.0_{-0.2}^{+0.2}$ & $47.8_{-1.1}^{+1.2}$ & $8.7_{-1.7}^{+2.4}$ & {-70.5}\\
$\lambda$ Orionis* & $3.3_{-0.1}^{+0.1}$ & $28.9_{-0.8}^{+0.8}$ & $11.6_{-0.6}^{+0.7}$ & $10.9_{-0.8}^{+0.8}$ & $0.5_{-0.1}^{+0.2}$ & {3.5} & {-23.7}& $41.0_{-0.4}^{+0.4}$ & $4.1_{-0.2}^{+0.2}$ & $1.7_{-0.1}^{+0.1}$ & $0.4_{-0.1}^{+0.1}$ & {-17.5} & $5.2_{-0.3}^{+0.3}$ & $35.9_{-1.5}^{+1.5}$ & $11.1_{-2.2}^{+3.2}$ & {-74.6}\\
\enddata
\tablecomments{Radial density profile fitting results for local clouds, computed over radial distances between 0 and 20 pc. Unlike Table \ref{tab:radial_profiles_excl2}, we do not exclude the inner 2 pc from the fit. (1) Name of the cloud. (2 - 6) Best-fit parameters for the two-component Gaussian fit, including the standard deviation and amplitude of the narrower Gaussian ($\sigma_1$, $a_1$), the standard deviation and amplitude of the broader Gaussian ($\sigma_2$, $a_2$), and the square of the modeled scatter in the density measurements ($\sigma_{n, \rm G2}^2$). (7) Ratio of the outer to inner Gaussian standard deviations.  (8) Logarithm of the evidence of the two-component Gaussian fit. (9-12) Best-fit parameters of the Plummer fit, including the peak amplitude $n_0$, the flattening radius $R_{\rm flat}$, the power index of the density profile $p$, and the square of the modeled scatter in the density measurements ($\sigma_{n,\rm P}^2$). (13) Logarithm of the evidence of the Plummer fit. (14-16) Standard deviation $\sigma$, amplitude $a$, and the square of the modeled scatter in the density measurements ($\sigma_{n,\rm G1}^2$) for the single component Gaussian fit. (17) Logarithm of the evidence of the single-component Gaussian fit. We report the single-component Gaussian results here for completeness only, as it is a poor fit to the data particularly at large radial distances. A machine readable version of this table is available at \url{https://doi.org/10.7910/DVN/V8NGLX}.}
\tablenotetext{a}{At small radial distances, Musca is largely unresolved in the \citet{Leike_2020} map, and at large radial distances, its structure is consistent with being prior-dominated, so we urge caution when interpreting the results.} 
\tablenotetext{*}{Clouds below the horizontal line -- Orion A, Orion B, and $\lambda$ Orionis -- should be treated with caution, as they lie at the very edge of the \citet{Leike_2020} 3D dust grid and are subjected to additional biases. Part of Orion A, toward the Orion Nebula Cluster, is also devoid of dust in the \citet{Leike_2020} 3D dust map despite being highly extinguished. See \S \ref{exposition} for full details.}
\end{deluxetable}
\end{rotatetable}

%% file: table2_samples.tex
\movetabledown=1.5in
\begin{rotatetable}
\begin{deluxetable}{@{\extracolsep{2pt}}ccccccccccccccccc}
\tablecaption{Radial Volume Density Fitting (Computed using All Realizations of 3D Dust Map)\label{tab:samples_results}}
\tabletypesize{\scriptsize}
\setlength{\tabcolsep}{3pt}
\def\arraystretch{2.0}
\colnumbers
\tablehead
{
\colhead{} &
  \multicolumn{7}{c}{Two-Component Gaussian}&
  \multicolumn{5}{c}{Plummer}&
  \multicolumn{4}{c}{Single Gaussian}\\
\cline{2-8} \cline{9-13} \cline{14-17}\\[-0.5em]
\small Cloud & \small $\sigma_1$ &  \small $a_1$ &  \small $\sigma_2$ &  \small $a_2$ &  \small $\sigma_{n, \rm G2}^2$  & \small $\frac{\sigma_2}{\sigma_1}$  & $lnZ_{\rm G2}$ &  \small $n_0$ &  \small $R_{\rm flat}$ &  \small $p$ &  \small $\sigma_{n, \rm P}^2$ &  $lnZ_{\rm P}$ & \small $\sigma$ &  \small $a$ &  \small $\sigma_{n, \rm G1}^2$  & $lnZ_{\rm G1}$ \\[0.1in]  &  \scriptsize pc & \scriptsize cm$^{-3}$ & \scriptsize pc &\scriptsize cm$^{-3}$ & \scriptsize cm$^{-3}$ & & & \scriptsize cm$^{-3}$ &  \scriptsize pc & &
 \scriptsize cm$^{-3}$ & &  \scriptsize pc & \scriptsize cm$^{-3}$ & \scriptsize cm$^{-3}$ &}
\startdata
Chamaeleon & $2.8_{-0.1}^{+0.1}$ & $29.2_{-0.6}^{+0.7}$ & $10.4_{-0.3}^{+0.4}$ & $8.0_{-0.4}^{+0.4}$ & $0.1_{-0.0}^{+0.0}$ & {3.7} & {-4.1}& $46.9_{-1.9}^{+1.7}$ & $2.6_{-0.1}^{+0.2}$ & $1.7_{-0.0}^{+0.1}$ & $0.1_{-0.0}^{+0.0}$ & {0.3} & $4.7_{-0.4}^{+0.4}$ & $28.3_{-2.3}^{+3.7}$ & $4.8_{-1.1}^{+1.6}$ & {-55.7}\\
Ophiuchus & $2.7_{-0.1}^{+0.1}$ & $34.0_{-1.5}^{+1.2}$ & $10.9_{-0.5}^{+0.5}$ & $10.3_{-0.6}^{+0.6}$ & $0.4_{-0.1}^{+0.1}$ & {4.1} & {-19.7}& $56.6_{-2.5}^{+2.7}$ & $2.3_{-0.1}^{+0.2}$ & $1.4_{-0.0}^{+0.0}$ & $0.2_{-0.0}^{+0.1}$ & {-8.2} & $5.2_{-0.4}^{+0.5}$ & $28.7_{-2.3}^{+2.3}$ & $7.9_{-1.8}^{+2.4}$ & {-63.0}\\
   Lupus & $3.5_{-1.4}^{+0.2}$ & $27.6_{-1.4}^{+14.6}$ & $10.0_{-1.4}^{+1.3}$ & $9.4_{-1.8}^{+4.0}$ & $0.4_{-0.1}^{+0.1}$ & {2.9} & {-19.5}& $42.3_{-3.8}^{+11.7}$ & $3.5_{-0.9}^{+1.3}$ & $1.8_{-0.2}^{+0.4}$ & $0.3_{-0.1}^{+0.1}$ & {-13.3} & $5.4_{-0.2}^{+0.3}$ & $29.5_{-1.5}^{+1.4}$ & $4.2_{-1.0}^{+1.2}$ & {-54.0}\\
  Taurus & $2.3_{-0.1}^{+0.1}$ & $36.2_{-1.5}^{+1.2}$ & $11.5_{-0.4}^{+0.6}$ & $8.7_{-0.4}^{+0.3}$ & $0.2_{-0.0}^{+0.1}$ & {5.0} & {-9.7}& $54.4_{-4.7}^{+7.2}$ & $1.6_{-0.2}^{+0.2}$ & $1.3_{-0.0}^{+0.0}$ & $0.2_{-0.0}^{+0.1}$ & {-10.2} & $5.2_{-0.7}^{+0.7}$ & $24.0_{-2.8}^{+4.1}$ & $7.9_{-1.8}^{+2.4}$ & {-63.0}\\
 Perseus & $3.7_{-0.1}^{+0.1}$ & $26.3_{-0.9}^{+0.9}$ & $11.4_{-0.6}^{+0.7}$ & $11.5_{-1.0}^{+1.0}$ & $0.4_{-0.1}^{+0.1}$ & {3.1} & {-20.3}& $40.8_{-0.9}^{+1.0}$ & $4.7_{-0.3}^{+0.3}$ & $1.8_{-0.1}^{+0.1}$ & $0.3_{-0.1}^{+0.1}$ & {-17.5} & $6.4_{-0.6}^{+0.4}$ & $31.1_{-1.7}^{+3.2}$ & $6.1_{-1.4}^{+1.9}$ & {-59.3}\\
   Musca\tablenotemark{\tiny{a}} & $2.9_{-0.1}^{+0.1}$ & $36.1_{-0.8}^{+0.8}$ & $20.9_{-2.9}^{+2.7}$ & $3.3_{-0.2}^{+0.3}$ & $0.4_{-0.1}^{+0.1}$ & {7.1} & {-17.1}& $145.3_{-22.9}^{+16.9}$ & $1.0_{-0.1}^{+0.2}$ & $1.6_{-0.0}^{+0.0}$ & $0.5_{-0.1}^{+0.2}$ & {-24.3} & $3.6_{-0.2}^{+0.2}$ & $34.5_{-2.5}^{+2.7}$ & $5.2_{-1.1}^{+1.6}$ & {-56.9}\\
    Pipe & $2.9_{-0.2}^{+0.4}$ & $29.5_{-9.3}^{+8.4}$ & $9.9_{-0.7}^{+0.8}$ & $8.3_{-1.1}^{+1.3}$ & $0.5_{-0.1}^{+0.2}$ & {3.4} & {-21.8}& $49.5_{-3.1}^{+3.4}$ & $2.6_{-0.2}^{+0.3}$ & $1.8_{-0.1}^{+0.1}$ & $0.3_{-0.1}^{+0.1}$ & {-16.0} & $6.6_{-0.2}^{+0.2}$ & $21.8_{-0.8}^{+0.8}$ & $2.3_{-0.5}^{+0.7}$ & {-44.9}\\
 Cepheus & $2.8_{-0.6}^{+0.1}$ & $41.6_{-1.4}^{+8.7}$ & $10.9_{-1.6}^{+1.4}$ & $6.7_{-0.9}^{+1.9}$ & $0.6_{-0.1}^{+0.2}$ & {3.9} & {-25.1}& $159.1_{-12.6}^{+8.0}$ & $1.0_{-0.0}^{+0.1}$ & $1.6_{-0.0}^{+0.0}$ & $0.2_{-0.0}^{+0.1}$ & {-6.5} & $3.9_{-0.2}^{+0.2}$ & $38.9_{-2.8}^{+2.7}$ & $6.5_{-1.4}^{+2.1}$ & {-61.0}\\\cline{1-17}
  Orion A* & $3.1_{-0.6}^{+0.2}$ & $48.0_{-14.4}^{+19.7}$ & $8.7_{-1.0}^{+3.3}$ & $7.0_{-3.5}^{+2.3}$ & $0.4_{-0.1}^{+0.1}$ & {2.8} & {-18.9}& $80.1_{-34.0}^{+3.4}$ & $4.2_{-0.3}^{+0.6}$ & $3.3_{-0.4}^{+0.3}$ & $0.4_{-0.1}^{+0.1}$ & {-17.7} & $3.9_{-0.1}^{+0.1}$ & $46.9_{-1.3}^{+1.5}$ & $2.1_{-0.5}^{+0.6}$ & {-43.4}\\
  Orion B* & $3.8_{-0.1}^{+0.1}$ & $35.5_{-1.2}^{+1.1}$ & $11.1_{-0.7}^{+0.8}$ & $11.7_{-1.3}^{+1.3}$ & $0.6_{-0.1}^{+0.2}$ & {2.9} & {-25.3}& $50.8_{-1.1}^{+1.1}$ & $5.4_{-0.3}^{+0.3}$ & $2.3_{-0.1}^{+0.1}$ & $0.5_{-0.1}^{+0.2}$ & {-24.4} & $6.0_{-0.2}^{+0.2}$ & $39.9_{-1.4}^{+1.5}$ & $5.0_{-1.2}^{+1.6}$ & {-57.0}\\
$\lambda$ Orionis* & $3.8_{-0.1}^{+0.1}$ & $30.3_{-0.7}^{+0.6}$ & $13.9_{-1.2}^{+1.6}$ & $5.9_{-0.6}^{+0.7}$ & $0.3_{-0.1}^{+0.1}$ & {3.7} & {-14.3}& $85.1_{-11.7}^{+22.9}$ & $1.2_{-0.3}^{+0.2}$ & $1.3_{-0.0}^{+0.0}$ & $0.2_{-0.0}^{+0.1}$ & {-8.7} & $5.7_{-0.3}^{+0.4}$ & $28.0_{-6.0}^{+2.7}$ & $3.5_{-0.8}^{+1.1}$ & {-50.5}
\enddata
\tablecomments{Radial density profile fitting results for local clouds, computed using the likelihood function shown in Equation \ref{eq:samples_likelihood} and described in Appendix \ref{appendix_samples}.  The likelihood in Equation \ref{eq:samples_likelihood}  utilizes all realizations of the 3D dust map distribution from \citet{Leike_2020}, rather than just the mean distribution, allowing us to quantify the effect of the underlying uncertainty in the 3D dust reconstruction on our results. (1) Name of the cloud. (2 - 6) Best-fit parameters for the two-component Gaussian fit, including the standard deviation and amplitude of the narrower Gaussian ($\sigma_1$, $a_1$), the standard deviation and amplitude of the broader Gaussian ($\sigma_2$, $a_2$), and the square of the modeled scatter in the density measurements ($\sigma_{n, \rm G2}^2$). (7) Ratio of the outer to inner Gaussian standard deviations.  (8) Logarithm of the evidence of the two-component Gaussian fit. (9-12) Best-fit parameters of the Plummer fit, including the peak amplitude $n_0$, the flattening radius $R_{\rm flat}$, the power index of the density profile $p$, and the square of the modeled scatter in the density measurements ($\sigma_{n,\rm P}^2$). (13) Logarithm of the evidence of the Plummer fit. (14-16) Standard deviation $\sigma$, amplitude $a$, and the square of the modeled scatter in the density measurements ($\sigma_{n,\rm G1}^2$) for the single component Gaussian fit. (17) Logarithm of the evidence of the single-component Gaussian fit. We report the single-component Gaussian results here for completeness only, as it is a poor fit to the data particularly at large radial distances. A machine readable version of this table is available at \url{https://doi.org/10.7910/DVN/94OKZD}.}
\tablenotetext{a}{At small radial distances, Musca is largely unresolved in the \citet{Leike_2020} map, and at large radial distances, its structure is consistent with being prior-dominated, so we urge caution when interpreting the results.} 
\tablenotetext{*}{Clouds below the horizontal line -- Orion A, Orion B, and $\lambda$ Orionis -- should be treated with caution, as they lie at the very edge of the \citet{Leike_2020} 3D dust grid and are subjected to additional biases. Part of Orion A, toward the Orion Nebula Cluster, is also devoid of dust in the \citet{Leike_2020} 3D dust map despite being highly extinguished. See \S \ref{exposition} for full details.}

\end{deluxetable}
\end{rotatetable}